\documentclass[12pt,preprint]{aastex}
\usepackage{color}
\usepackage{natbib}
\usepackage{caption}
\def\halpha{\ifmmode {\rm H}\alpha \else H$\alpha$\fi}
\def\civ{\ifmmode {\rm C{\sc iv}} \else C~{\sc iv}\fi}
\def\gsim{\lower 2pt \hbox{$\, \buildrel {\scriptstyle >}\over
{\scriptstyle \sim}\,$}}
\def\lsim{\lower 2pt \hbox{$\, \buildrel {\scriptstyle <}\over
{\scriptstyle \sim}\,$}}

\slugcomment{Submitted to ApJ}

\shortauthors{Bonning et al.}
\shorttitle{SMARTS/Fermi Cycle 1 paper}

\begin{document}

\title{SMARTS Optical and infrared monitoring of 12  gamma-ray
bright blazars}

\author{E.~W. Bonning\altaffilmark{1},
C. M. Urry\altaffilmark{1},
C. Bailyn\altaffilmark{2}, 
M. Buxton\altaffilmark{2}, 
R. Chatterjee\altaffilmark{2},
P. Coppi\altaffilmark{2}, 
G. Fossati\altaffilmark{4}, 
J. Isler\altaffilmark{1},
L. Maraschi\altaffilmark{3}}

\altaffiltext{1}{Department of Physics and Yale Center for Astronomy
and Astrophysics, Yale University, PO Box 208121, New Haven, CT 06520-8120;
erin.bonning@yale.edu}
\altaffiltext{2}{Department of Astronomy and Yale Center for Astronomy
and Astrophysics, Yale University, PO Box 208101, New Haven, CT 06520-8101}
\altaffiltext{3}{INAF - Osservatorio Astronomico di Brera, V. Brera 28, I-20100 Milano, Italy}
\altaffiltext{4}{Department of Physics and Astronomy, Rice University, Houston, TX 77005}

\begin{abstract}

We present multiwavelength data for twelve blazars observed from
2008-2010 as part of an ongoing optical-infrared photometric
monitoring project. Sources were selected to be bright, southern
($\delta < 20^\circ$) 
blazars observed by the {\em Fermi} Gamma-Ray Space Telescope, with
daily and weekly gamma-ray fluxes made available from the start of the
{\em Fermi} mission. Light curves are presented for the twelve blazars
in BVRJK (0.4 to 2.2 $\mu$m) at near-daily cadence.  We find that
optical and infrared fluxes are well correlated in all sources, with
no measured lag between bands. Gamma-ray bright flat spectrum radio quasars
(FSRQs) in our sample have optical/infrared emission correlated with
gamma-rays consistent with inverse
Compton-scattering leptonic models for GeV emission. In FSRQs, the
variability amplitude decreases towards optical/IR wavelengths,
consistent with the presence of a thermal emission component from the
accretion disk varying on significantly longer timescales than the jet
synchrotron emission.  In BL Lac objects, variability is mainly
constant across wavelengths, consistent with a weak or radiatively
inefficient disk. Also consistent with this picture, FSRQs have redder
optical-infrared colors when they are brighter, while BL Lac objects
show no such trend. Several objects show complicated color-magnitude
behavior: AO~0235+164 appears in two different states depending on
whether it is gamma-ray bright or not. OJ~287 and 3C~279 show some
hysteresis tracks in their color-magnitude diagrams. Individual flares
may be achromatic or otherwise depart from the trend, suggesting
different jet components becoming important at different times. We
present a time-dependent spectral energy distribution of the bright
FSRQ 3C~454.3 during December 2009 flare, which is well fit by an
external Compton model in the bright state, although day to day
changes in the course of the flare pose challenges to a simple
one-zone model. All data from the SMARTS monitoring program are
publicly available on our website.

\end{abstract}

\keywords{galaxies: active --- quasars: general --- black hole
physics --- BL Lacertae objects}

\section{Introduction}
\label{sec:intro} 
Blazars form a sub-class of active galactic nuclei (AGN) with a
bright, relativistic jet viewed closely along our line of sight
\citep{urry95}. Blazars are often very luminous and violently variable
over a large range of wave bands from radio to gamma-rays. Spectral
energy distributions (SEDs) of blazars are characterized by two broad
components: one peaking anywhere from infrared to X-ray frequencies,
and a second peak at higher energies, from hard X-rays to TeV gamma-rays.
The radio to optical/UV emission
in blazars is interpreted as synchrotron radiation by the energetic
electrons in the jet \citep{konigl81,urry82} while the mechanism of
the high energy emission (i.e., X-rays and gamma-rays) is less
certain. It may be due to the inverse-Compton scattering of seed
photons by the same relativistic electrons responsible for the
synchrotron radiation
\citep[the so-called leptonic models; e.g.][]{bot07} or due to
synchrotron radiation of protons co-accelerated with the electrons in
the jet, interactions of these highly relativistic protons with
external radiation fields, or proton-induced particle cascades
\citep[hadronic models; e.g.][]{mucke01,mucke03}.

Both leptonic and hadronic models can successfully explain the SEDs
observed so far, but they have very different implications for the
kinetic power of the jet and hence how it is produced and its influence
on its environment. In the case of leptonic models,
the low-energy seed photons may be the synchrotron photons produced
within the jet \citep[synchrotron self-Compton, or SSC; ][]{jones74} or
thermal emission from outside the jet. External Compton (EC)
scenarios produce high energy emission from upscattering photons from
the accretion disk, broad line region (BLR), or dusty torus
\citep{sikora94, dermer93, ghisellini96, tavecchio00, ghiscelotti01}. 
The jet plasma may consist of electrons and protons, electrons and
positrons, or a combination of the two
\citep{ghistavecchio10, sikora00}.

Knowing the composition of the jet is necessary to deduce its kinetic
power, which in turn reflects how it is launched, accelerated, and
collimated. Studying the SED and its variation with time allows us to
determine the radiation mechanism, and thus the physical parameters of
the emission region, such as the magnetic field, particle number
density, and bulk velocity of the plasma.

Until recently, blazar SED studies occurred primarily when the
brightness of a blazar significantly increased in one or multiple wave
bands. Due to the difficulty of coordinating large multiwavelength
campaigns, data in other wavebands were often non-simultaneous. The
near-continuous monitoring activity of the Large Area Telescope (LAT)
instrument on board the {\em Fermi Gamma-Ray Space Telescope},
launched in 2008, provides the opportunity to study the variable SEDs
of a large sample of blazars with truly simultaneous multi-frequency
data. Although many blazars radiate most of their energy in the
gamma-ray band, it is the characterization of both broad components
of the blazar SEDs and their relative variation with time that allows
us to infer the physics of these sources. Specifically, correlations
between variations in gamma-ray flux and those at lower energies
are useful indicators of the relative locations of emission regions
and the radiation mechanism(s).

{\em Fermi}/LAT provides regular and well sampled gamma-ray light
curves of a large sample of blazars. Obtaining the same quality data
at optical and infrared wavelengths is equally important. We use the
meter-class telescopes of the Small and Moderate Aperture Research
Telescope System (SMARTS) to carry out photometric monitoring of 
bright southern gamma-ray blazars on a regular cadence, at both
optical and near-infrared wavelengths.

In this paper, we report on the Yale/SMARTS blazar monitoring program
for the years 2008-2010. We present light curves for the blazars
with the greatest coverage, and refer the reader to our 
website\footnote{http://www.astro.yale.edu/smarts/fermi} for the full data
set. In addition, we report on gamma-ray/optical-IR
cross-correlation functions for several blazars and show an example of
how a variable SED has the potential to constrain the physics of a jet
and/or accretion disk. Finally, we discuss the color-magnitude
diagrams for blazars and how these can yield broad inferences about
particle acceleration and radiative losses in blazar jets.

In {\S}~\ref{sec:obs} we present the sample selection and data
reduction and present the multi-wavelength light curves. In
{\S}~\ref{sec:results},  we 
discuss the variability characteristics of the blazars in our sample,
including cross-correlations between gamma-ray and optical-IR
variations, frequency-dependent variability amplitude, and
color-magnitude relations for our sample of blazars. We discuss SED
fits for a sample blazar flare in {\S}~\ref{sec:sed}.
In {\S}~\ref{sec:conclusions} we present the summary and
conclusions.

\section{Sample Selection and Data Analysis}
\label{sec:obs}

The SMARTS blazar sample was initially (in 2008) defined to include
all LAT-monitored blazars on the initial public release list with
declination $< 20^\circ$.  Prior to the launch of Fermi, the list of
sources which were to have fluxes made publicly available consisted
mainly of bright blazars observed by EGRET. Additional sources were
added to our monitoring campaign as they were added to the public LAT
source list or were the target of a multi-wavelength campaign. We
observed the twelve sources presented here with a cadence of
approximately once every three days. Brighter or flaring sources were
observed nightly. The SMARTS source list, including positions,
redshifts, and observation timeframe, is given in Table~1. Each of the
blazars is identified by its spectral class: flat-spectrum radio
quasar (FSRQ), low frequency-peaked BL Lac object (LBL), or high
frequency-peaked BL Lac (HBL). These form a rough sequence of
decreasing continuum luminosity and emission line luminosity
\citep{fossati98}. The most luminous blazars (FSRQ and LBL) have their
synchrotron peak at IR/optical wavelengths and the inverse Compton
peak in GeV gamma-rays so they form the bulk of our monitored sample.

The 12 blazars were observed with the SMARTS 1.3m telescope and
ANDICAM instrument \citep{depoy03}. ANDICAM is a dual-channel imager
with a dichroic that feeds an optical CCD and an IR imager, which can
obtain simultaneous data from 0.4 to 2.2 $\mu$m. Observations were
taken in B, V, R, J, and K bands, except for two sources: PKS~0528+134
and 3C~273, for which K-band images were not obtained. The former was
below the K-band detection limit; the latter, being very bright, had
very short exposure times in optical bands. Since IR images are taken
simultaneously with optical, there was only sufficient time for J-band
images to be obtained for 3C~273. Additionally, spectra were obtained
for a number of the brighter sources, using the SMARTS
RCSPEC+1.5-meter telescope.  These data will be discussed in
forthcoming paper (Isler et al. {\em in prep}).

Optical data were bias-subtracted, overscan-subtracted, and flat
fielded using the CCDPROC task in IRAF. Infrared data were
sky-subtracted, flat fielded, and dithered images combined using
in-house IRAF scripts. Slight blemishes reflecting the dither pattern
are apparent in the final images, but introduce at most a 0.1\%
photometric error.

Optical and infrared aperture photometry was performed using the PHOT
task in IRAF.  Non-variable comparison stars with comparable
magnitude to the blazar were chosen in each field.  The raw
photometry of comparison stars in the field of 
the blazar was calibrated using photometric zeropoints that were
measured from ANDICAM observations during 2008-2009 of optical
\citep{landolt92} and near-infrared \citep{persson98} primary standards for
each filter, correcting for atmospheric extinction derived from all
the standards taken together. The number of photometric nights available
for the calibration for each field and each filter differs, but ranges
from 40-128 nights in the optical, and 13-98 nights for the
near-infrared. The average of the comparison stars was used as a
basis of differential photometry with respect to the blazar for all
observations. 

We compared our calibrated comparison star magnitudes to values
reported in the literature where available. In the large majority of
cases, our values agree with previous photometric sequences measured
for these fields to within a typical 1-$\sigma$ error of $\sim$0.05
mag with an occasional discrepancy of $\sim$0.1 mag. Our optical
magnitudes for the comparison stars in the field of AO~0235+164 are
consistent with values reported by \citet{smith85},
\citet{fiorucci98}, and \citet{gonzalez01}. In the field of
PKS~0528+134, our comparison star magnitudes match extremely well with
those published by \citet{gonzalez01} with the exception of $V$-band
for star 1, where we differ by 0.1 mag. When compared to the values
published by \citet{raiteri98}, we
agree in $R$-band; however our $B$ and $V$ magnitudes are fainter by 0.1 mag in B
and 0.2 mag in V compared to those of Raiteri et al. In the case of
OJ~287, our comparison stars are in 
good agreement with \citeauthor{gonzalez01}, with the largest
descrepancy being $\sim$0.07 mag.  Likewise, our $V$ and $R$
measurements for the  
comparison stars for OJ~287 are in close agreement with those measured
by \citet[][,who do not report $B$-band.]{fiorucci96} The comparison
star of 3C~273 is within
uncertainties reported by \citet{smith85}. 3C~279 is in close 
agreement with the values published by \citet{smithbalonek98} with
$R$-band being the only one to differ by more than 3-$\sigma$ (0.1
mag). Our comparison stars for
PKS~1510-089 generally agreed within the uncertainities with the values
published by \citet{gonzalez01}, with the exception of $B$-band for
our star 1, which differed by 0.1 mag. In the field of
PKS~1622-297, our stars 1 and 2 are stars 10 and 14 of
\citet{gonzalez01}, and they are in excellent agreement in $V$ and $R$
band. (\citeauthor{gonzalez01} do not report $B$-band for this
source).  The comparison star in our field for PKS~2155-304 is in
excellent agreement with the photometry of \citet{hamuy89}. 
In the case of 3C~454.3, comparison star magnitudes have been
reported for this well-studied source also by \citet{angione71} and 
\citet{fiorucci98}. We agree extremely well with the results of
\citet{fiorucci98} in $V$ and $R$, and again find a $\sim$0.1 mag
(fainter) discrepancy with \citet{raiteri98} in $B$, a discrepancy
which only occurs for some stars reported by \citet{angione71}.

In general, we find that in those sources with multiple photometric
sequences available (PKS1510-089, 3C~273, 3C~454.3), published
comparison star values frequently differ on the level of $\sim$0.1
magnitude. This is consistent with a comparison of our data to the
literature.  For our remaining sources, PKS~0208-512, PKS~1406-076,
and PKS~1730-130, existing photometric values for comparison
stars (given, for example, by the finding charts at
http://www.lsw.uni-heidelberg.de/projects/extragalactic/charts) are
taken from the USNO B1.0 catalog \citep{monet03}. Our calibrated
magnitudes are generally fainter than USNO 
by 0.2-0.4 mag.  The USNO B1.0
catalog was designed to be a proper-motion catalog. Although
photometry was obtained, it has an accuracy of 0.3 mag \citep{monet03}
which is significantly greater than the errors found for our
photometry.

The optical and IR light curves for the monitored blazars through July
2010 are shown in Figure~1.  (At the time of writing, some third-year
SMARTS data are already available online but the basic scientific
results do not change for the 12 blazars discussed here.)  These
figures include daily and weekly gamma-ray fluxes from the 
{\em Fermi}/LAT public light 
curves for those sources bright enough in the gamma-ray that these are
measureable\footnote{List of {\em Fermi}/LAT monitored sources is at
http://fermi.gsfc.nasa.gov/ssc/data/access/lat/msl\_lc/}. The optical
and infrared light curves and 
calibrated magnitudes are available online at the Yale SMARTS Blazar
site$^{5}$. Optical and infrared finding charts are shown in 
Appendices A and B. Calibrated magnitudes for the comparison stars in
each blazar field are given in Appendix C.

The error in calibrating the secondary star magnitudes was found by
calculating the 1-$\sigma$ standard error of the mean over the number
of photometric nights mentioned above. Results that were greater than
$\pm$ 3$\sigma$ from the mean were rejected and the mean and $\sigma$
were recalculated. This procedure was repeated until no more
rejections were made. The resulting errors are given in Tables
C.1. and C.2. These errors do not account for systematic errors
associated with effects such as the difference in effective filter
responses between SMARTS and the standard system. Such systematics
are likely to contribute a few hundredths of a magnitude of
calibration error.

Previous work with SMARTS photometry (Buxton et al. 2011, {\em submitted})
has led to an understanding of the errors in differential photometry
for point sources as a function of count rate.  Based on this work,
we find errors as low as 0.01 magnitudes in the optical and 0.02
magnitudes in the IR for bright sources ($<$16 mag in the optical
and $<$13 mag in IR), and up to ten times that for sources near the
detection limit, which varies depending on the exposure time.  We note
that these errors are random errors in the individual points, rather
than systematic offsets in the magnitude system, as is the case for
the calibration errors.

\newcommand{\newfootnotemark}[1]{\footnote{addtocounter}{#1}
\footnotemark[\value{footnote}]}
\newcommand{\newfootnotetext}[2]{\addtocounter{footnote}{#1}
\footnotetext[\value{footnote}]{#2}}

\begin{table}[htbp]
\hspace{-3mm}
 \begin{tabular}{lcccc}
\hline
 Name & R.A.~\&~Dec.\ (J2000) & Class & $z$ & Dates \\
      & [h:m:s]~~~[d:m:s] & & & (MJD)\\
\hline
PKS\, 0208$-$512    &  02:10:46.2   $-$51:01:01  & FSRQ & 1.003 & 54640
-- 55282\\
AO\,0235+16     &  02:38:38.9   +16:36:59  & LBL  & 0.940 & 54662 -- 55487\\
PKS\,0528+134   &  05:30:56.4   +13:31:55  & FSRQ & 2.060 & 54701 -- 55511\\
OJ\,287         &  08:54:48.8   +20:06:31  & LBL  & 0.306 & 54777 -- 55500\\
3C\,273         &  12:29:06.7   +02:03:09  & FSRQ & 0.158 & 54603 -- 55412\\
3C\,279         &  12:56:11.1 $-$05:47:21  & FSRQ & 0.536 & 54603 -- 55416\\
PKS\,1406$-$076   &  14:08:56.5 $-$07:52:27  & FSRQ & 1.494 & 54501 -- 55385\\
PKS\,1510$-$08   &  15:12:50.5 $-$09:06:00  & FSRQ & 0.360 & 54603 -- 55464\\
PKS\,1622$-$29    &  16:26:06.0 $-$29:51:27  & FSRQ & 0.815 & 54501 -- 55481\\
PKS\,1730$-$130       &  17:33:02.6 $-$13:04:49  & FSRQ & 0.902 & 54603 -- 55494\\
PKS\,2155$-$304   &  21:58:52.0 $-$30:13:32  & HBL  & 0.112 & 54603 -- 55557\\
3C\,454.3       &  22:53:57.7   +16:08:54  & FSRQ & 0.859 & 54640 -- 55545\\
\hline
  \end{tabular}
 \label{tab:sample}
\caption{The SMARTS  blazar sample. Coordinates and
redshifts from NED$^{7}$. Dates indicate limits of
$R$-band observations reported here. Other bands may have slightly
different date ranges. Full data are available online.}
\end{table}
\newfootnotetext{1}{http://ned.ipac.caltech.edu/ The NASA/IPAC
Extragalactic Database
(NED) is operated by the Jet Propulsion Laboratory, California
Institute of Technology, under contract with the National Aeronautics
and Space Administration.}

\section{Results and Discussion}
\label{sec:results}

\subsection{Multiwavelength Cross-Correlation Results}
\label{ssec:cc}

Every blazar in our sample is highly variable in gamma-rays and, in
many cases, in the optical/IR as well; only 3C273 shows minimal
variations in the optical/IR.  The optical, near-infrared and
gamma-ray light curves are shown in Figure 1.

 We investigate the correlation between optical and near-IR bands
through the discrete correlation function 
\citep[DCF,][]{edelson88} with the corrections of 
\citet{white94}. Figure 2 shows the $B-J$ band DCF
for each source in our sample. (We also show the $R-J$ band DCF for
PKS~0528+134. This is the only source for which the DCF was
significantly different in shape between bands.) Examining the DCFs
for each source we find that, apart from 3C~273 which has no strong
variations, all DCFs have a peak at zero lag. No source shows a
significant peak at any non-zero lag.  Note that given our daily
cadence, we are sensitive only to lags longer than a couple of days.
Figure 3 shows two examples of the correlation between optical
($B$-band) and IR ($J$-band) for two blazars: PKS~2155-304 (an HBL,
Fig. 3a), and 3C~279 (an FSRQ, Fig.~3b).  Fig.~3b also shows
distinct tracks that reflect changes in the optical/IR spectral shape
of 3C~279 over time ({\em cf.} Fig. 5b).

Six blazars --- four FSRQs, one LBL, and one HBL --- were bright
enough during 2008--2010 to be regularly detected by Fermi LAT in
one-day time bins.  The gamma-ray - infrared discrete correlation
functions for the these sources are shown in Fig.~4).  In three cases,
the optical-IR variability has been shown to be reasonably
well-correlated with the gamma rays: 3C454.3
\citep{bonning09}, PKS~1510-089 \citep{marscher10}, and AO~0235+164
\citep{agudo11}.   The FSRQs 3C~273 and 3C~279 show weak (if any) correlations
between infrared and gamma-rays, while the optical/IR emission of the
HBL PKS~2155-304 is clearly uncorrelated with gamma-ray emission. 
The strongly disk-dominated FSRQ 3C~273, shows little to no
variability in the optical/IR bands (fluctuations on the order of 1\%)
compared to the variations by a factor of several to 10 in gamma-rays.
The small fractional variation at optical/IR frequencies in 3C~273 may
be due to the relatively large luminosity of the accretion disk
\citep{ramos97}, since the disk probably does not vary in the
optical/IR over time scales as short as the gamma rays.  For 3C~279,
optical/IR-gamma ray correlations have been reported by
\citet{abd10_279} who note the coincidence of a gamma-ray flare with a
strong change in optical polarization angle. However, our SMARTS data
show larger amplitude changes (by several orders of magnitude) than
the gamma-ray data, over both the first and second years of {\em
Fermi} observations.  The one HBL with high enough gamma-ray flux to
appear consistently in the LAT daily fluxes, PKS~2155-304, shows
variations of a factor of 2-3 in gamma-rays, along with
longer-timescale variations in the optical-IR, but the two appear to
be uncorrelated.

The close correlation of optical/IR and gamma-ray fluxes strongly
favors leptonic models over hadronic models. In the latter, the two
broad components of the SED should vary almost independently, while in
leptonic models, the two peaks are should vary together.  In the
FSRQs, the peak of the low energy component of the SED is in the
far-IR, longward of the SMARTS bands, while the high-energy component
peaks in or near the Fermi LAT energy range. In terms of simple
homogeneous leptonic models, the SMARTS data generally sample emission
from higher energy electrons than the Fermi LAT data. \citep[this can
obviously change if the SED peak moves, as in a major acceleration
event, e.g.][ for the case of Mrk 501]{pian98} More realistic electron
distributions and/or geometries will complicate this picture, but in
any case, as discussed in Section 4, there are multiple plausible
reasons for variations in the strength of the optical-IR versus
gamma-ray correlations among the FSRQs in our sample.

\subsection{Excess Variance}

Most blazars are strongly variable at gamma-ray energies and in the
optical/IR.  To quantify and compare the degree of variability in
different blazars, and in different bands for the same blazar, we
calculate the ``excess variance," which is the fractional
root-mean-squared variability amplitude normalized by the mean flux:
\begin{equation}
F_{\rm var}=\sqrt{{\frac{\rm S^2-\overline{\sigma_{\rm err}^2}}{\overline{x}^2}}}~~ ,
\end{equation}
where $\rm S$ is the variance, $\sigma_{\rm err}$ is the observational
uncertainty, and $\overline{x}$ is the mean of the data
\citep{nandra97,edelson02,vaughan03}. 
The values of $F_{\rm var}$ calculated for SMARTS and Fermi LAT data 
are listed in Table 2.
In cases where $K$-band observations did not cover
the same time periods as the other filters, we omit the $K$-band
values.

The IR/optical/UV emission from blazars is a combination of thermal
emission from a (possibly weak) accretion disk and relativistically
beamed nonthermal emission from a jet. The disk emission peaks in the
UV band so its relative contribution in optical/IR bands decreases
toward longer wavelengths. Since the disk emission comes from $\gtrsim
10^4$ r$_{g}$ \citep{malkan83,sunmalkan89}, it is almost certainly
less rapidly variable than from the jet.   Therefore, on the
timescales considered here, we expect the variability amplitude for
FSRQs (blazars with strong emission lines implying a radiatively
efficient accretion disk) in the optical/IR to increase with
wavelength, i.e., infrared bands to vary more than the optical. In
general, this effect is seen in Table
\ref{tab:fvar}. For the FSRQs, J-band is more variable than B-band in
all objects except 3C~279, in which the trend is reversed. In Section
\ref{sec:colmag} and Figure 4f, it is shown that the changes in
spectral shape of 3C~279 are also not consistent with a simply varying
jet component overlaid on a constant blue accretion disk. The BL Lac
objects do not show greater variability towards the IR: the two LBLs
are more variable towards the B-band, and the HBL PKS~2155-304 shows
no frequency-dependent variability changes.  BL Lacs, having weak
emission lines, are likely to have weakly accreting or radiatively
inefficient disks, so it is not surprising that they do not behave as
the majority of our FSRQs. However, this is a small sample, which
includes the unusual source AO~0235+164 in which broad emission lines
have been observed, the strength of which calls into question its
status as an LBL \citep{cohen87, nilsson96}. Nevertheless, the SED of
AO~0235+164 shows a strong and variable UV component in addition to the
synchrotron and inverse-Compton peaks \citep{raiteri08c}, which may
explain its variability behavior observed here.

\begin{table}[htbp]
\hspace{-3mm}
   \begin{tabular}{llcccccc}
\hline
 Name & Class & K & J & R & V & B & $\gamma$ \\ \hline
PKS\, 0208-512 & FSRQ	& 0.48 & 0.52 & 0.45 & 0.40 & 0.43 & -- \\
AO\,0235+16    & LBL  & 1.10  & 1.22 & 1.33 & 1.31 & 1.33 & 0.38\\
PKS\,0528+134  & FSRQ &  --    & 0.53 & 0.28 & 0.24 & 0.35 & -- \\
OJ\,287        & LBL &  0.45 & 0.37 & 0.43 & 0.45 & 0.46 & -- \\
3C\,273        & FSRQ &  -- & 0.13 & 0.041 & 0.044 & 0.042 & 0.77 \\
3C\,279        & FSRQ & 0.49  & 0.79 & 0.82 & 0.87 & 0.86 & 0.38 \\
PKS\,1406-076  & FSRQ &  --  & 0.61 & 0.48 & 0.49 & 0.46 & -- \\
PKS\,1510$-$08 & FSRQ & --   & 1.04 & 0.94 & 0.94 & 0.83 & 0.56 \\
PKS\,1622-29   & FSRQ &  0.68 & 0.54 & 0.36 & 0.29 & 0.27 & -- \\
PKS\,1730-130  & FSRQ &  0.58 & 0.47 & 0.37 & 0.34 & 0.35 & -- \\
PKS\,2155-304  & HBL &  0.28 & 0.33 & 0.33 & 0.33 & 0.33 & 0.25 \\
3C\,454.3      & FSRQ &  0.68 & 0.58 & 0.51 & 0.47 & 0.44 & 1.40\\
\hline
   \end{tabular}
\caption{Fractional variability amplitude (as defined in Eqn.(1) for observed IR, optical, and
gamma-rays. The gamma-ray values are listed only for blazars bright
enough to be detected in daily {\em Fermi}/LAT averages. }\label{tab:fvar}
\end{table}

\section{Color-magnitude variations}
\label{sec:colmag}

We present in Figure 4 the relation between IR ($J$-band) magnitude
and optical-IR spectral shape of the low-energy peak of the SED (given by $B-J$
color, such that a {\em larger} value corresponds to a {\em redder} color).
Overall, luminous blazars (FSRQs) are redder when brighter and bluer
when fainter. A typical example is shown in Fig.~4i (PKS~1622-29). As
the source varies in brightness, the color changes along a narrow
locus in the color-magnitude plane, such that when the source is
brighter, the color gets redder. Conversely, when the source is
fainter, the color gets bluer (although the SED remains
synchrotron-dominated and therefore red). This 'flattening' of the SED
in faint states suggests a strong, blue accretion disk component is
mixed with redder jet emission, which may only be visible as a typical
'blue bump' in extremely faint states
\citep[e.g. as observed in 3C~454.3 by][]{raiteri07}.  Most of our FSRQs
(with exceptions noted below) 
follow this redder-when-brighter track, consistent with FSRQs having
luminous accretion disks as also evidenced by their typically strong
emission lines. 

Although the FSRQs mostly show redder-when-brighter trends, they can
also show more complicated behavior.  For example, the May 2010 flare
in PKS~1510-089 (indicated by arrows in Fig.~4h) was essentially
achromatic and, unlike its other flares, had a much larger amplitude
than the associated gamma-ray flare (cf. Fig.~1h). Such anomalous
flares indicate the interplay of distinct components in the source; in
the case of PKS~1510-089, the sharp achromatic flare could be caused
by a plasma blob hitting a transverse shock in the jet
\citep{marscher10}.  (More speculatively, achromatic flares could be
caused by gravitational microlensing in the halo of our galaxy, as in
the MACHO experiment \citep{alcock00}, since these blazar jets are
effectively point sources, but the probability of such an event should
be vanishingly small.) Similarly, the December 2009 flare of 3C~454.3
(Fig.~4l) moved off the standard red-bright track, although it was not
as obviously achromatic.

In contrast to FSRQs, the lower luminosity BL Lac objects do not show
the same color-magnitude trend, consistent with BL Lacs having weak
accretion disks. The HBL PKS~2155-304 is shown in Fig.~4k;
its variations are more random with respect to color, and smaller in
amplitude.

Somewhat more complicated is the color behavior of the LBL OJ~287
(Fig.~4d), which shows some redder-when-brighter changes but also
variations that are bluer-when-brighter, and still other color changes
at fixed magnitude.  OJ~287 is known for a striking series of
double-peaked outbursts occuring approximately every 12 years over the
last century \citep{sillanpaa88,lehtovalt96}. Our
observing period begins after the last double outburst in
2005-2007. We find that OJ~287 is brighter and more variable post-burst
than it was between the 2005 and 2007 peaks. As it was during the
double-peaked bursts, OJ~287 remained highly polarized during the
early 2009 observing season \citep[see Figure 1 of ][]{villforth10},
with a strong, stably, underlying optically polarized core. Villforth
et al. also found an overall bluer-when-brighter color trend over the time
period 2005-2009. Additionally, observations in 1993-1994 showed
constant optical colors over a range in optical flux
\citep{sillanpaa96, hagenthorn98}. They suggest that these changes in spectral
shape are due to injections of high energy electrons into the jet
emitting region. Alternatively, it may be that the accretion disk
component in OJ~287 is variable and comparable in magnitude to the jet
component. This is suggested by analogy to X-ray binaries, in which
this complicated color-magnitude behavior is a known phenomenon, and
hysteresis in the X-ray flux--spectral index plane has been
interpreted in terms of accretion state changes \citep{smith07,
maccarone03}.  (Due to their hotter accretion disks, the thermal
peak in X-ray binaries lies in the soft X-ray, compared to the
rest-frame UV for LBLs and FSRQs.)  On the other hand, hysteresis
behavior has also been been reported for the HBL PKS~2155-304
\citep{kataoka00}, in which the X-rays come from the jet and no disk
has been detected. This hypothesis can be tested by observing changes
in emission lines; if the disk is varying, the lines should respond in
intensity. On balance, it is not obvious what causes the unusual
color-magnitude trends in OJ~287. 

Still more anomalous behavior is exhibited by AO~0235+164
(Fig.~4b). At the beginning of our monitoring
program, for example, increasing intensity was characterized by bluer
(not redder) colors. This was also when AO~0235+164 was detected by
Fermi in the daily light curve (i.e., it was gamma ray-bright, cf. Fig.~1b).
Later, when the source was much fainter in gamma rays, the
color-magnitude trend reverted to the usual redder-when-brighter
relation. 

 AO~0235+164 is an unusually situated source. An AGN sits
2$^{\prime\prime}$ to the south of the blazar as well as an
intervening galaxy at 
$z=0.525$ \citep{yanny89, burbidge96, nilsson96}, which we do not
resolve in our imaging. The AGN has a   
fairly blue spectrum ($B=21.4$,$V=20.9$,$R=20.5$) which has not been
observed to vary strongly \citep{raiteri05}.  The extinction due to the
intervening galaxy is nearly a magnitude greater than the Galactic
extinction \citep[$A_B=1.268$ vs $0.341$, ][]{junkkarinen04}, resulting
in a very intrinsically red spectrum.  We removed the
contribution from this AGN in the B band before plotting the
color-magnitude relation in Fig.~4b, so the bluer-when-fainter
behavior of AO~0235+164 is not due to the underlying blue colors of
the nearby AGN.

If the observed optical-IR emission is a combination of jet and
accretion disk, and if the gamma rays are produced by inverse Compton
scattering of disk or line photons, then the two trends can be
explained self-consistently. When the jet becomes brighter, the color
of the combined emission gets redder since the jet synchrotron
emission is intrinsically redder than that of the blue accretion disk.
Alternatively, if the jet is constant and the disk emission increases,
then the summed emission becomes bluer. Moreover, an increase in disk
emission would be accompanied by a strong increase in gamma rays, due
to the increased scattering of disk/line photons, hence the
association of a bright gamma-ray state with the bluer-when-brighter
trend. In addition, when the jet is quite bright ($J\gsim$14 in this
case), further brightening of the jet may be due to emission from an
even higher energy population of electrons; this would also result in
a bluer color. 3C~279 (Fig. 4b) has a very similar (i.e., hybrid)
color-magnitude diagram, roughly achromatic in the optical when bright
in gamma rays but much fainter in gamma rays when it moves along the
more usual FSRQ bluer-when-fainter trend.

In the examples of AO~0235+16 (Fig. 4b) and 3C~279 (Fig. 4f),
different loci in the optical/IR color-magnitude diagram are
associated with different gamma-ray intensities.  This bimodal
behavior has some similarity to the Galactic X-ray binaries. 
For example, the X-ray binary GX 339-4 has been observed to
change from a bluer-when-brighter state (associated with the high/soft
X-ray state) to a redder-when-brighter state (associated with the hard
X-ray state --- see Buxton et al. 2011 submitted). 
X-ray binary emission is a combination of soft X-rays from a disk and
hard X-rays from a power-law component (likely from a hot corona).  In
the canonical ``high-soft" state, the X-ray spectrum is disk-dominated
and hence, softer, but the intensity is higher, while in the canonical
``low-hard" state the intensity is lower but the spectrum is dominated
by the power-law component, hence, harder \citep{remillard06}. There
is a third so-called ``very high state" in which the XRB spectrum is
dominated by the power-law component at a very high intensity. For
these anomalous FSRQs, the gamma-ray bright times may correspond to
the ``very high state," in which case the optical emission will also
be very bright and will get bluer when brighter because it is totally
jet dominated.

\section{The time-dependent spectral energy distribution}
\label{sec:sed}

Figure~5 shows the variable broadband spectral energy distribution for
3C~454, with quasi-simultaneous data (within hours) for UT 2009
Dec. 03 ({\it blue}) and UT 2009 Dec. 04 ({\it red}), when the source
was near the peak of its large December 2009 flare. A representative
SED for a non-flaring state is shown for UT 2009
Aug. 12 when the source was relatively faint ({\it black}).  The X-ray
({\em Swift} XRT and BAT) and gamma-ray ({\em Fermi}) data points shown are taken
from the analysis in the paper of \citet{bonnoli11}.  The
NIR-optical points shown are the SMARTS monitoring data presented in
this paper, dereddened using the extinction relations in \citet{cardelli89},
with the value for A$_{\rm B}$ given by \citet{schlegel98}. Magnitudes
were converted into flux densities using the zero-point fluxes given by
\citet{bessell98} and \citet{beckwith76}   The observing conditions 
were photometric with good seeing on the nights shown in Fig.~5. The
near-IR part of the 
spectrum shows a significant change between the two nights, turning
over between  $J$ and $K$ bands. We searched the  light curve of
3C~454.3 for similar behavior and found  similar turnovers in the NIR
SED on 12 nights, generally also with good observing conditions. This
flattening of the SED occured in faint states as well as bright
states.

The gamma-ray (inverse Compton) and optical-NIR spectra
(presumably the high-energy end of the synchrotron emission component)
both varied significantly during the 3-4 Dec 2009 flare, but in
opposite senses. That is, the gamm-ray spectrum flattened and declined
slightly while the optical/IR steepened and increased in flux. This
definitely points to complex physical changes beyond a simple electron
acceleration event.
Note that the SMARTS data are extended enough in energy coverage that
one can see that the synchrotron spectrum is not consistent with a
power law and that the synchrotron emission peak likely moved.  On the
other hand, if one ignored the NIR ($J, K$ band) points, then one
might conclude from the small spectral changes in the optical
and gamma-rays  that the underlying electron energy
distribution and source parameters were roughly similar. This was the
conclusion of \citet{bonnoli11}, who focused on fitting the X-rays
and gamma-rays and only roughly matched the optical intensity of the
source.  Thus the ability of SMARTS to cover both optical and IR
bandpasses constrains the physical interpretation in ways that
gamma-ray or optical monitoring alone does not.

The SEDs on the two nights of the December 2009 flare were modeled using
the one zone code of Coppi (1992). This model injects electrons with a
power-law distribution in energy, turning over at the high energy
end, giving effectively a maximum and minimum $\gamma$, as also used
in the models of Ghisellini et al. (cf. \citet{ghisellini07}). We
assume a soft photon field with a black body temperature of
kT$\sim$1-10 eV that is isotropic in the jet frame. The electron
distribution is convolved with the synchrotron spectrum and thermal
photon field to produce gamma-rays. The key difference between the 
\citet{coppi92} code and others is that the electron spectrum changes
through both 
synchrotron and Compton cooling. In the Klein-Nishina limit, Compton
cooling becomes inefficient while synchrotron cooling is not affected,
thus producing more power in the high-frequency end of the
synchtrotron radiation. This may be what is seen in the $J-K$ band
peak.

If we assume the optical-NIR emission is produced in the same source
region as the gamma-rays (which is not unreasonable as this is a very
bright flare and the gamma-ray and optical variations are correlated),
then our preliminary SED modeling of the NIR to gamma-ray data (shown
by the solid red and blue curves in Fig.~5) indicates that the source
variations must be significant. For source
parameters similar to \citet{bonnoli11} ($R_{source}=\sim 10^{17}$cm,
$\delta\sim 20,$ $B\sim 1$, $\gamma_{peak} \sim 500$), we find that
the electrons responsible for the optical-NIR emission are likely
higher in energy than those responsible for the gamma-ray emission,
and that Klein-Nishina effects are thus important for them.  This
means that in order to produce a {\it steeper} yet higher amplitude
synchrotron spectrum, the external photon field seen by the emitting
electrons must have changed to include more lower-temperature photons
than the BLR; see \citet{moderski2005} and
\citet{moderskierr2005} for a discussion of the dependence of
Klein-Nishina effects on the
external photon field. Varying the ambient photon field alone,
however, is not enough to explain all the observed spectral changes:
in the context of traditional one-zone synchrotron-Compton models,
some combination of peak electron energy, the magnetic field, and the
Doppler beaming factor must have changed by factors of two or more.

 We note here that this is not a systematic exploration of
paramater space, but merely a demonstration that the short-term
variations in the SED cannot be produced merely through changing the
maximum value of $\gamma$, or the strength of the magnetic field
alone. The change in model parameters has to be more
complicated. Further exploration of multiwavelength spectral
variability will be presented in a 
forthcoming paper (Chatterjee et al. {\em in prep.}).We further note
that if our preliminary conclusions are correct, namely that 
gamma-ray and optical emission comes from somewhat different energy
electrons, and that the mapping from electron energy to synchrotron and
Compton photon energy may change during the flare, then this might
explain why we see general but not exact correlations between the
variability in  gamma-ray and lower energy bands. One-zone models
may still have some relevance during bright flares, if one region of
the jet  dominates the overall emission.  Finally, this implies that
the lore that Fermi spectra do not change much during flares appears
to be at best a coincidence in 3C 454.3: the underlying electron
and target photon spectra probably change significantly.

\section{Conclusions}
\label{sec:conclusions}

Systematic monitoring of southern blazars with the SMARTS
optical+infrared ANDICAM has generated BVRJK light curves for a dozen
blazars, all of which were contemporaneously observed with the Fermi
gamma-ray satellite. All of the SMARTS blazar photometry is available
on a publicly accessible web site,
http://www.astro.yale.edu/smarts/fermi.  This paper reported on the 12
blazars that have well defined optical/IR light curves over the first
2 years of the Yale/SMARTS blazar monitoring project, which commenced
in August 2008.

Not all twelve blazars are significantly detected in gamma-rays on
$\sim$day timescales. For the first two years of data (through July 2010), six
SMARTS-monitored blazars were bright enough to be well detected in
one-day averages at gamma-ray energies:  the FSRQs 3C~454.3, 3C~273,
3C~279, and PKS~1510-089; the LBL AO~0235+164; and the HBL PKS~2155-304. 
Detailed multiwavelength analysis was presented for these six.

We find the optical/IR and gamma-ray light curves for these six
southern blazars are generally well correlated, with lags of less than
1 day for most FSRQs (roughly the sampling interval for the SMARTS and
Fermi data). 
This strongly supports leptonic models in which ambient photons ---
perhaps from the accretion disk or broad emission-line clouds --- are
upscattered to gamma-ray energies by synchrotron-emitting electrons in
the relativistic blazar jet.  Hadronic models, in which the gamma-rays
come from energetic protons that ultimately decay into
synchrotron-emitting electrons, do not have the same natural
explanation for correlated optical/IR and gamma-ray variability.

The multiwavelength variability for the six blazars shows two types of
behavior.  In luminous blazars like FSRQs, the amplitude of optical/IR
variability decreases towards shorter wavelengths, as if that
wavelength range had a significant underlying contribution from a more
slowly varying thermal disk peaking in the UV.  In BL Lac objects, the
optical/IR variability is similar in different bands, consistent with
little or no contribution from a (radiatively inefficient) disk.

Color-magnitude variations in these six blazars can be explained in
part by highly variable jet emission mixed with slower varying disk
emission.  Gamma-ray flares can result from particle acceleration in
the jet or from secular increases in disk/line emission, hence the
somewhat complicated trajectories of blazars in color-magnitude space.
Analogous multiple-state behavior has been observed in Galactic X-ray
binaries.  Meanwhile, individual jet flares in a given source can also
follow different color-color tracks because of, for example, standing
shocks  in different parts of the jet.

The best-studied blazar, the FSRQ 3C~454.3, has had a strong flare in
each year of SMARTS-{\em Fermi} monitoring. Its optical/IR and
gamma-ray light curves are well correlated, with no measureable lag
longer than one day. 
Its broad-band SED in the bright state is well fit by an external
Compton model; in its faint state, the disk contributes a larger
fraction of the light and the SED is flatter.  Day-to-day changes are
difficult to explain with a simple one-zone model, however, unless there are
large changes (factors of two or more) in Doppler factor, electron
energy, and/or magnetic field.  More generally, our simple analysis of
the 3C~454.3 SED suggests that optical/IR radiation comes from
slightly higher energy electrons than the GeV gamma rays, and that
Klein-Nishina effects are important in shaping both SED components.

SMARTS monitoring of southern blazars continues, with the goal of  producing
additional rich multiwavelength data sets that will advance our
further understanding of blazar emission mechanisms and energy
transport. In more than three years of monitoring several dozen
blazars, only 6 sources - in three different classes - have produced
multiwavelength data of sufficient quality for detailed analysis. Only
some of these showed distinct flares over this period. The rarity of
blazars being bright while visible from ground (often there have been dramatic
gamma-ray flares while the sources had not yet risen in the night sky)
suggests that continued dedicated optical-infrared monitoring is
essential during the lifetime of the {\em Fermi} gamma-ray observatory.

\acknowledgments
SMARTS observations of LAT-monitored blazars are supported by Fermi GI
grant 011283 and by Yale University. CDB, MMB and the SMARTS 1.3m
observing queue receive support from NSF grant AST-0707627. RC is
supported by Fermi GI grant NNX09AR92G. GF is supported by Fermi GI
grant NNX10A042G. JI is supported by the NASA Harriet Jenkins
Fellowship program and NSF Graduate Research Fellowship under Grant No
DGE-0644492. This research has made use of the NASA/IPAC Infrared
Science Archive, which is operated by the Jet Propulsion Laboratory,
Californina Institute of Technology, under contract with the National
Aeronautics and Space Administration.

\clearpage

%
%


\begin{figure}[]
\begin{center}
\plotone{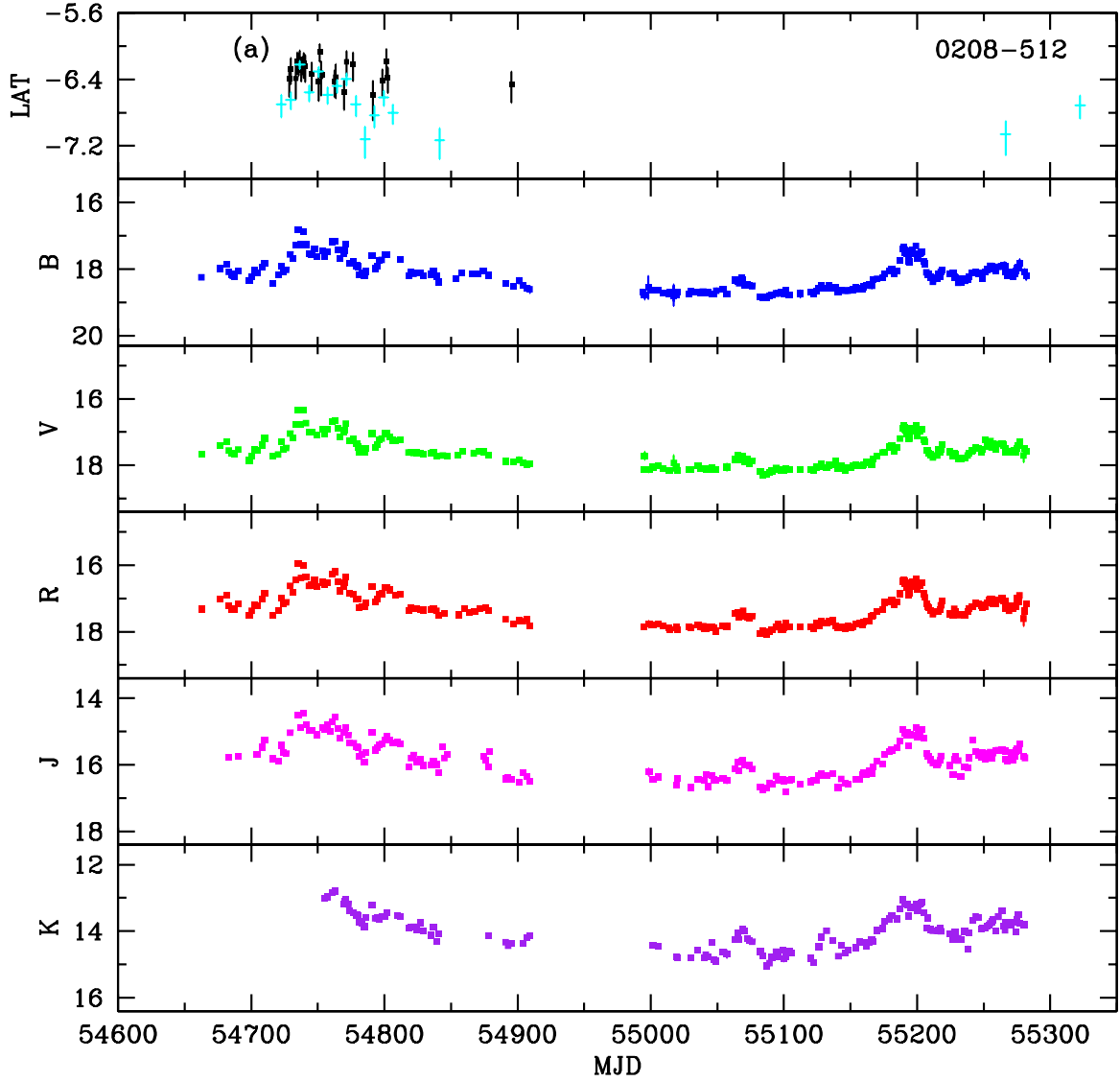}
\caption*{Figure 1a: SMARTS optical and near-infrared light curves for
PKS~0208-512. {\em Fermi}/LAT fluxes are taken from the public daily
(black points) and weekly (cyan points) light curves and are in units
of log (photons~sec$^{-1}$~cm$^{-2}$). }
\label{fig:lc}
\end{center}
\end{figure}

\begin{figure}[]
\begin{center}
\plotone{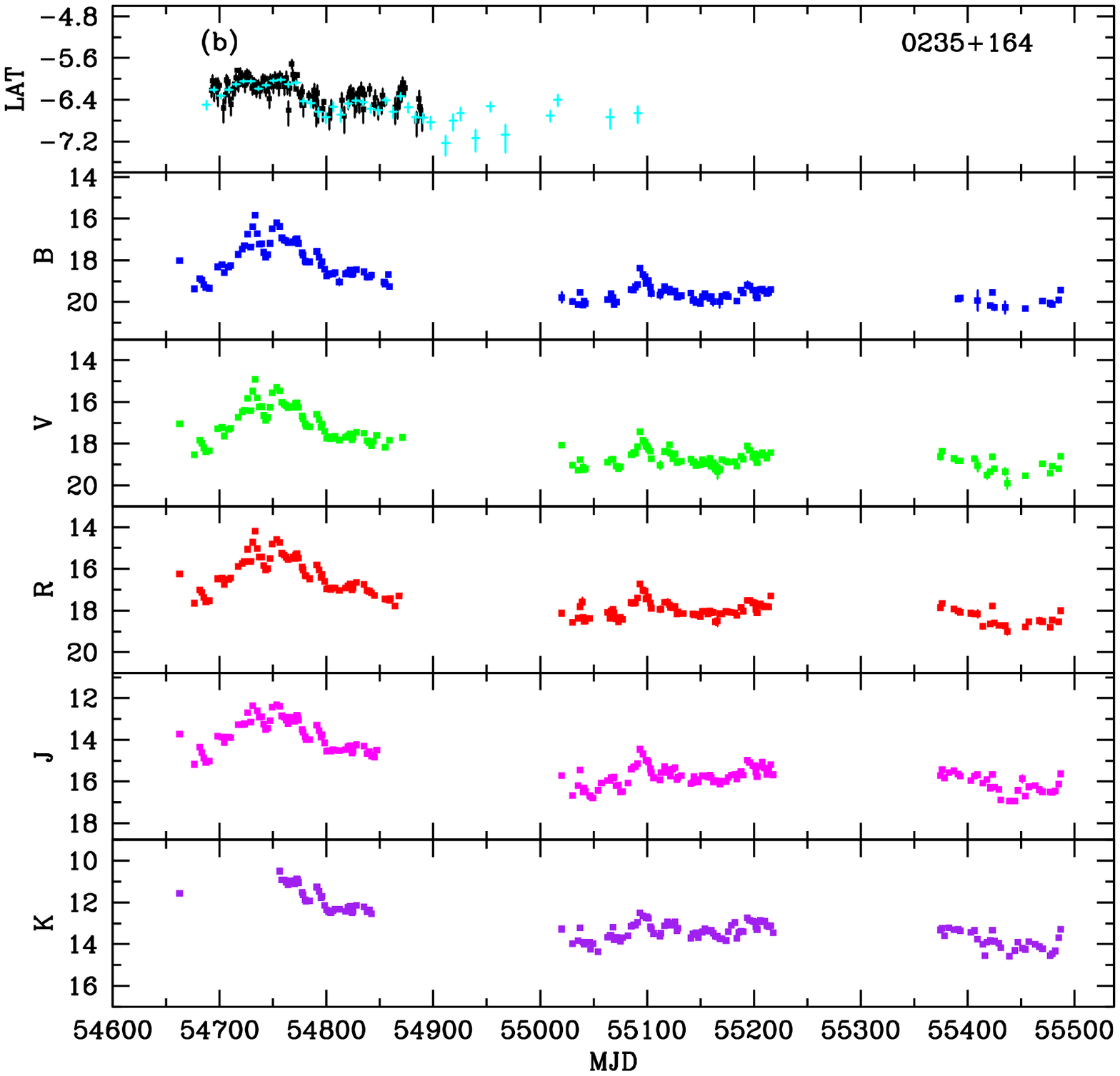}
\caption*{Figure 1b: SMARTS optical and near-infrared light curves for
AO~0235+164. {\em Fermi}/LAT fluxes are taken from the public daily
(black points) and weekly (cyan points) light curves and are in units
of log (photons~sec$^{-1}$~cm$^{-2}$). }
\end{center}
\end{figure}

\begin{figure}[]
\begin{center}
\plotone{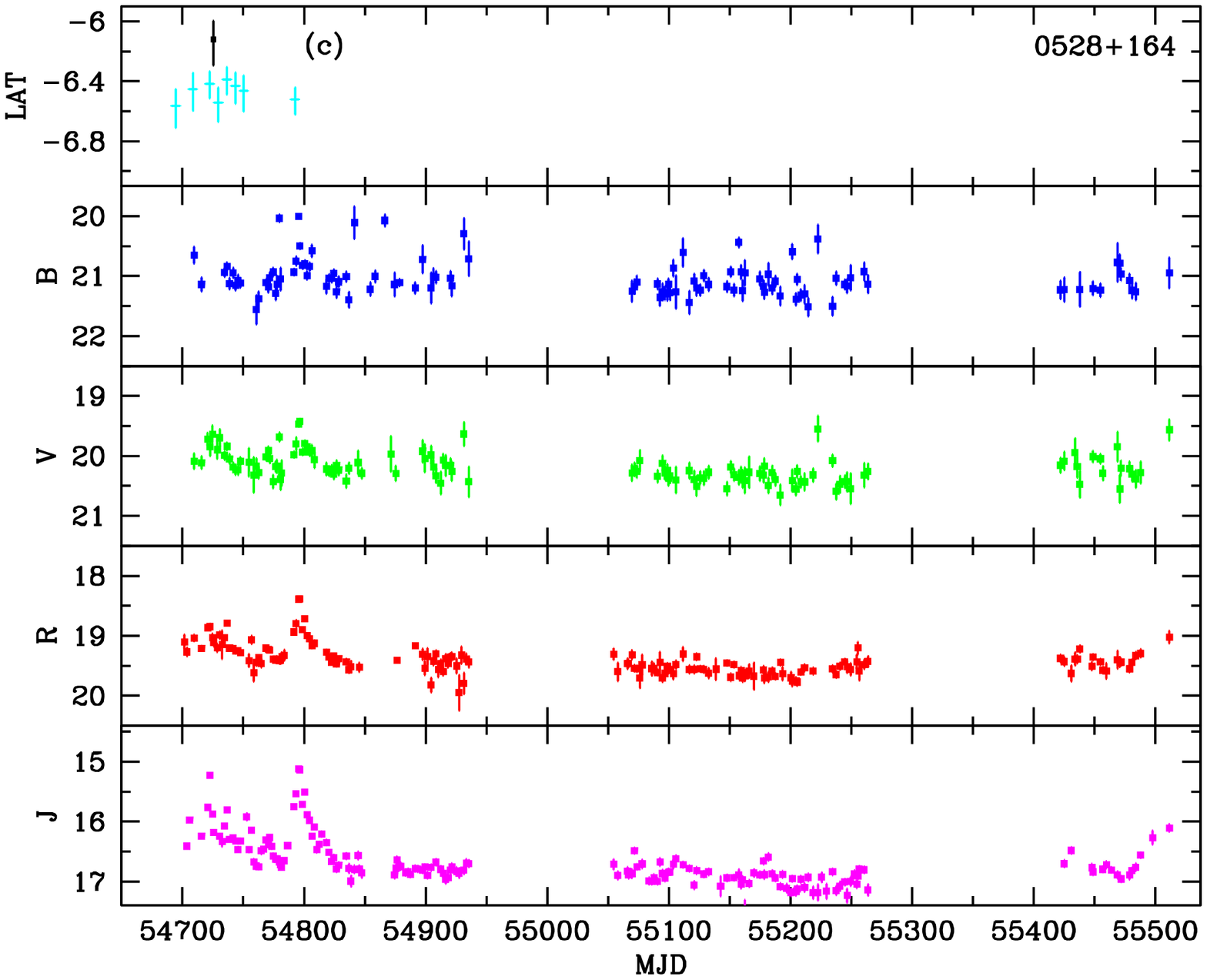}
\caption*{Figure 1c: SMARTS optical and near-infrared light curves for
PKS~0528+134. {\em Fermi}/LAT fluxes are taken from the public daily
(black point) and weekly (cyan points) light curves and are in units
of log (photons~sec$^{-1}$~cm$^{-2}$). }
\label{fig:lc0528}
\end{center}
\end{figure}

\begin{figure}[]
\begin{center}
\plotone{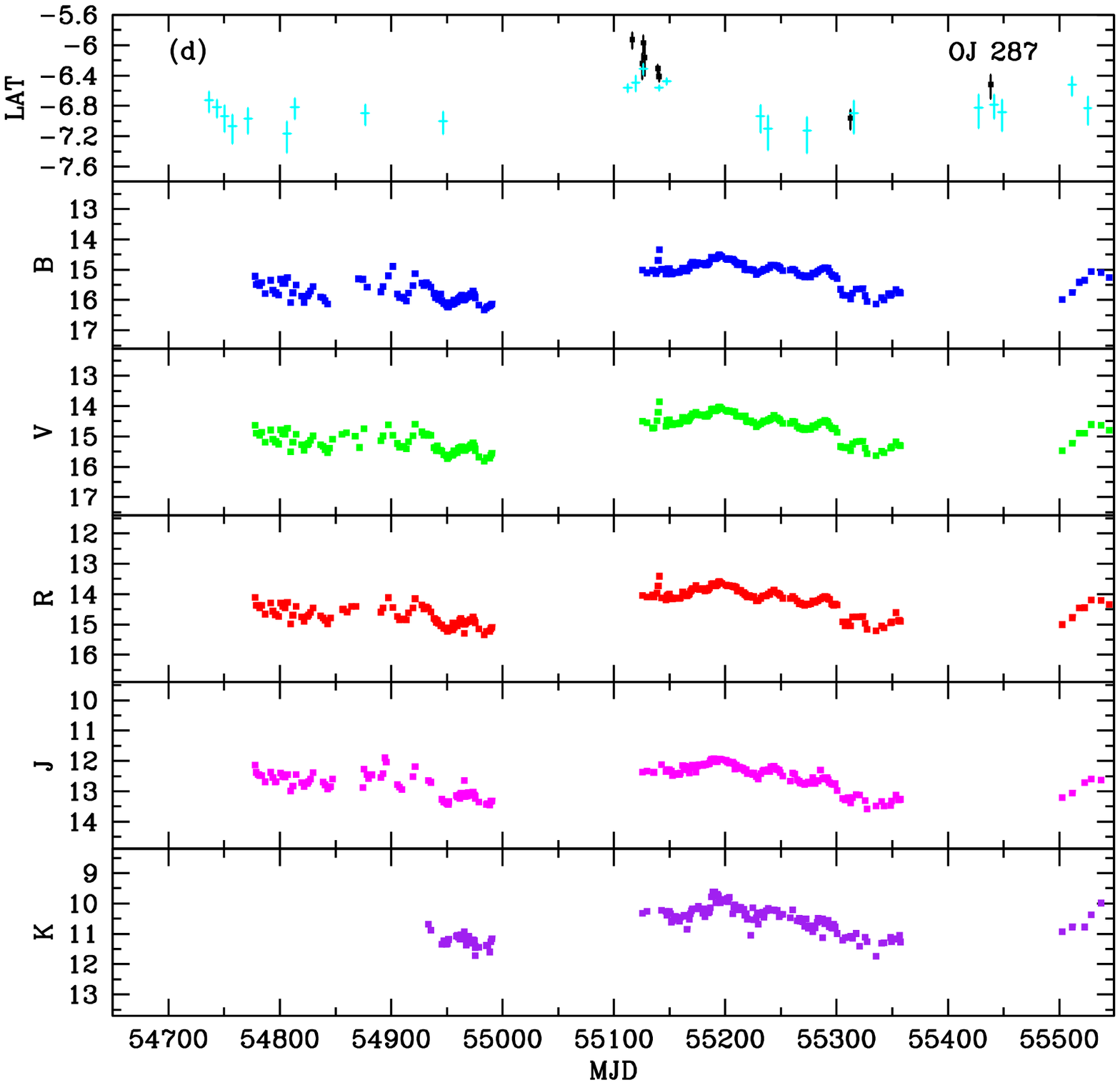}
\caption*{Figure 1d: SMARTS optical and near-infrared light curves for
OJ~287. {\em Fermi}/LAT fluxes are taken from the public daily
(black points) and weekly (cyan points) light curves and are in units
of log (photons~sec$^{-1}$~cm$^{-2}$). }
\label{fig:lcOJ287}
\end{center}
\end{figure}

\begin{figure}[]
\begin{center}
\plotone{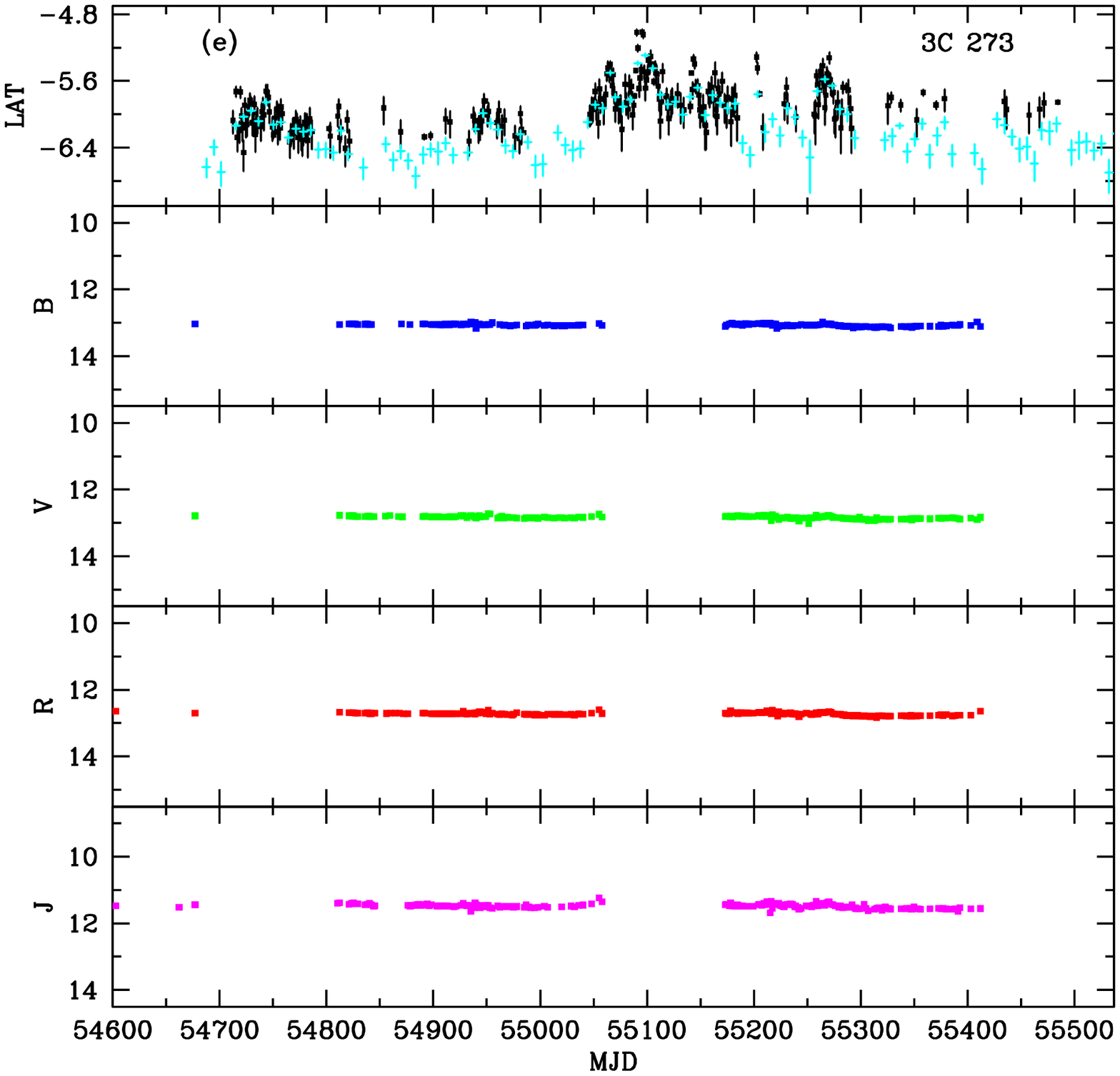}
\caption*{Figure 1e: SMARTS optical and near-infrared light curves for
3C~273. {\em Fermi}/LAT fluxes are taken from the public daily
(black points) and weekly (cyan points) light curves and are in units
of log (photons~sec$^{-1}$~cm$^{-2}$). }
\label{fig:lc3C273}
\end{center}
\end{figure}

\begin{figure}[]
\begin{center}
\plotone{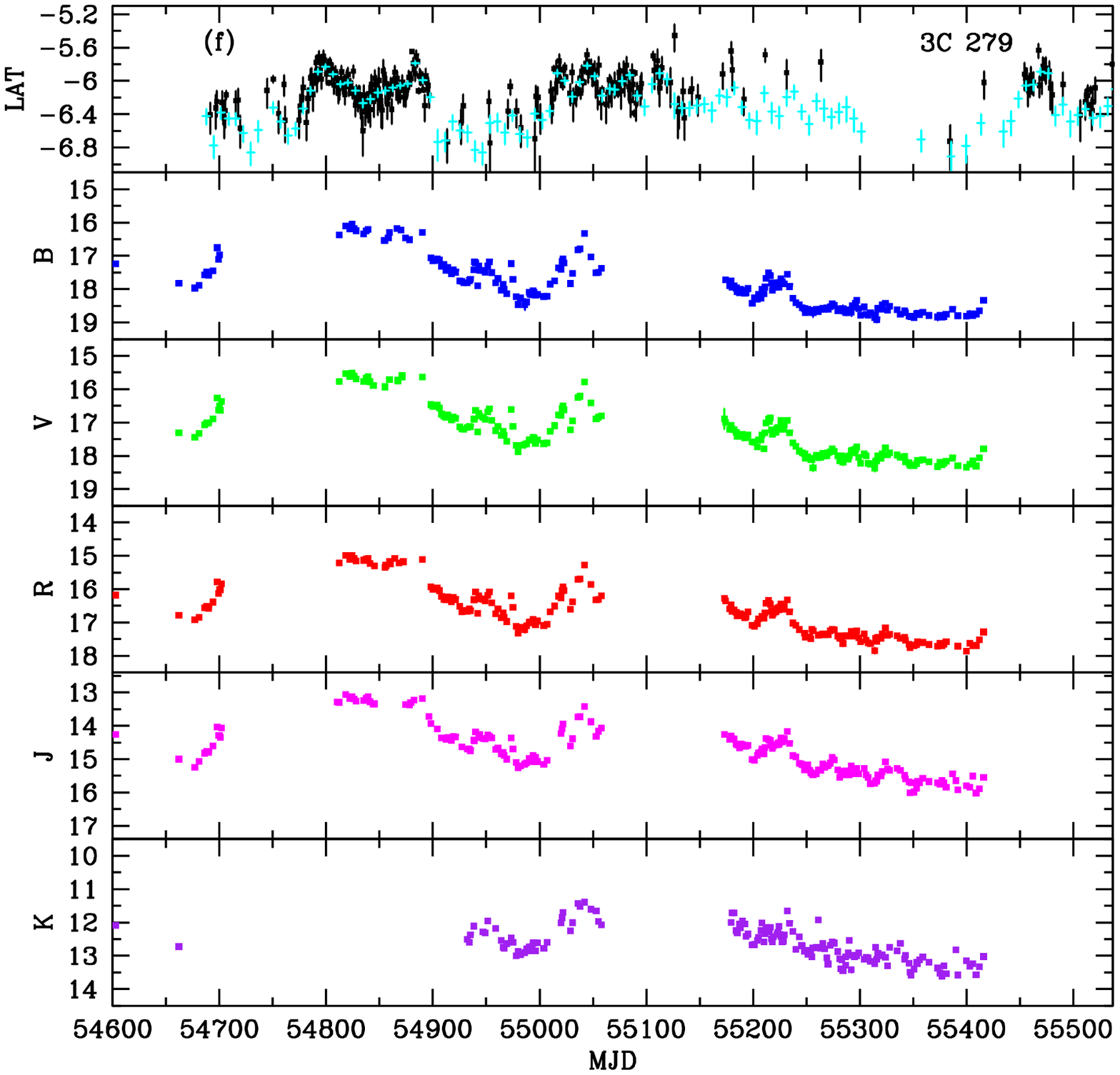}
\caption*{Figure 1f: SMARTS optical and near-infrared light curves for
3C~279. {\em Fermi}/LAT fluxes are taken from the public daily
(black points) and weekly (cyan points) light curves and are in units
of log (photons~sec$^{-1}$~cm$^{-2}$). }
\label{fig:lc3C279}
\end{center}
\end{figure}

\begin{figure}[]
\begin{center}
\plotone{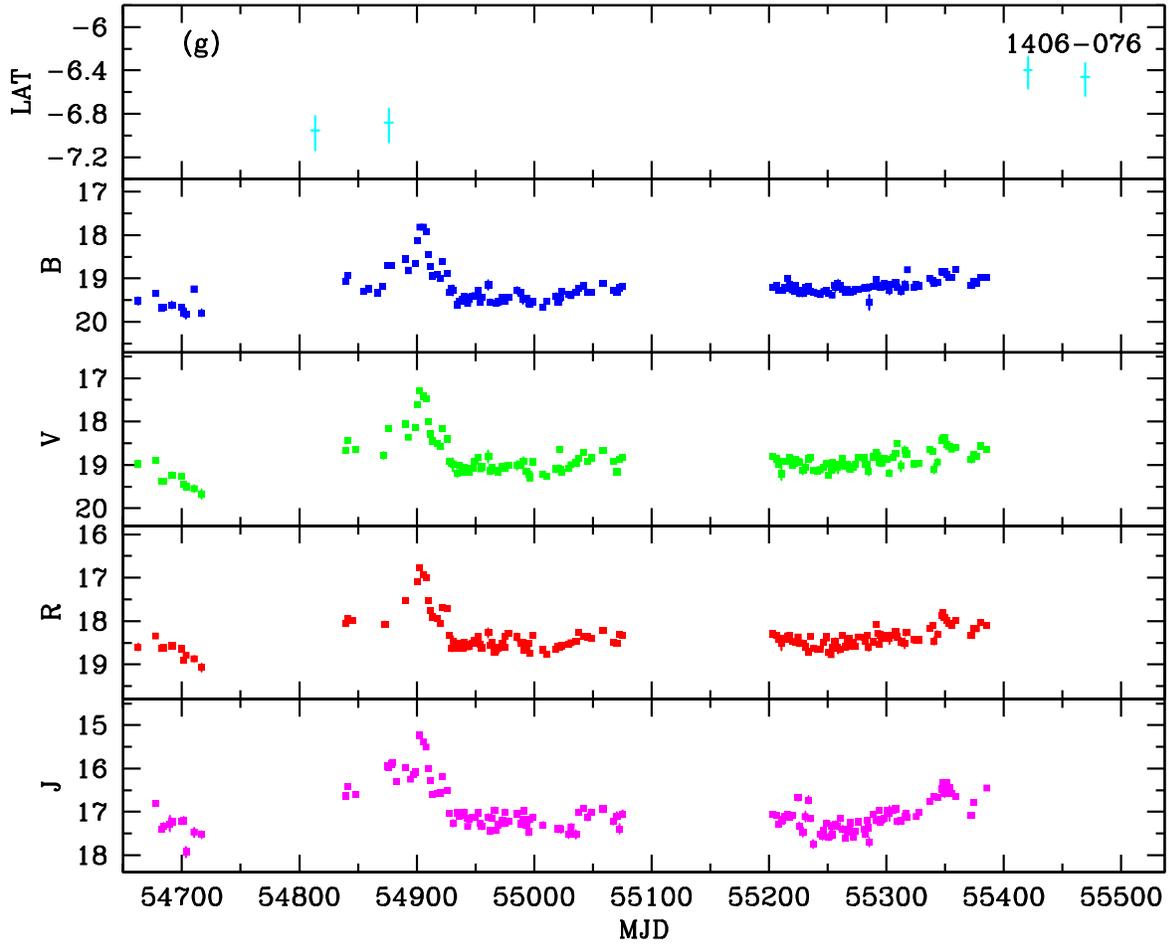}
\caption*{Figure 1g: SMARTS optical and near-infrared light curves for
PKS~1406-076. {\em Fermi}/LAT fluxes are taken from the public weekly
(cyan points) light curves and are in units 
of log (photons~sec$^{-1}$~cm$^{-2}$). }
\label{fig:lc1406}
\end{center}
\end{figure}

\begin{figure}[]
\begin{center}
\plotone{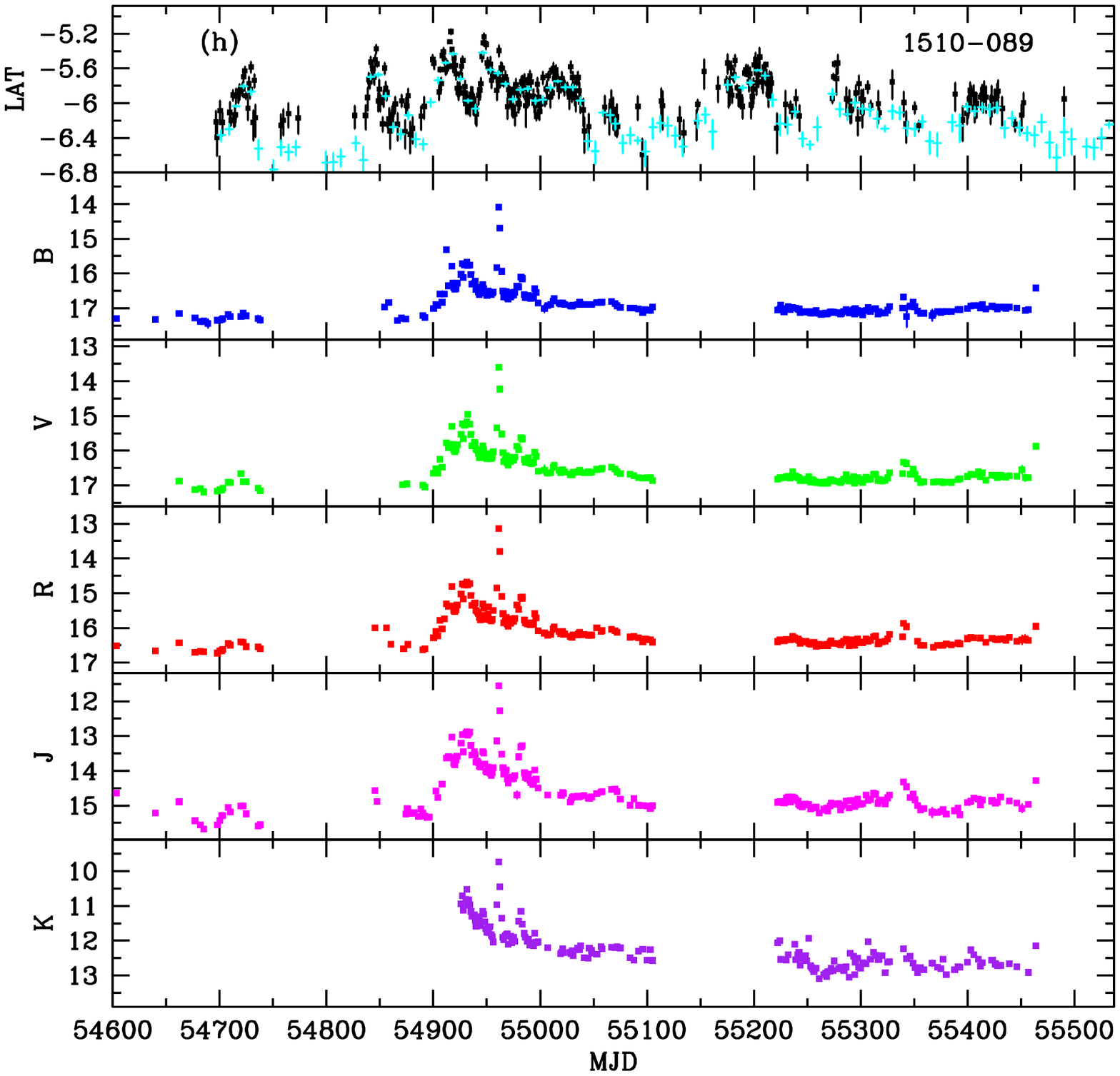}
\caption*{Figure 1h: SMARTS optical and near-infrared light curves for
PKS~1510-089. {\em Fermi}/LAT fluxes are taken from the public daily
(black points) and weekly (cyan points) light curves and are in units
of log (photons~sec$^{-1}$~cm$^{-2}$). }
\label{fig:lc1510}
\end{center}
\end{figure}

\begin{figure}[]
\begin{center}
\plotone{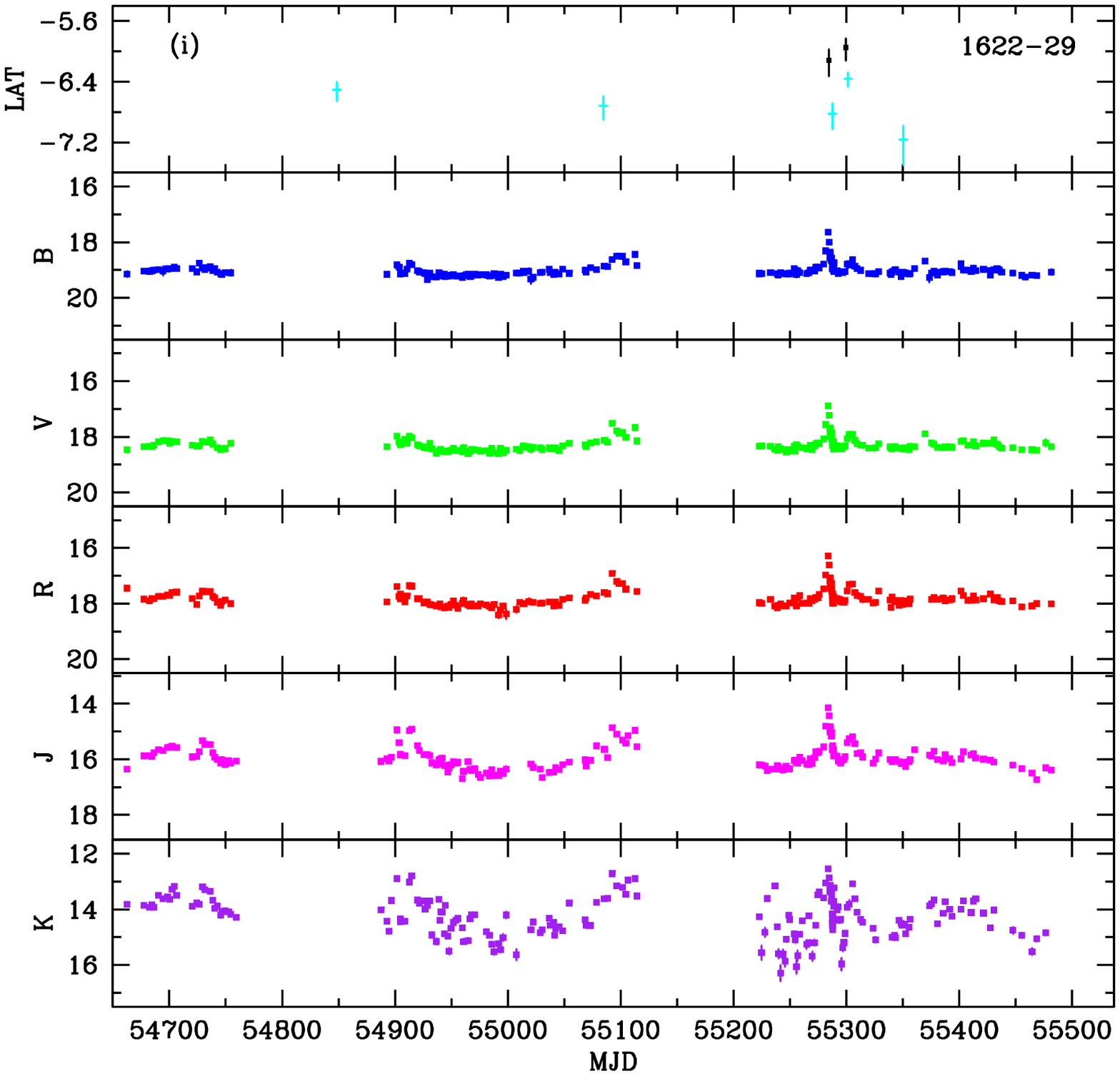}
\caption*{Figure 1i: SMARTS optical and near-infrared light curves for
PKS~1622-297. {\em Fermi}/LAT fluxes are taken from the public daily
(black points) and weekly (cyan points) light curves and are in units
of log (photons~sec$^{-1}$~cm$^{-2}$). }
\label{fig:lc1622}
\end{center}
\end{figure}

\begin{figure}[]
\begin{center}
\plotone{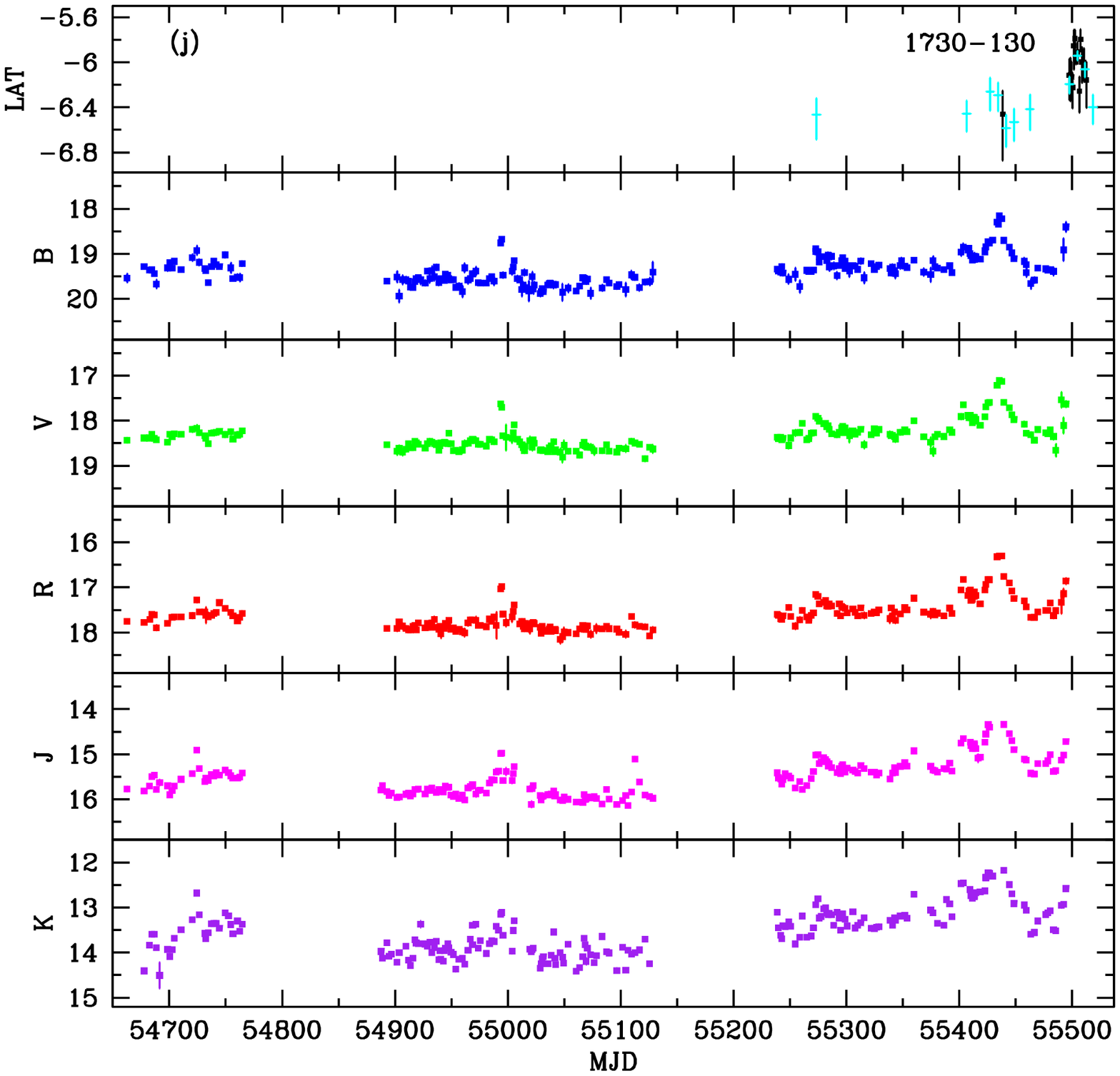}
\caption*{Figure 1j: SMARTS optical and near-infrared light curves for
PKS~1730-130. {\em Fermi}/LAT fluxes are taken from the public daily
(black points) and weekly (cyan points) light curves and are in units
of log (photons~sec$^{-1}$~cm$^{-2}$). }
\label{fig:lc1730}
\end{center}
\end{figure}

\begin{figure}[]
\begin{center}
\plotone{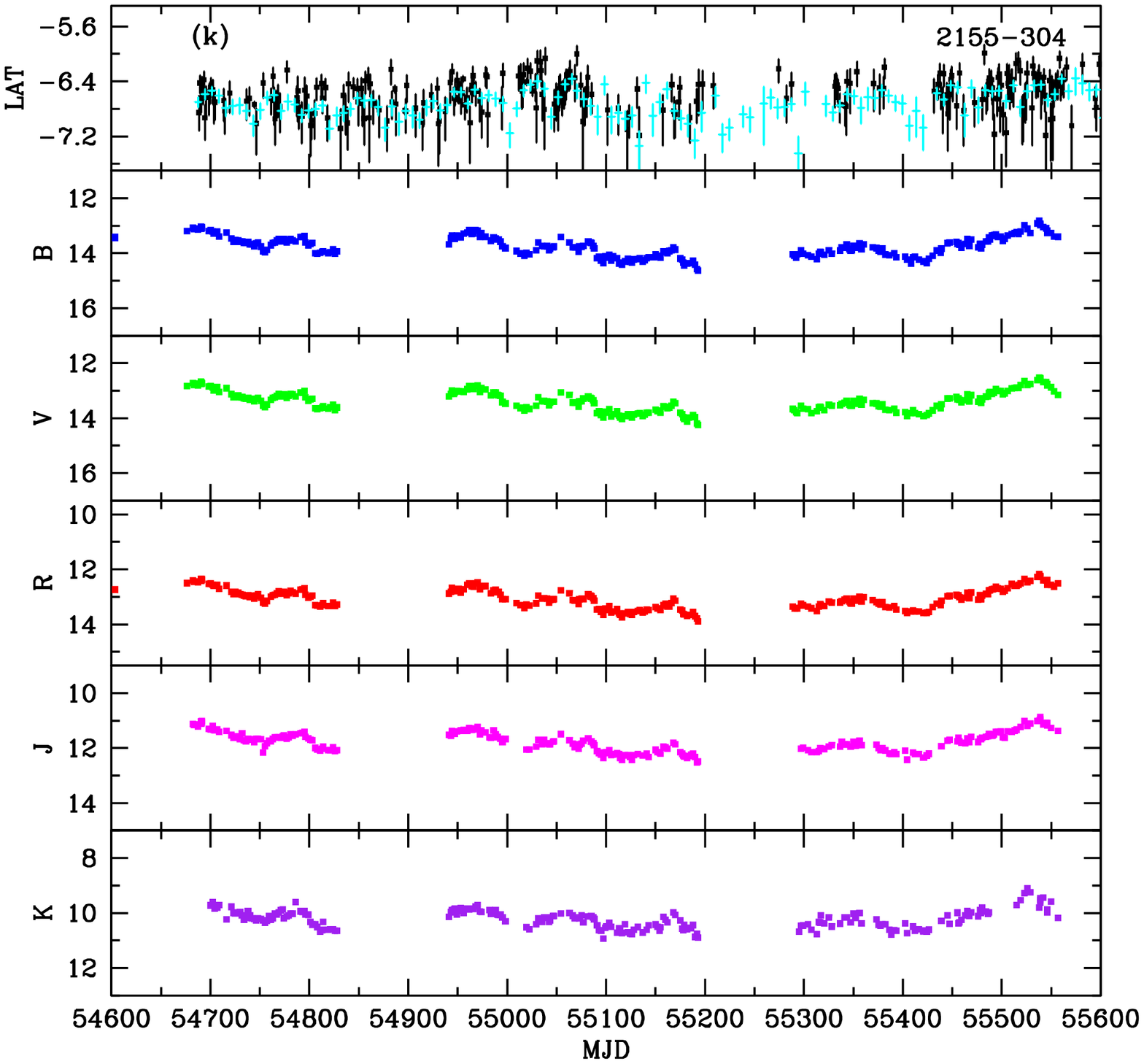}
\caption*{Figure 1k: SMARTS optical and near-infrared light curves for
PKS~2155-304. {\em Fermi}/LAT fluxes are taken from the public daily
(black points) and weekly (cyan points) light curves and are in units
of log (photons~sec$^{-1}$~cm$^{-2}$). }
\label{fig:lc2155}
\end{center}
\end{figure}

\begin{figure}[]
\begin{center}
\plotone{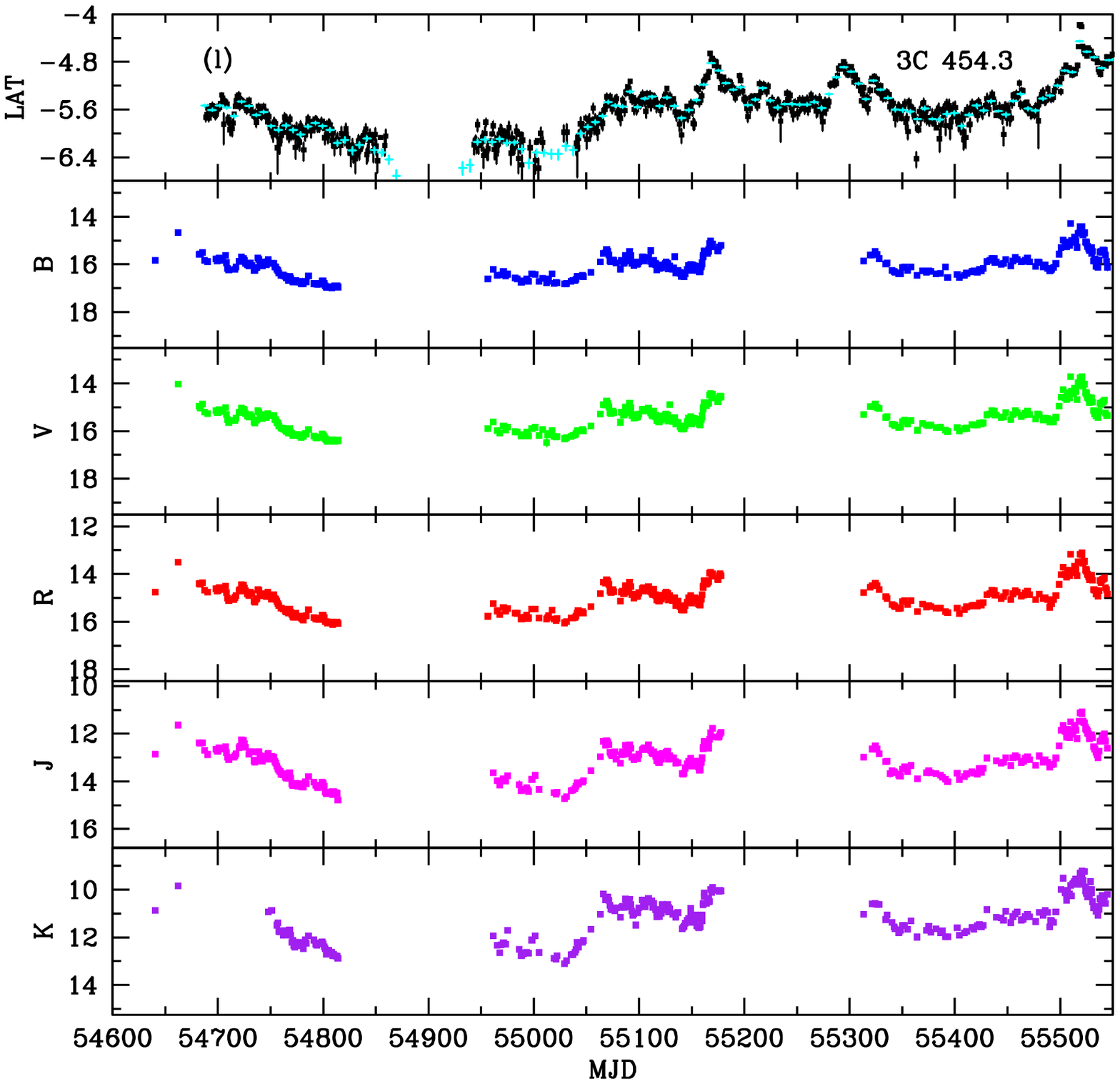}
\caption*{Figure 1l: SMARTS optical and near-infrared light curves for
3C~454.3. {\em Fermi}/LAT fluxes are taken from the public daily
(black points) and weekly (cyan points) light curves and are in units
of log (photons~sec$^{-1}$~cm$^{-2}$). }
\label{fig:lc3C454}
\end{center}
\end{figure}


\begin{figure}[]
\begin{center}
\plotone{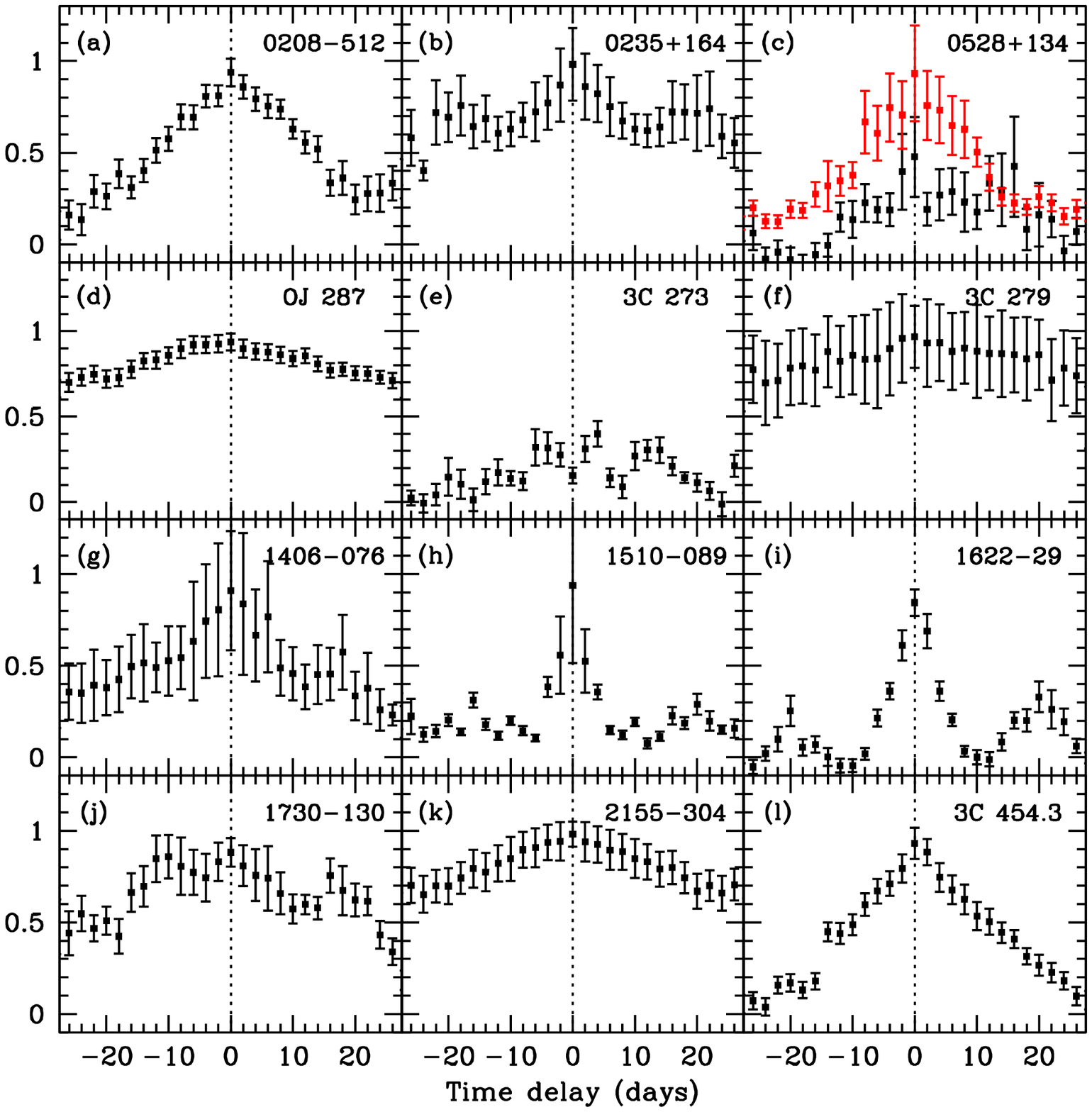} 
\caption*{Fig. 2: Discrete correlation functions (DCF) for $B$ vs $J$-band
light curves shown in Figure 1. With the exception of the minimally
varying quasar 3C~273, all DCFs show a maxmimum at zero lag. Figure
2c also shows the $R$ vs. $J$ DCF for PKS~0528+134 (in red). It is the only
source for which the $R$ vs. $J$ DCF has a significantly different shape
than $B-J$, although it too peaks at zero lag. }\label{fig:dcf} 
\end{center}
\end{figure}

\begin{figure}[]
\begin{center}
\plotone{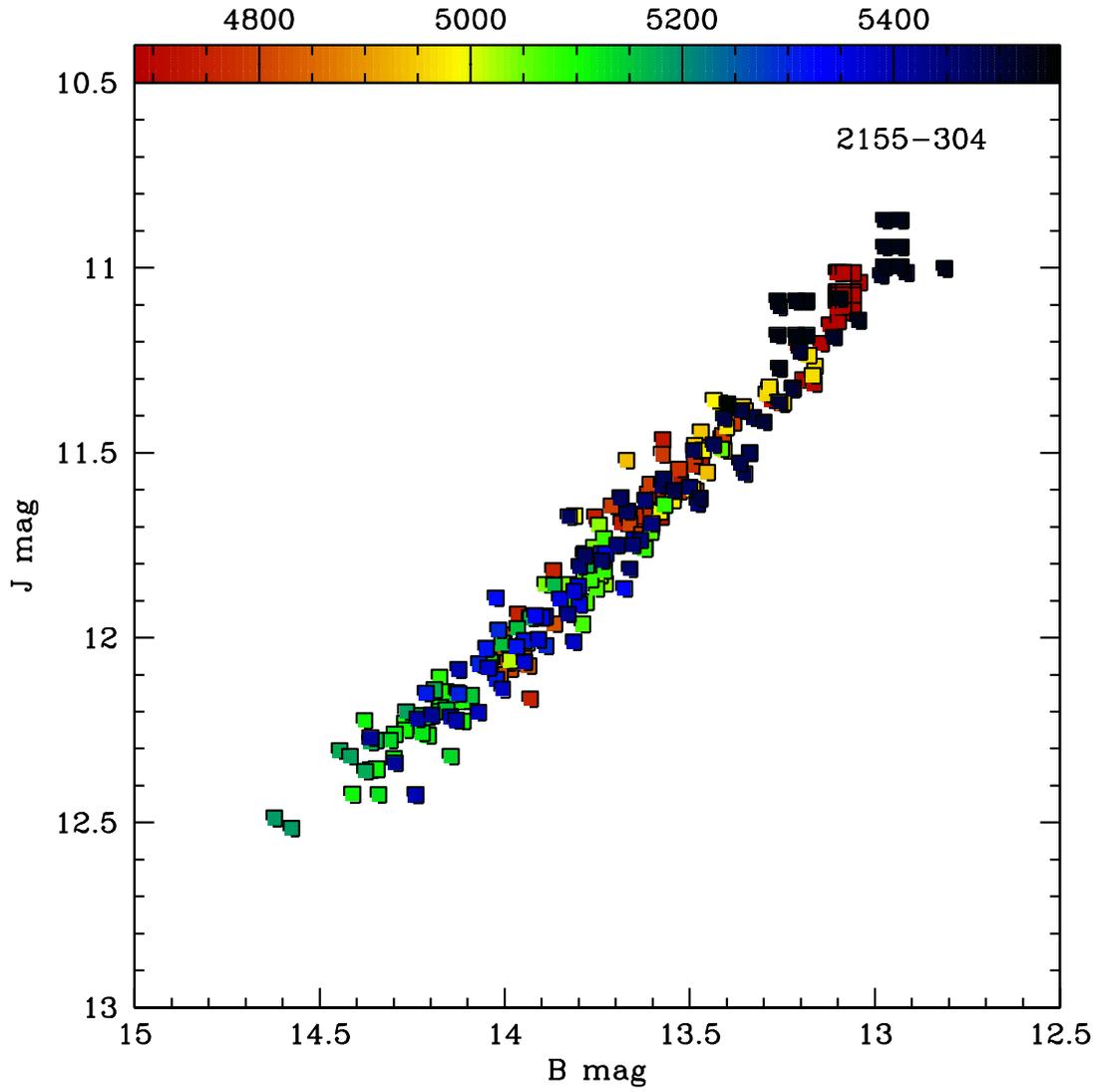} 
\caption*{Fig. 3a: $J$-band vs $B$-band magnitude for the HBL
PKS~2155-304. Color indicates the date of the observation, as shown in
the top bar in MJD-50000. The tight correlation indicates that the
optical and infrared emission vary together, with no significant
delay. }\label{fig:2155} 
\end{center}
\end{figure}

\begin{figure}[]
\begin{center}
\plotone{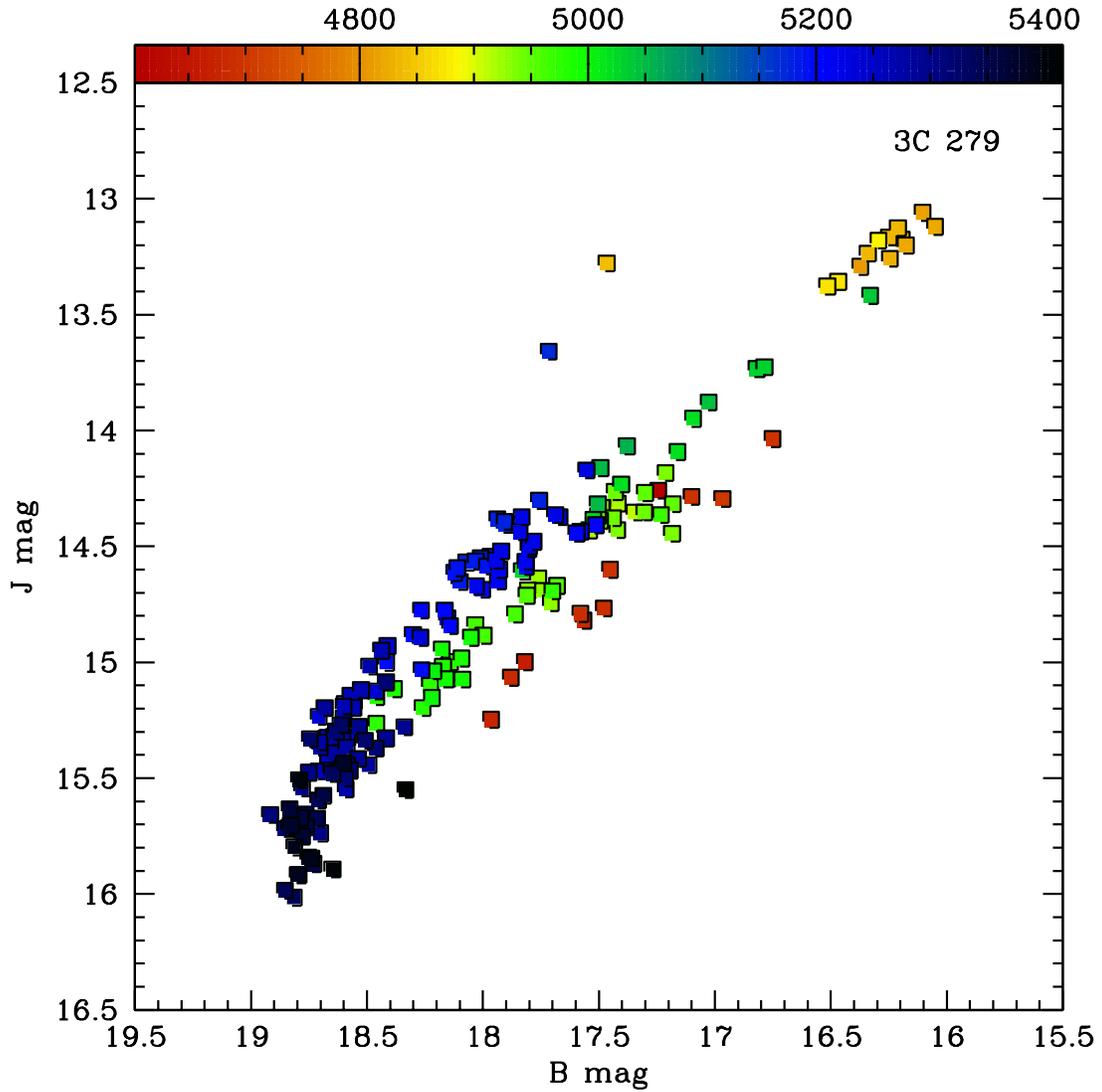} 
\caption*{Fig. 3b: $J$-band vs $B$-band magnitude for the luminous FSRQ
3C~279, with colors as in Fig. 3a. As in all sources observed, optical
and infrared 
emission varies together. The distinct tracks for the red, green, and
blue reflect changes optical/IR spectral shape over time ({\em cf.}
Fig. 5b). }\label{fig:3c279}
\end{center}
\end{figure}

\begin{figure}[]
\begin{center}
\plotone{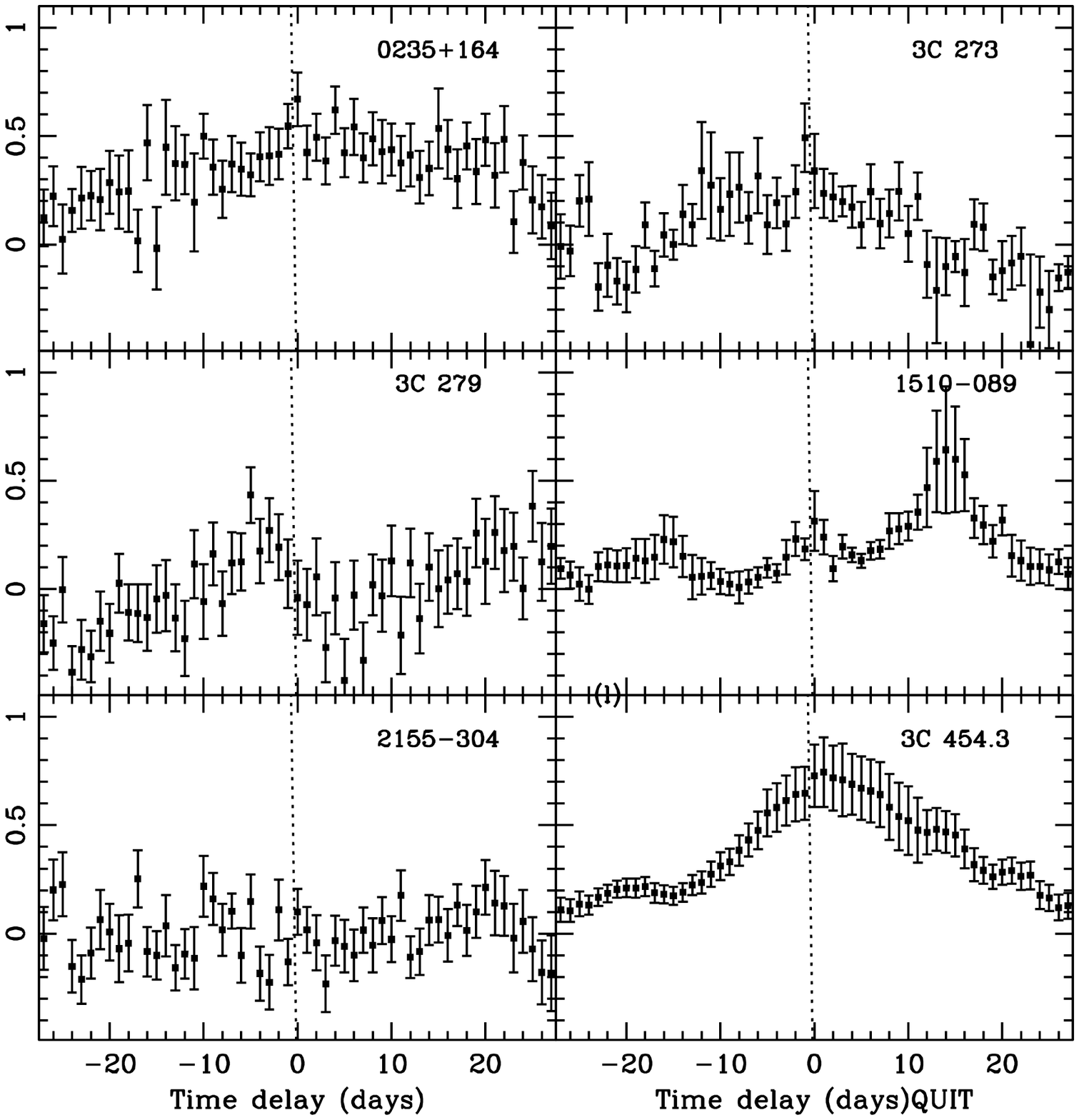} 
\caption*{Fig. 4: Discrete correlation functions (DCF) for $\gamma -
J$-band light curves shown in Figure 1. A positive time delay
corresponds to $J$ band lagging the gamma-rays. In several sources (3C
454.3, most clearly) the emission is well correlated with little or no
lag. In others, the DCF is flat, suggesting no strong correlation
(PKS~2155-304). In PKS 1510-089, the gamma-rays appear to
lead the J band by a couple of weeks.    }\label{fig:dcf} 
\end{center}
\end{figure}



\clearpage

\begin{figure}[]
\begin{center}
\plotone{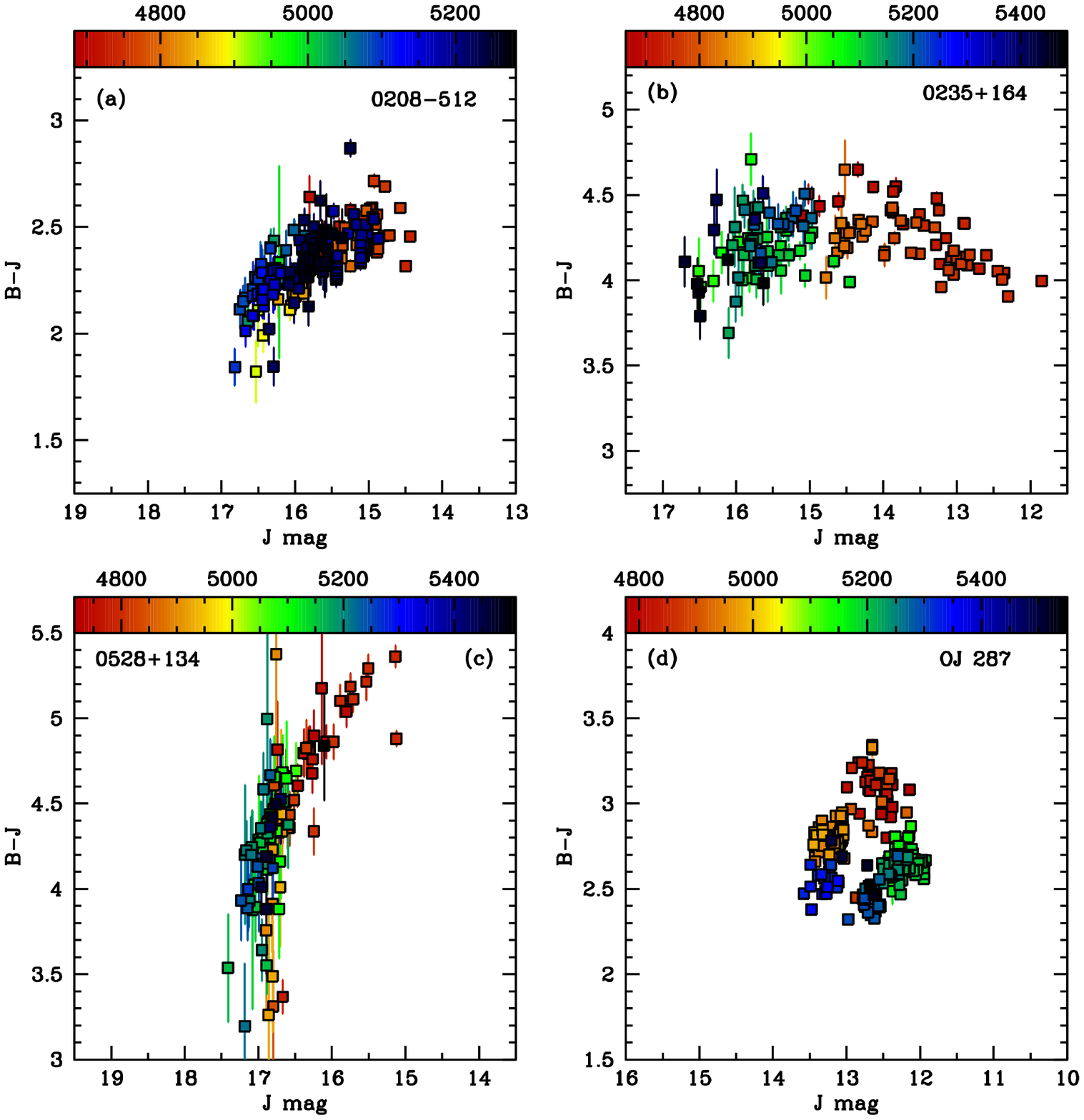}
\caption*{Fig. 5: ($B-J$) color vs. $J$-band magnitude for FSRQs PKS~0208-512
and PKS~0528+134, and low-frequency peaked BL Lac objects AO~0235+164
and OJ~287. Color indicates the date of the observation, as shown in
the top bar in MJD-50000. The FSRQs show an overall tendancy to become
bluer when fainter, which suggests the presence of a steady blue
accretion disk component underlying the more variable jet
emission. The two LBLs 
show more complicated behavior: AO~0235+164 shows some
redder-when-brighter trend (blue-black points) but at
early times (red-orange points) shows almost the opposite trend,
getting bluer when brighter. During this bluer-when-brighter period,
AO~0235+164 was very bright in gamma-rays and was easily detected by
Fermi on daily time scales. When AO~0235+164 became fainter in the
optical and was not detected in the daily LAT light curve, it shifted
to the ``normal" (for FSRQs) bluer-when-fainter trend. OJ~287 tends to
move around on a circular locus in color-magnitude space where trends
with brightness (including gamma-ray intensity) are not as easily
definable. }\label{fig:fig5a}
\end{center}
\end{figure}


\begin{figure}[]
\begin{center}
\plotone{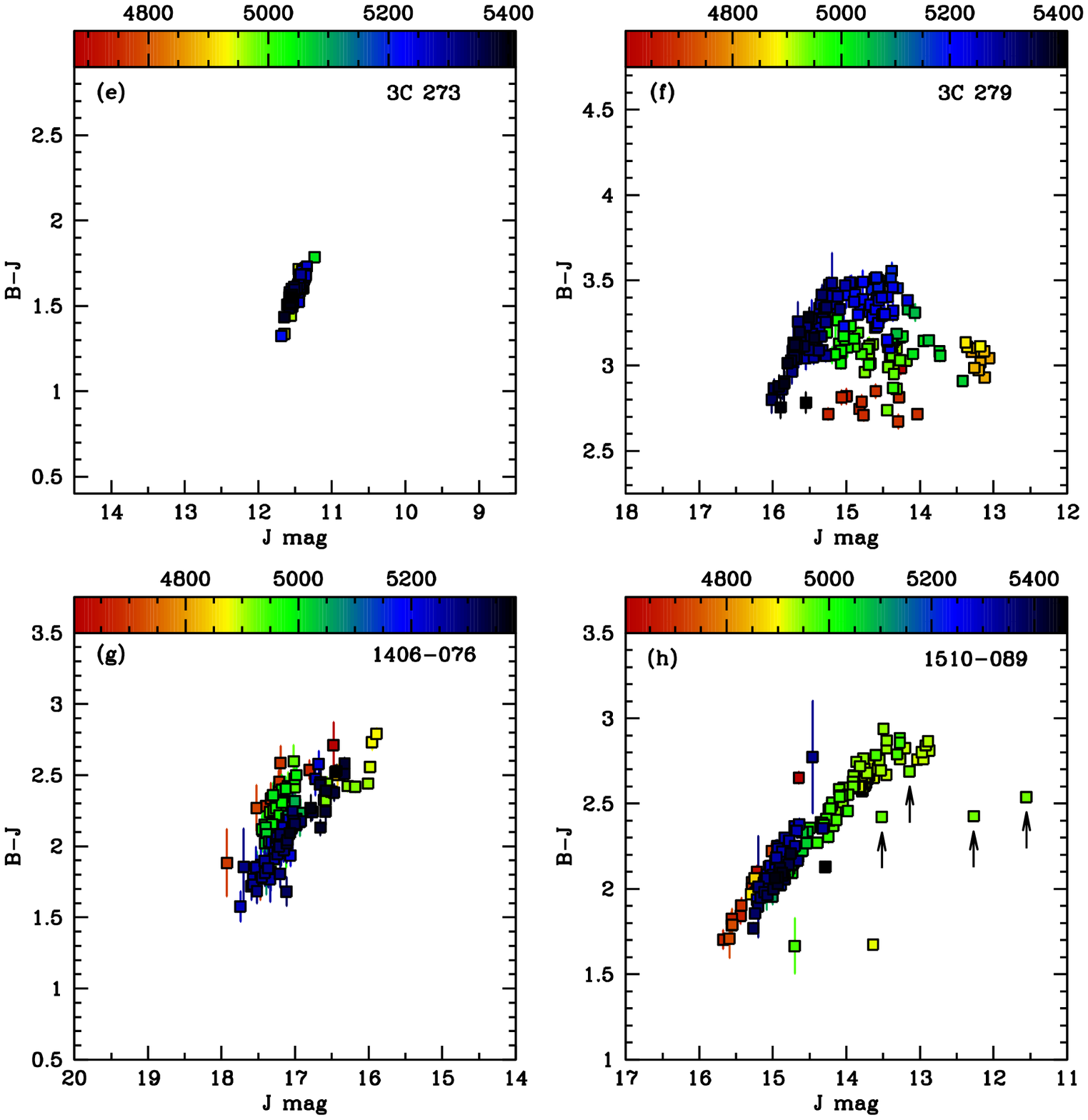} 
\caption*{Fig. 5 (cont'd): ($B-J$) color vs. $J$-band magnitude for
FSRQs 3C~273, 3C~279, PKS~1406-076, and PKS~1510-089. Color indicates
the date of the observation, as shown in the top bar in MJD-50000. All
of these FSRQs show the overall redder-when-brighter
(bluer-when-fainter) trend, consistent with a steady blue accretion
disk component underlying the jet emission.  Individual flares can
behave differently; for example, the flare in PKS~1510-089 in May 2010
is achromatic (arrows indicate the 4 green points tracing a horizontal
trend). The quasar 3C~279 shows compound behavior similar to
AO~0235+164 (Fig. 5b), with the red-light green points tracing a
horizontal loop in the color-magnitude figure before and just up to a
period of bright gamma-ray activity (cf. Fig. 1f).}
\end{center}
\end{figure}

\begin{figure}[]
\begin{center}
\plotone{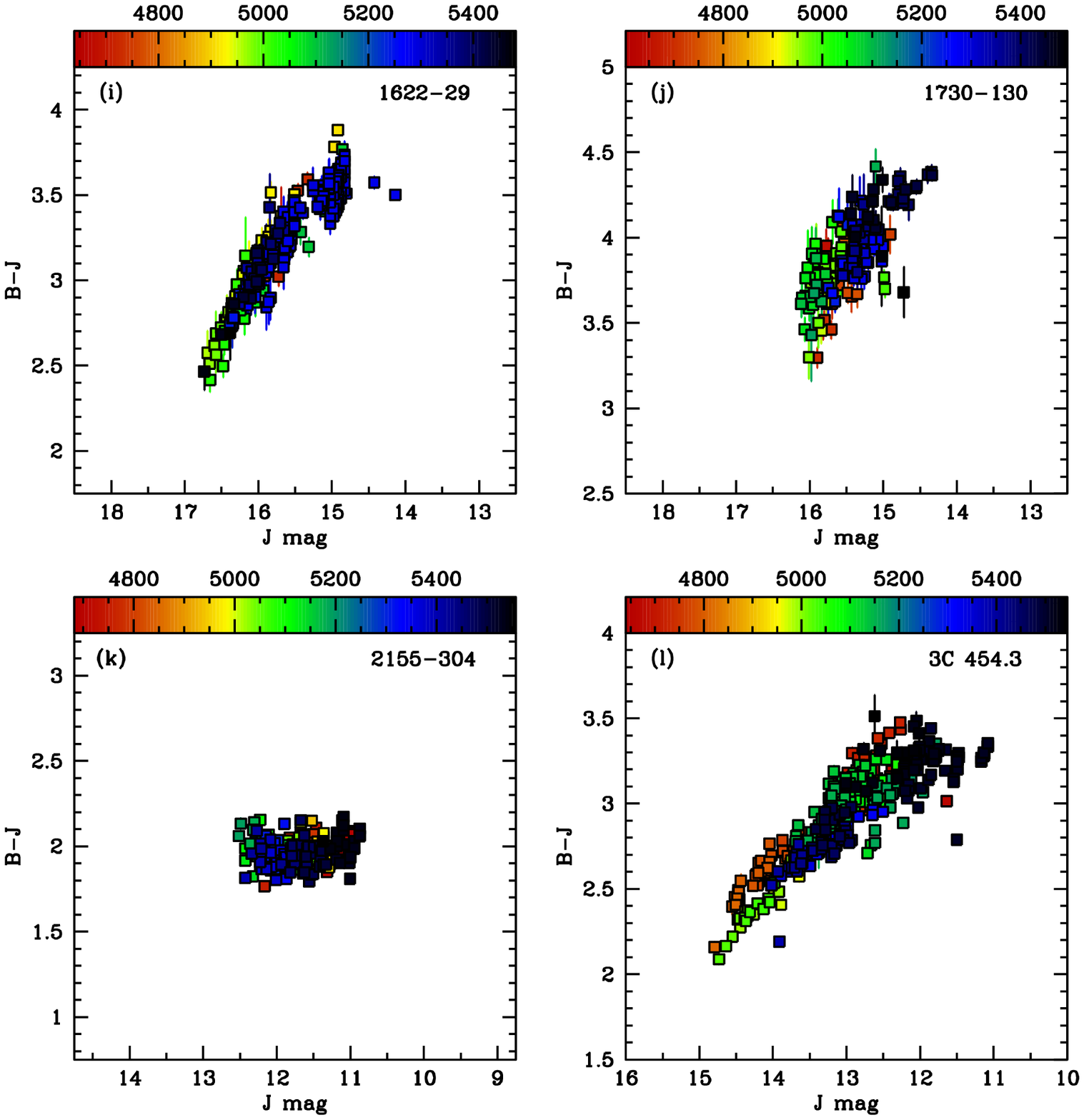} 
\caption*{Fig. 5 (cont'd): ($B-J$) color vs. $J$-band magnitude for FSRQs PKS~1622-29,
PKS~1730-130, and 3C~454.3, and the HBL PKS~2155-304. Color indicates
the date of the observation, as shown in the top bar in
MJD-50000. The FSRQs show the redder-when-brighter trend. However, for the
HBL, the brightness and spectral changes are relatively small, and
there is no trend of color with magnitude. This may be because the HBL
both lacks a luminous accretion disk and the optical jet emission in
this object is below the synchrotron peak, thus optically thick and
not highly variable.}
\end{center}
\end{figure}

\clearpage

\begin{figure}[]
\begin{center}
\plotone{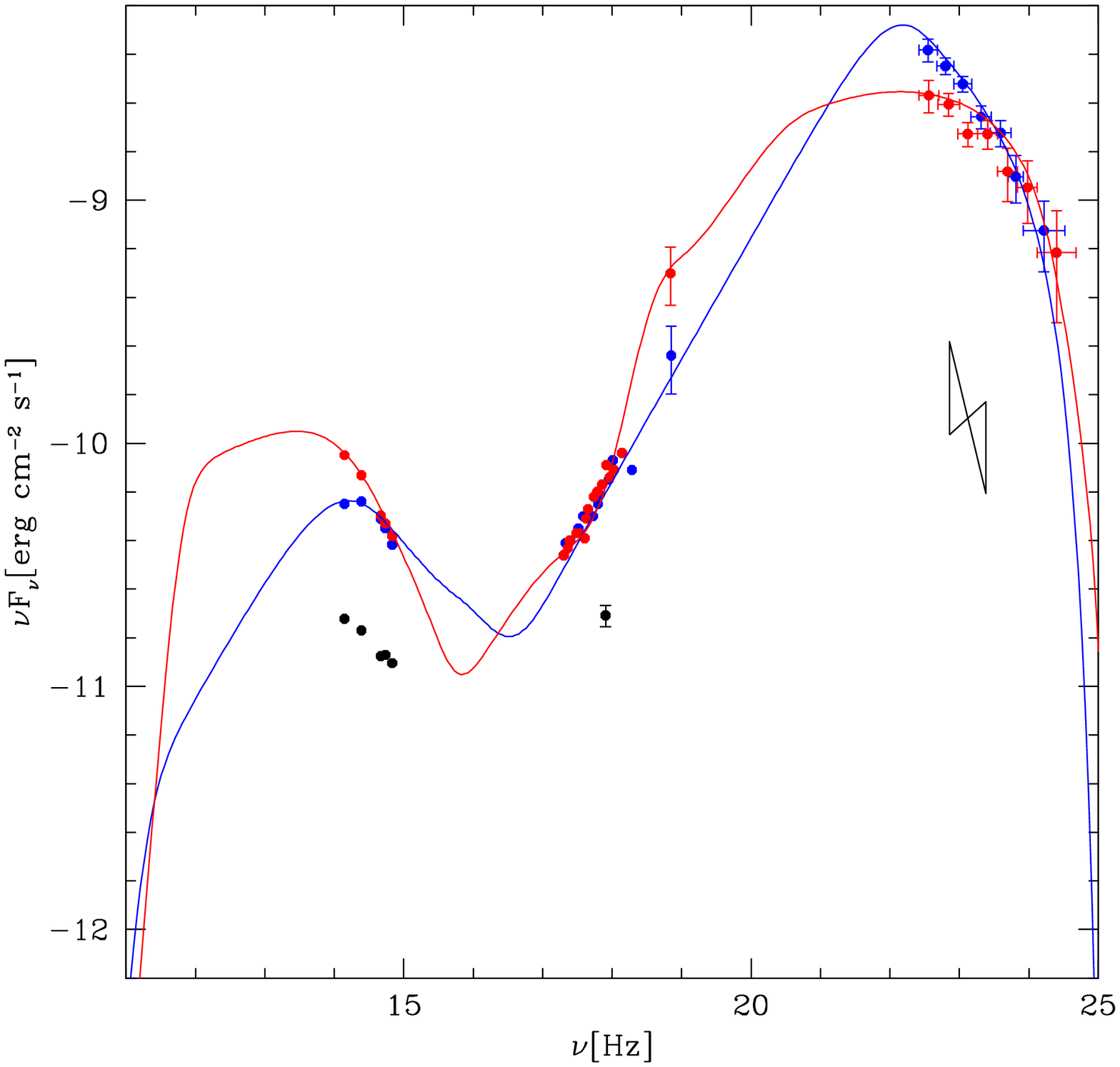} 
\caption*{Fig. 6: Spectral energy distribution of 3C~454.3 from data obtained
during the bright flare on UT Dec 03, 2009 ($blue$ $points$) and UT
Dec 04, 2009 ($red$ $points$, and in a low state on UT Aug 12, 2009
(black points). Blue and red $lines$ are model fits to the entire
dataset using the one-zone code of \citet{coppi92}. Error bars on the
optical and NIR points are smaller than the plotting symbol. The
infrared SMARTS data clearly shows a change in synchrotron peak energy
between December 3 and 4; modeling this together with the changing
gamma-ray peak and slope poses challenges for simple single-zone
models. However, these simple fits are sufficient to indicate that
Klein-Nishina effects are likely important and that the optical-IR is
produced by higher energy electrons than are producing the gamma
rays. 
}\label{fig:3c454sed} 
\end{center}
\end{figure}

\clearpage
\appendix

\section{Optical Finding Charts}

We present optical finding charts for each of the 12 SMARTS-monitored
blazars discussed in this paper. All figures are $V$-band.  Field of
view is approximately 6$'\times$ 6$'$.  North is at top of image, East is to
the left. Comparison stars are numbered; calibrated magnitudes for our
comparison stars are given in Appendix C.

\begin{figure}[]
\begin{center}
\plotone{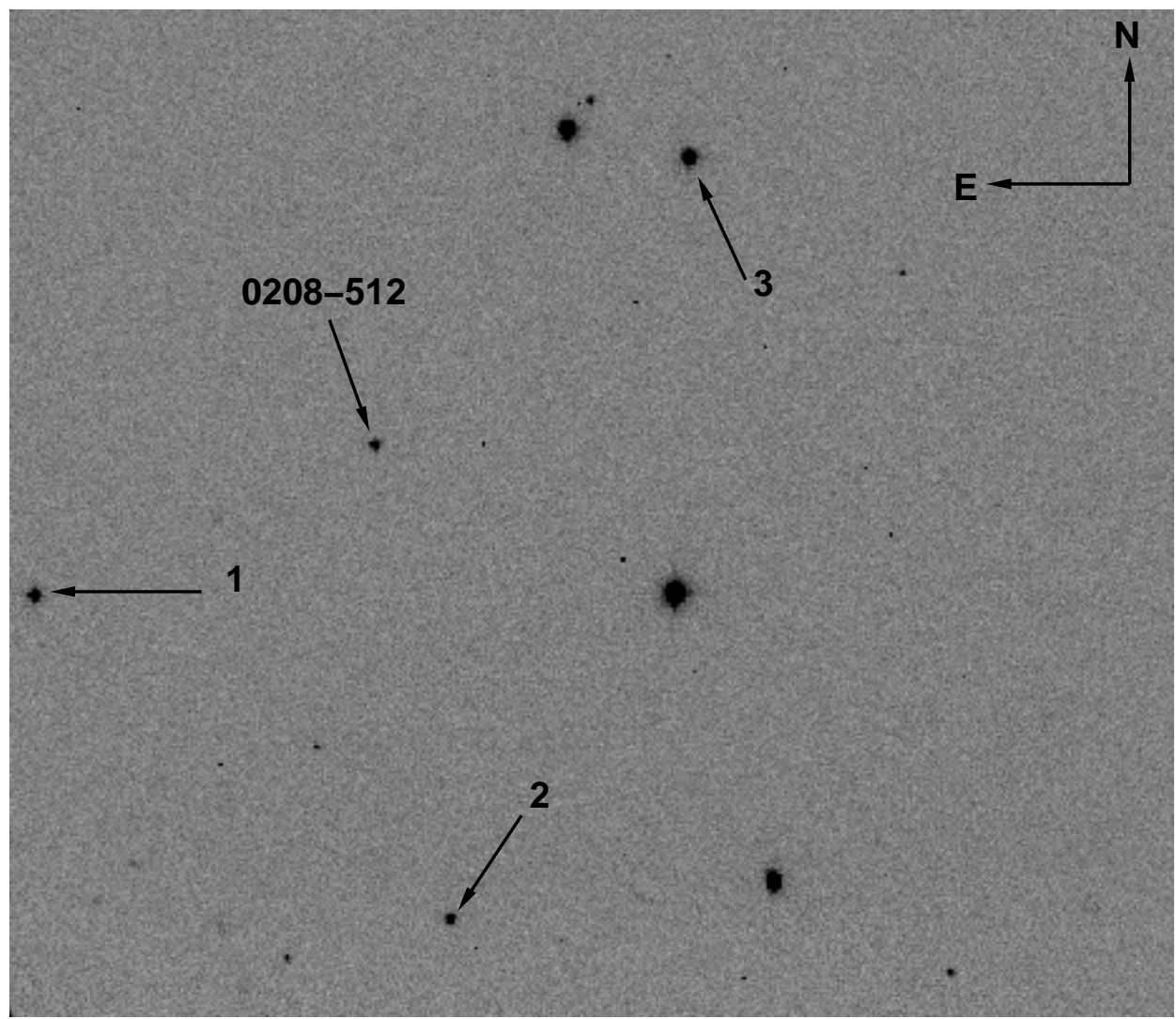} 
\caption*{A.1: PKS~0208$-$512.  }\label{fig:0208opfc}
\end{center}
\end{figure}

\begin{figure}[]
\begin{center}
\plotone{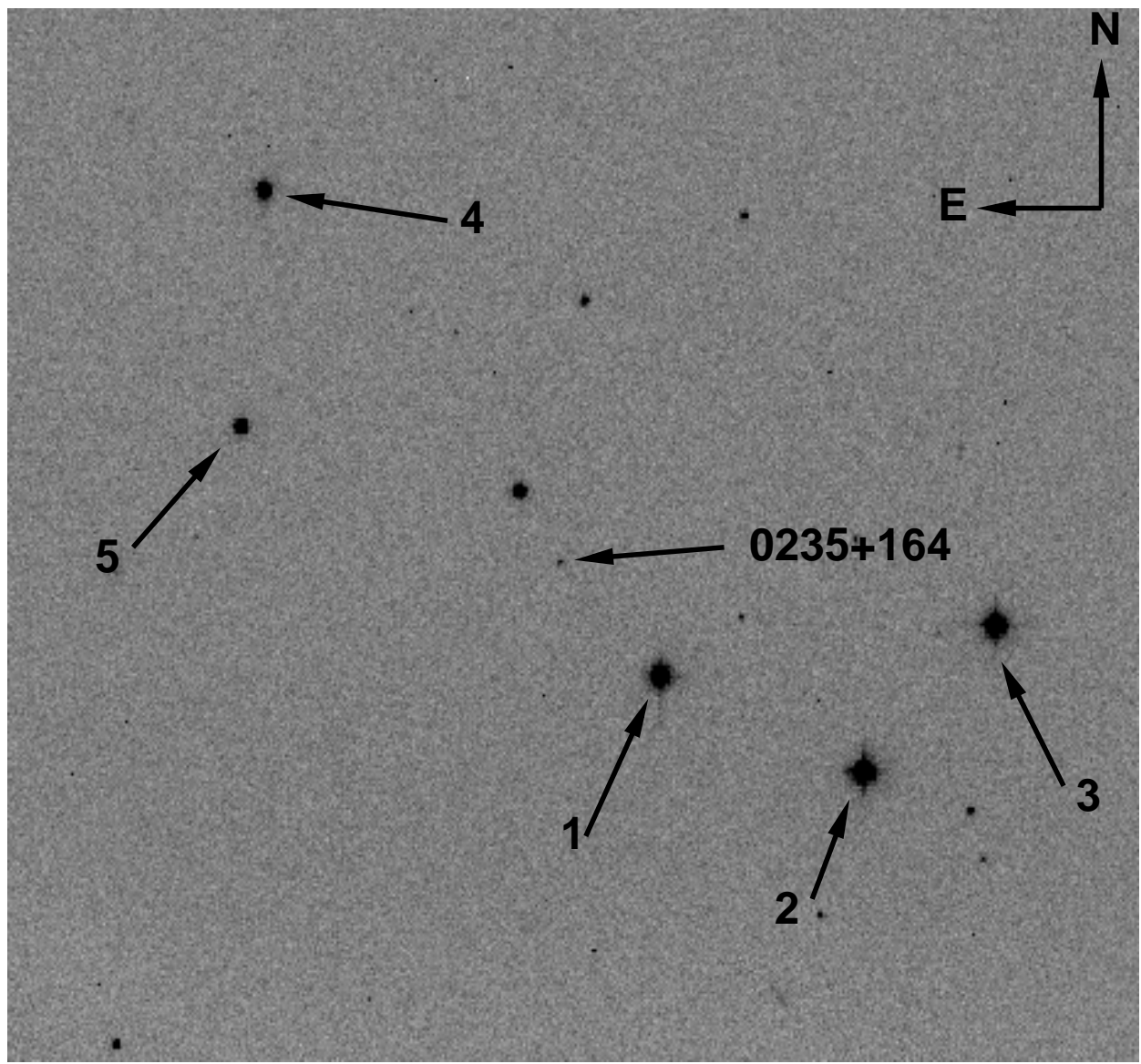} 
\caption*{A.2: AO~0235+164.  }\label{fig:0235opfc}
\end{center}
\end{figure}

\begin{figure}[]
\begin{center}
\plotone{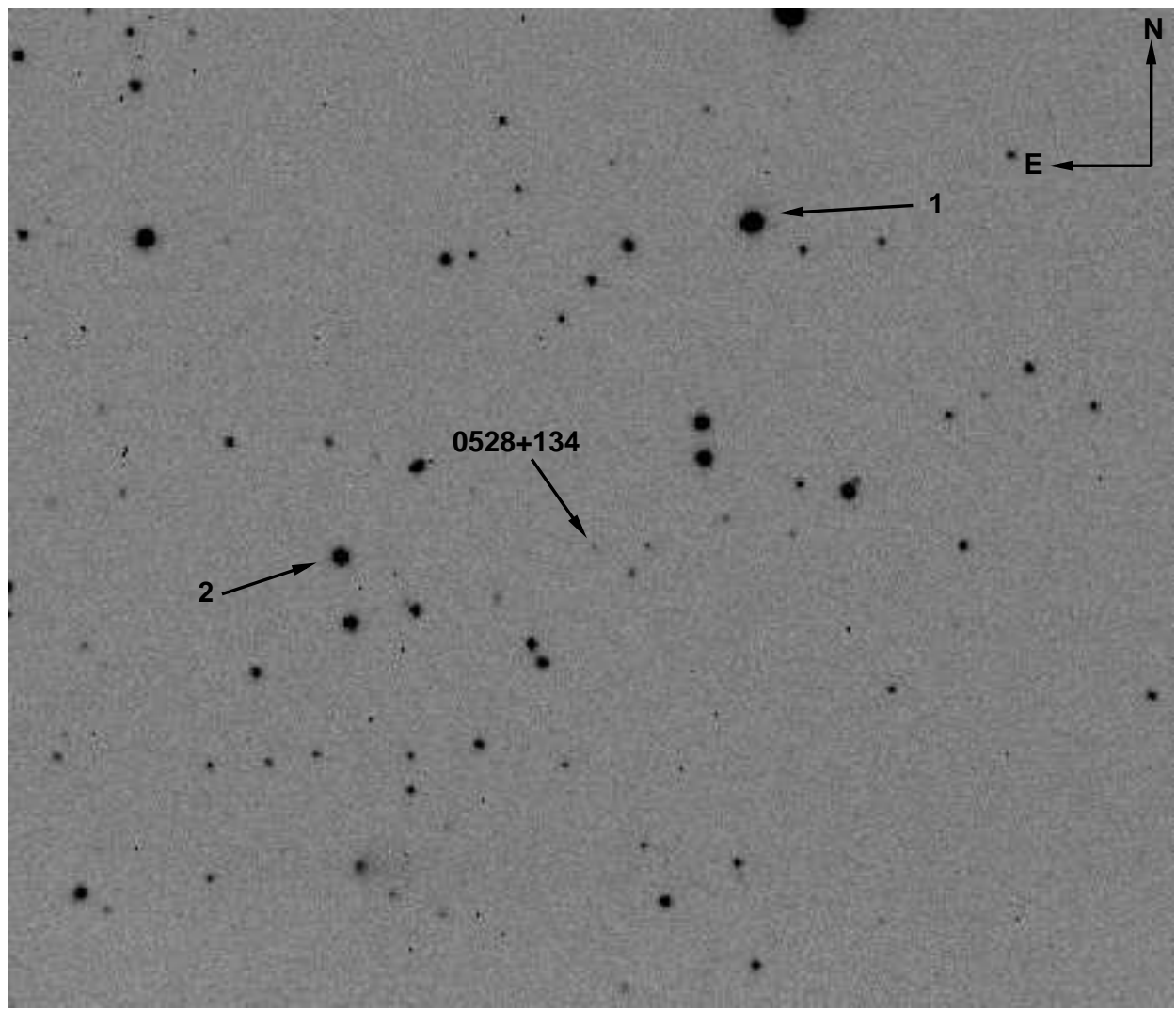} 
\caption*{A.3: PKS~0528+134.  }\label{fig:0528opfc}
\end{center}
\end{figure}

\begin{figure}[]
\begin{center}
\plotone{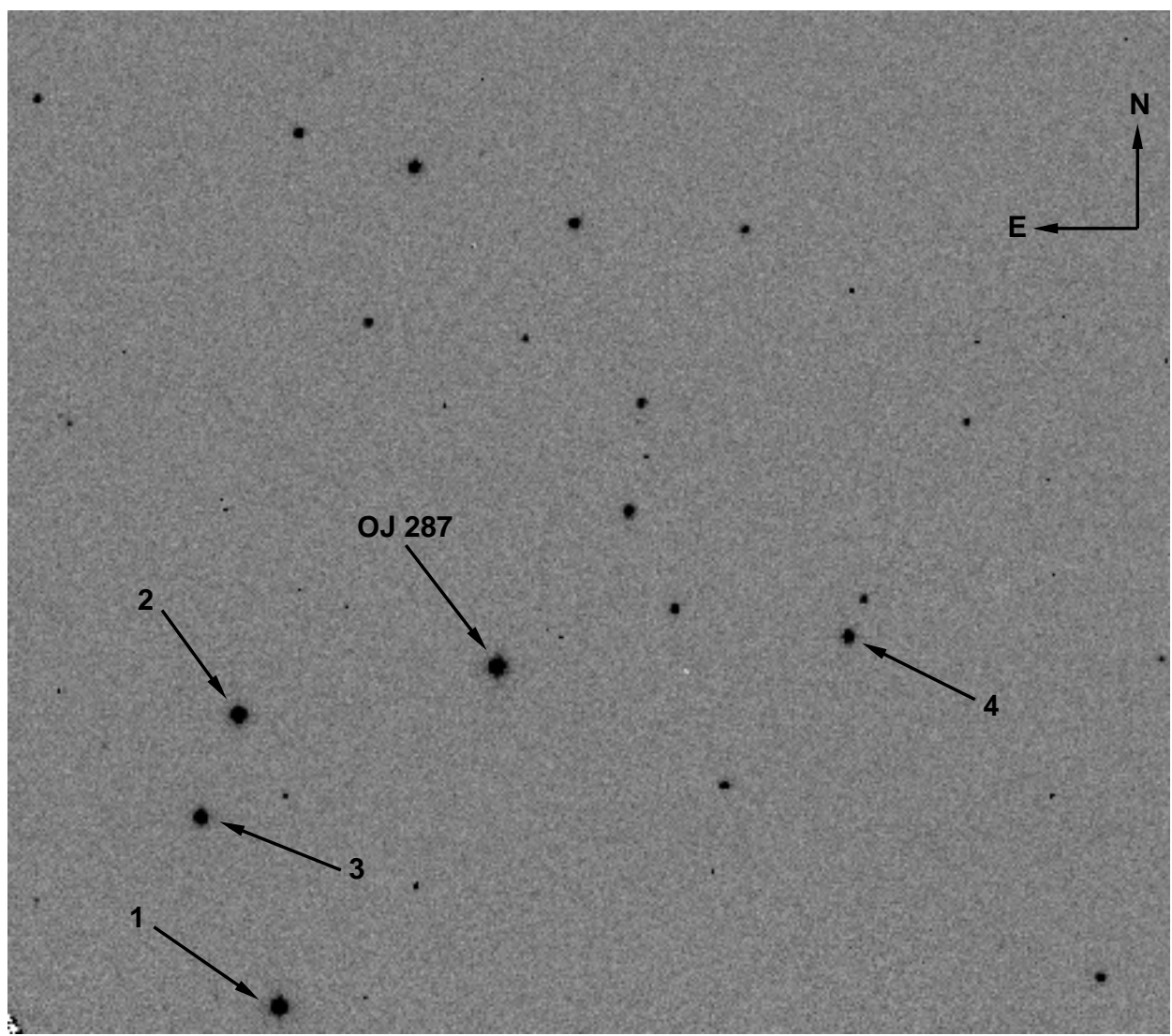} 
\caption*{A.4: OJ~287.  }\label{fig:OJ287opfc}
\end{center}
\end{figure}

\begin{figure}[]
\begin{center}
\plotone{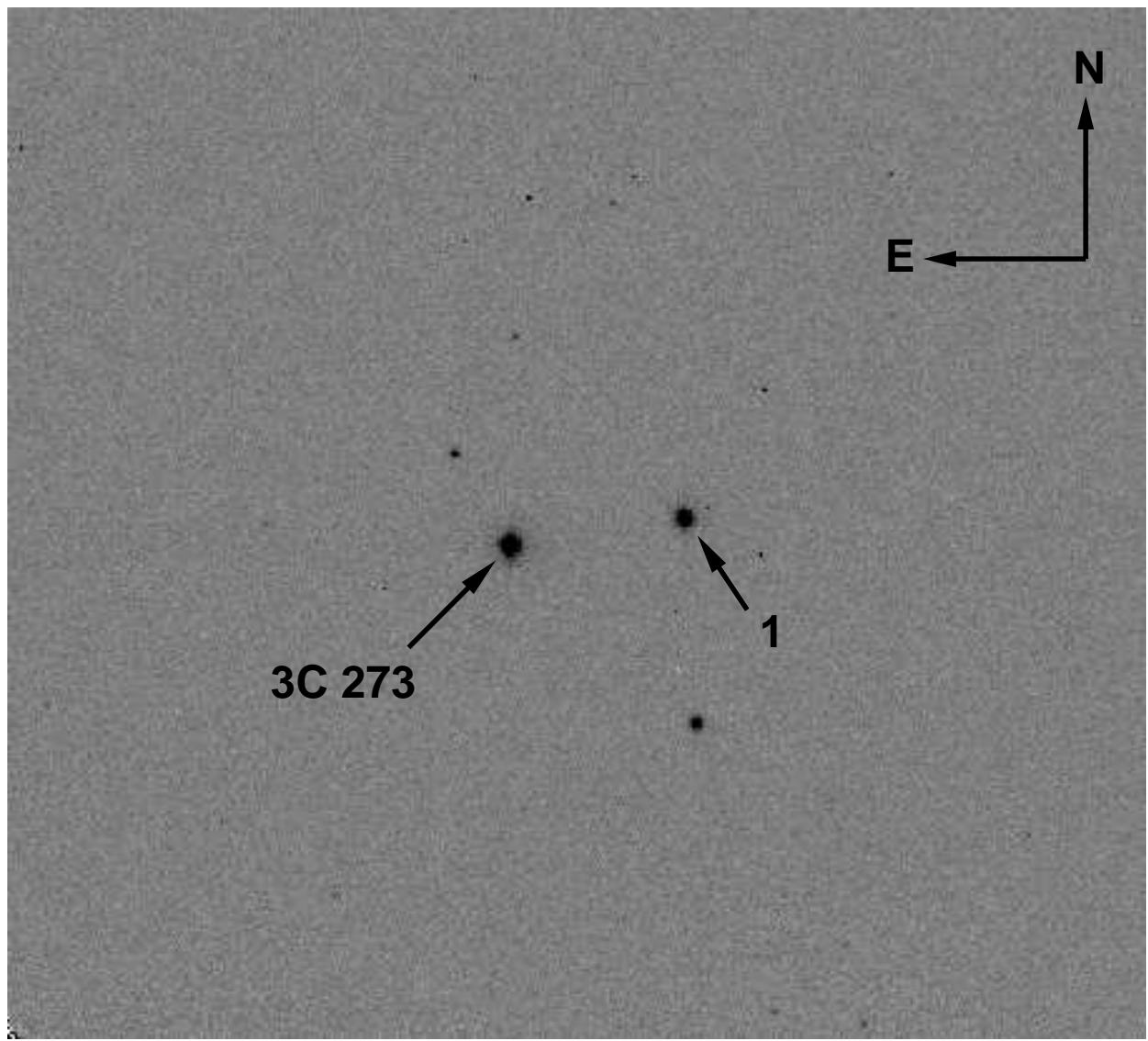} 
\caption*{A.5: 3C~273.  }\label{fig:3C273opfc}
\end{center}
\end{figure}

\begin{figure}[]
\begin{center}
\plotone{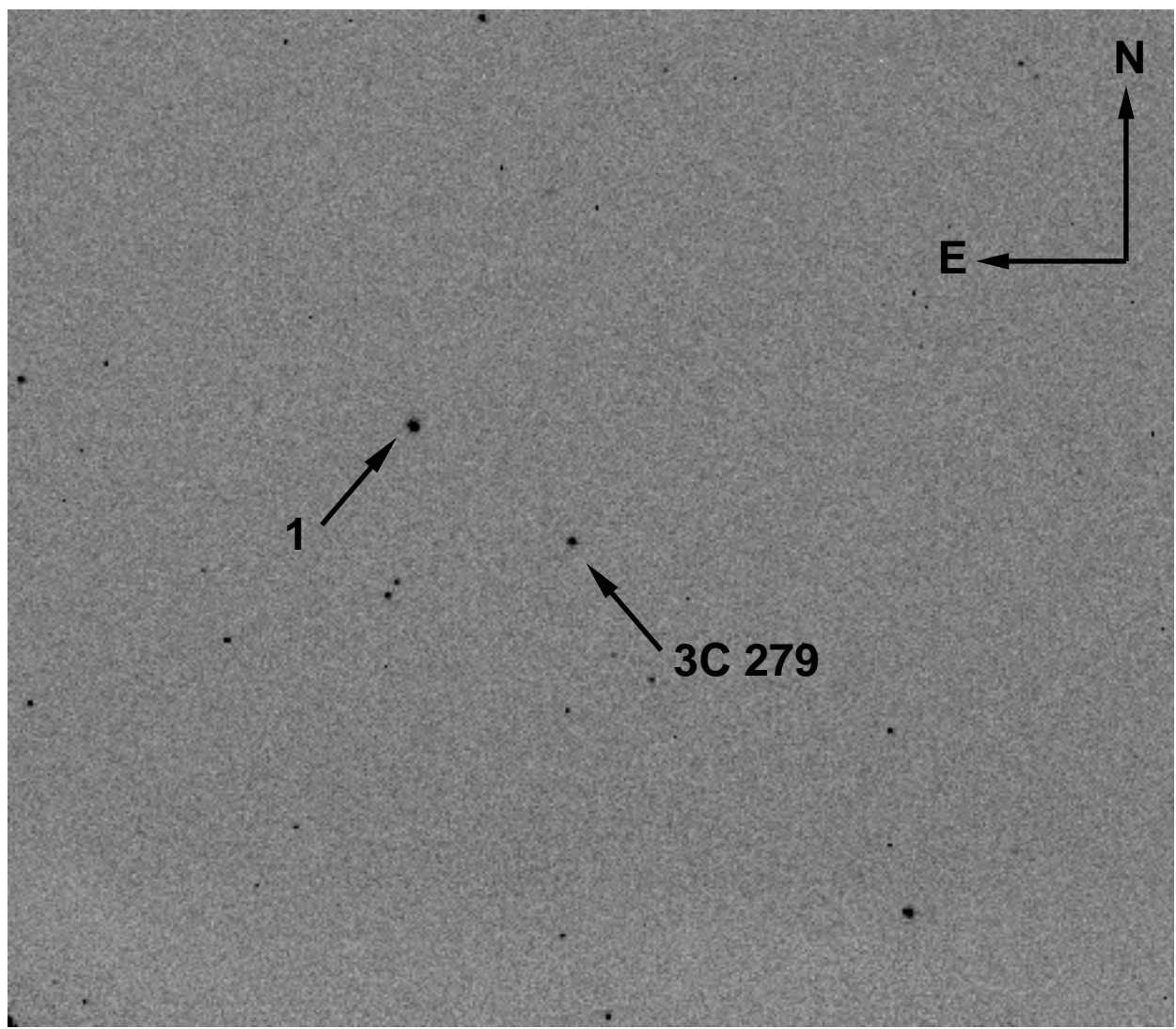} 
\caption*{A.6: 3C~279.  }\label{fig:3C279opfc}
\end{center}
\end{figure}

\begin{figure}[]
\begin{center}
\plotone{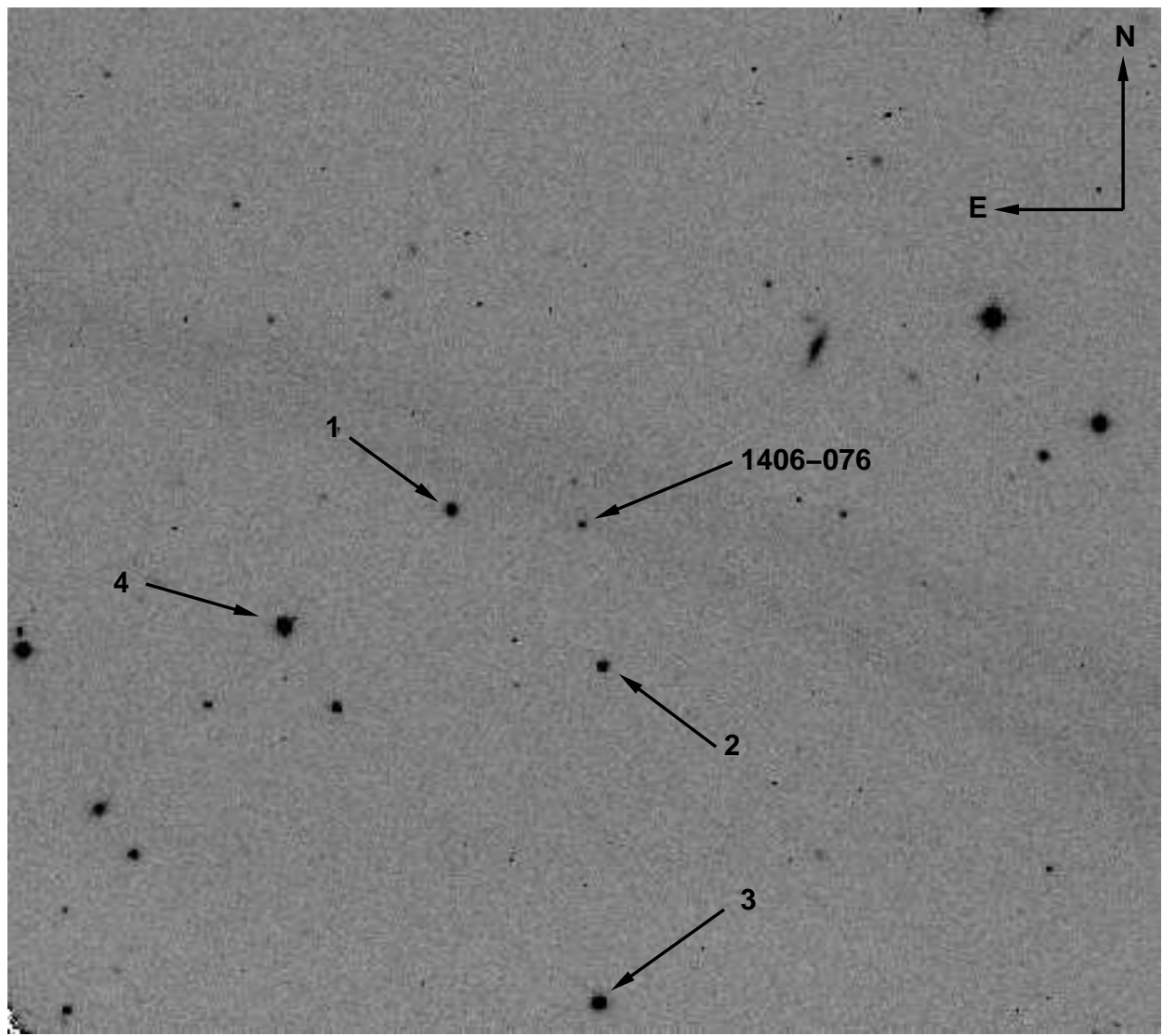} 
\caption*{A.7: PKS~1406$-$076.  }\label{fig:1406opfc}
\end{center}
\end{figure}

\begin{figure}[]
\begin{center}
\plotone{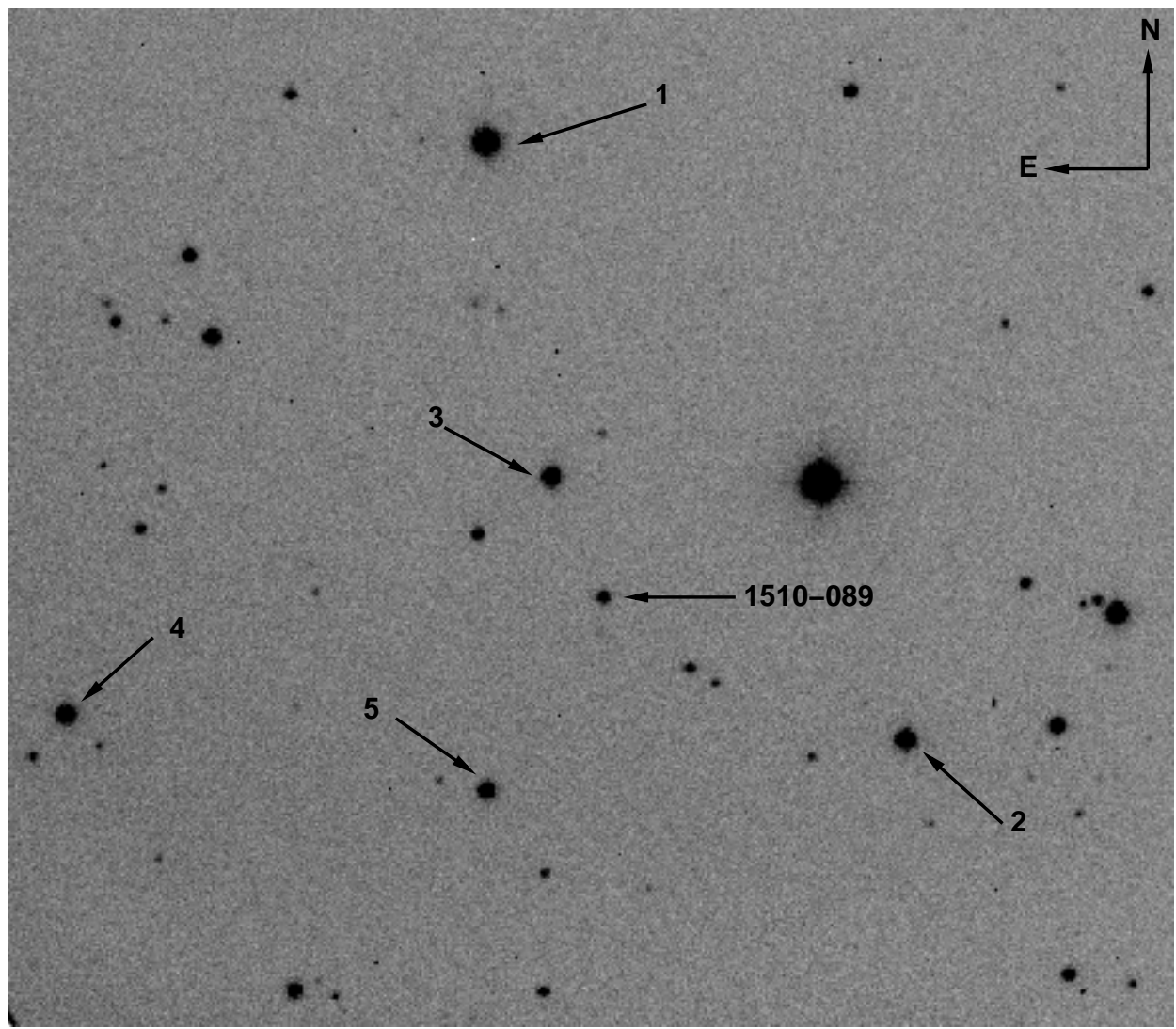} 
\caption*{A.8: PKS~1510$-$089.  }\label{fig:1510opfc}
\end{center}
\end{figure}

\begin{figure}[]
\begin{center}
\plotone{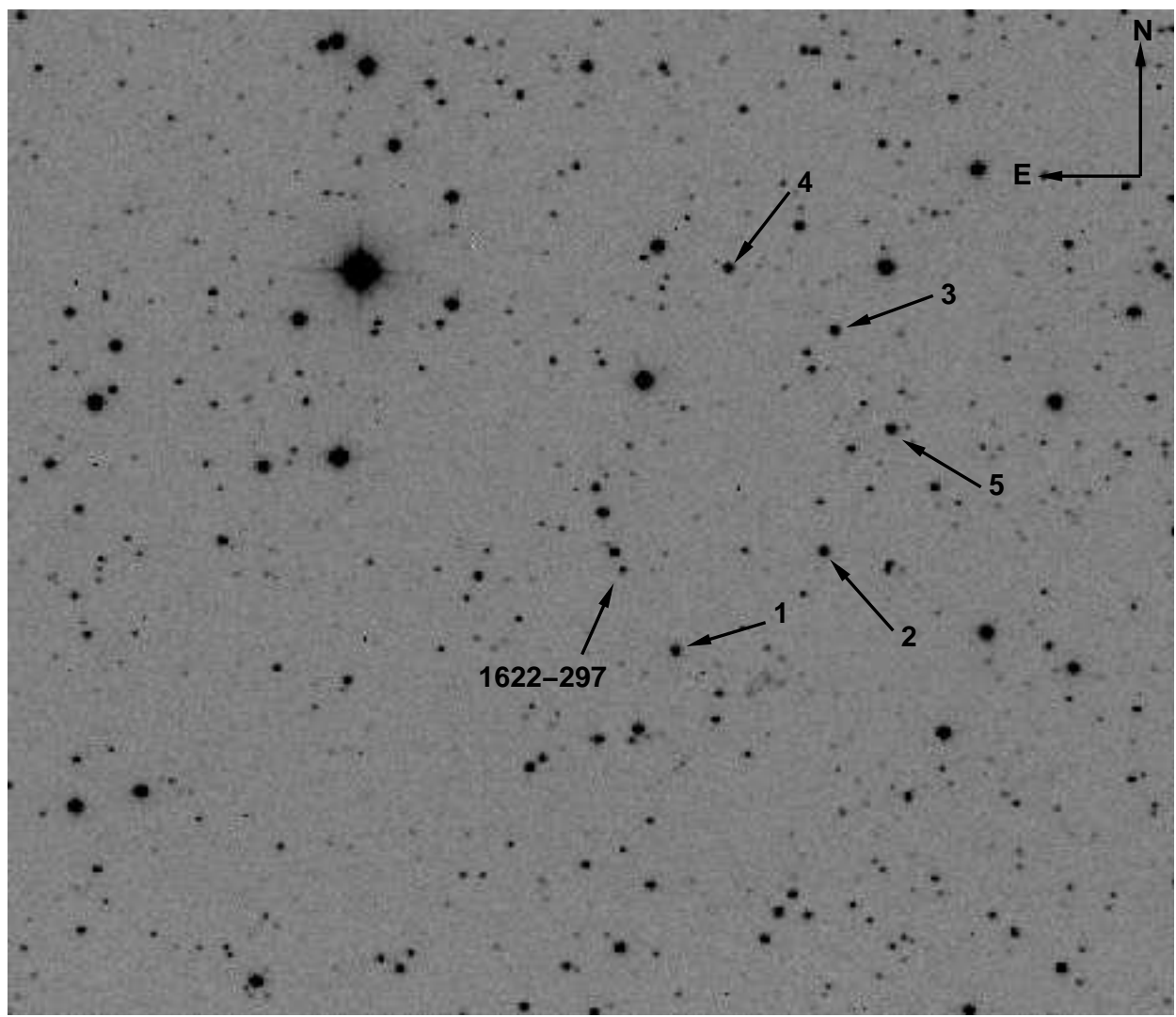} 
\caption*{A.9: PKS~1622$-$297. }\label{fig:1622opfc}
\end{center}
\end{figure}

\begin{figure}[]
\begin{center}
\plotone{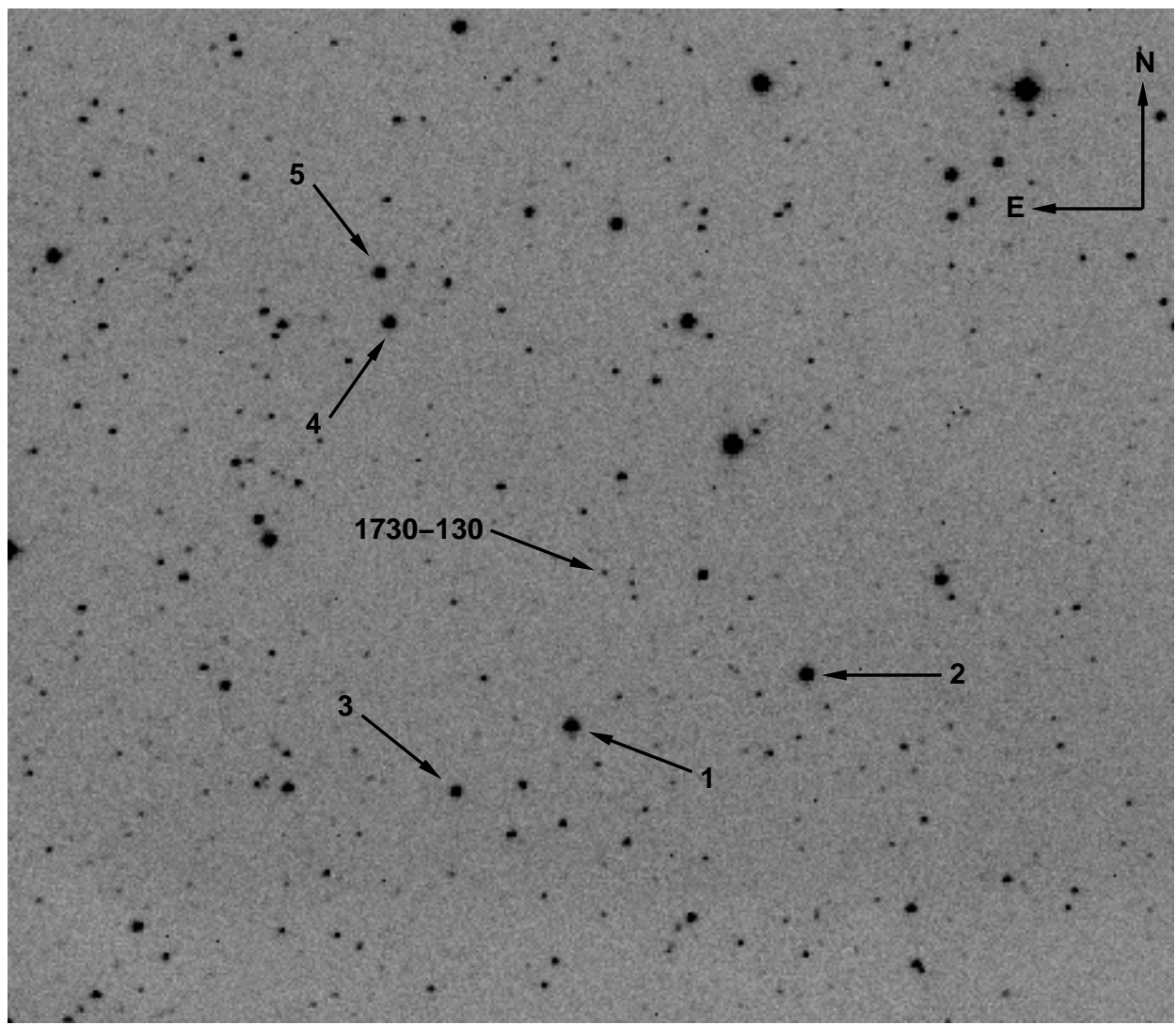} 
\caption*{A.10: PKS~1730$-$130.  }\label{fig:1730opfc}
\end{center}
\end{figure}

\begin{figure}[]
\begin{center}
\plotone{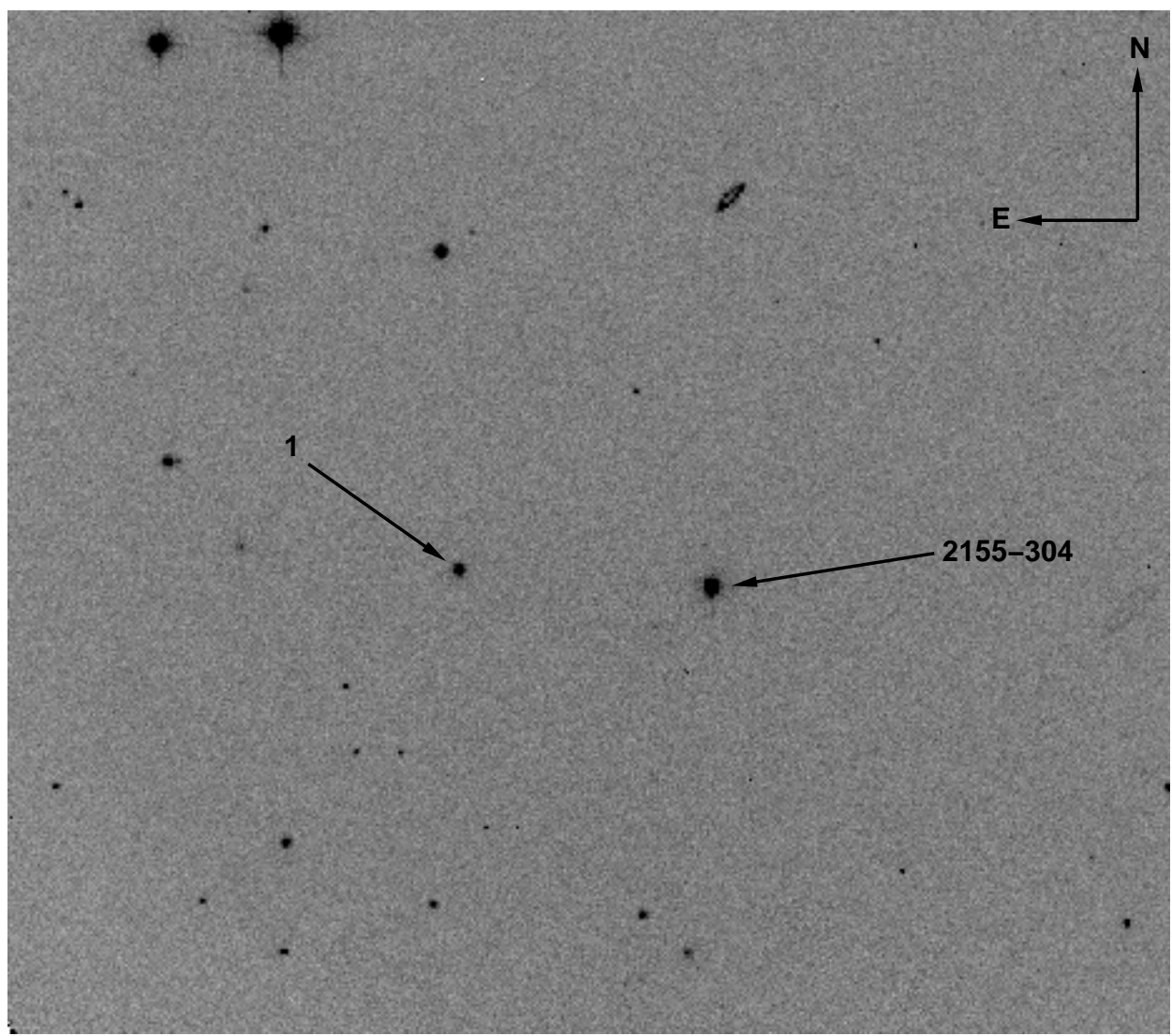} 
\caption*{A.11: PKS~2155$-$304.  }\label{fig:2155opfc}
\end{center}
\end{figure}

\begin{figure}[]
\begin{center}
\plotone{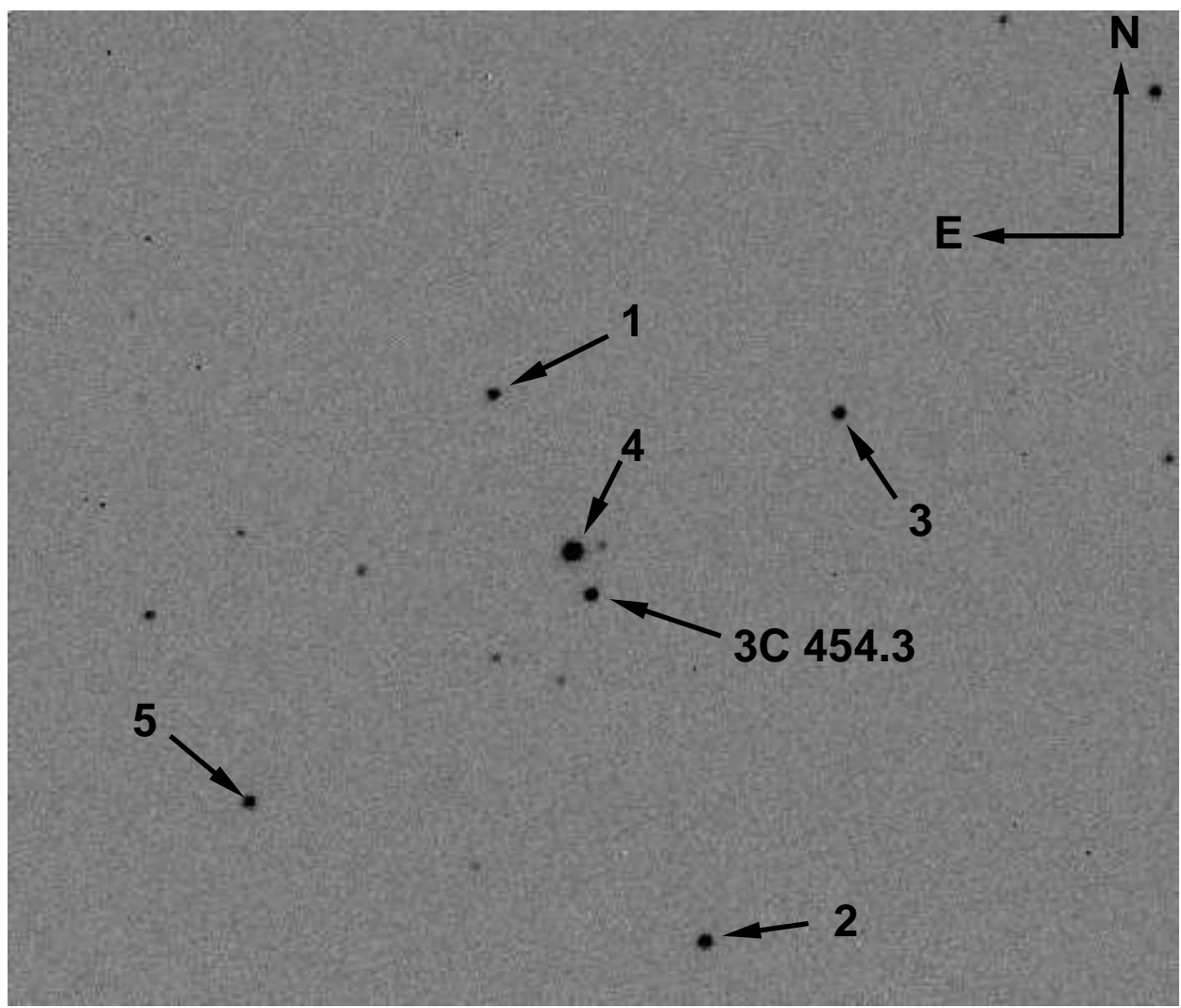} 
\caption*{A.12: 3C~454.3.  }\label{fig:3C454opfc}
\end{center}
\end{figure}

\clearpage

\section{Near-Infrared Finding Charts}

We present near-infrared finding charts for each of the 12
SMARTS-monitored blazars discussed in this paper. All figures are
$J$-band.  Field of view is approximately 2$' \times$ 2$'$.  North is
to the right of image, East is at the top. Comparison stars are
numbered; calibrated magnitudes for our comparison stars are given in
Appendix C.

\begin{figure}[]
\begin{center}
\plotone{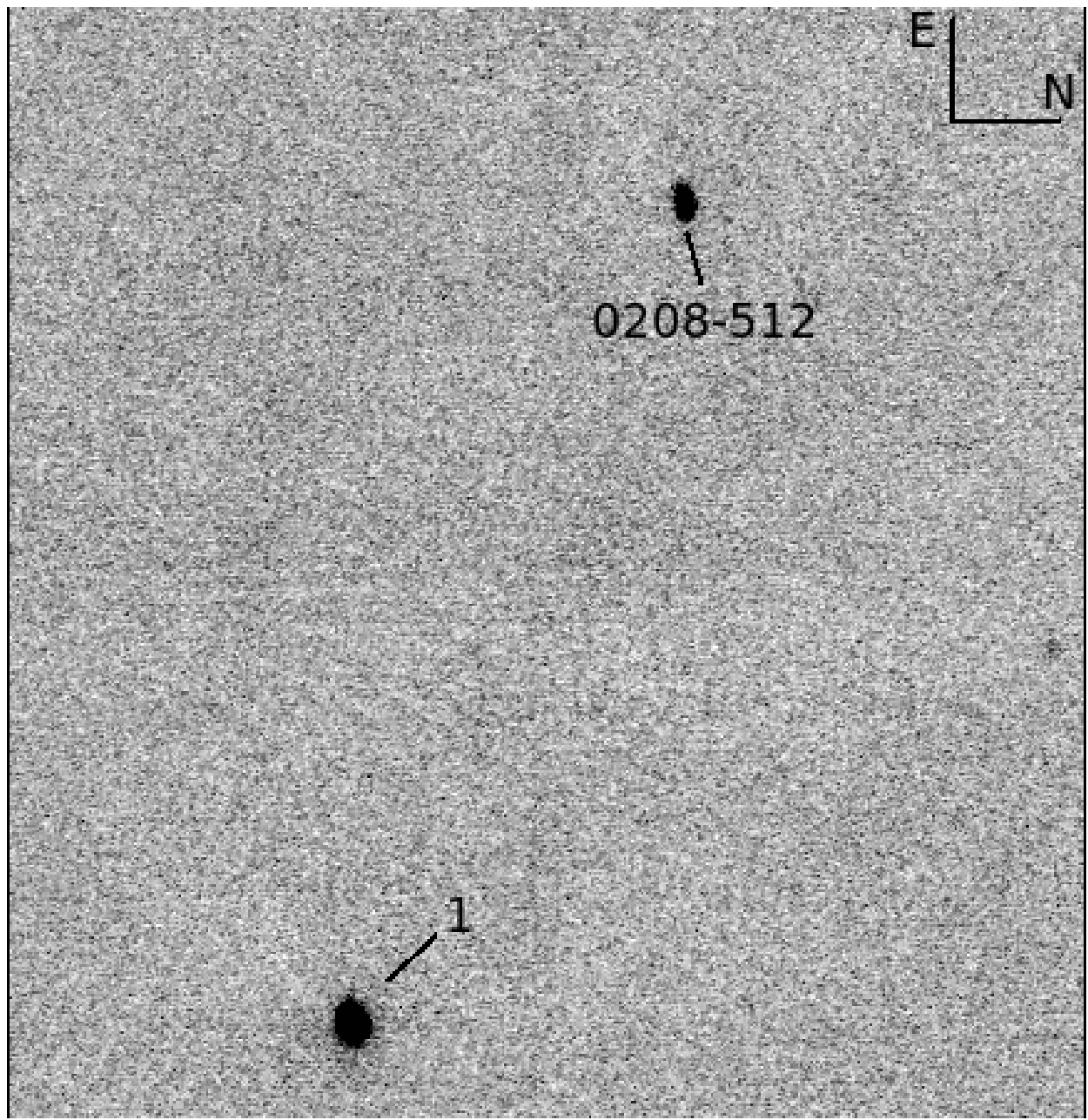} 
\caption*{B.1: PKS~0208$-$512.}\label{fig:0208irfc}
\end{center}
\end{figure}

\begin{figure}[]
\begin{center}
\plotone{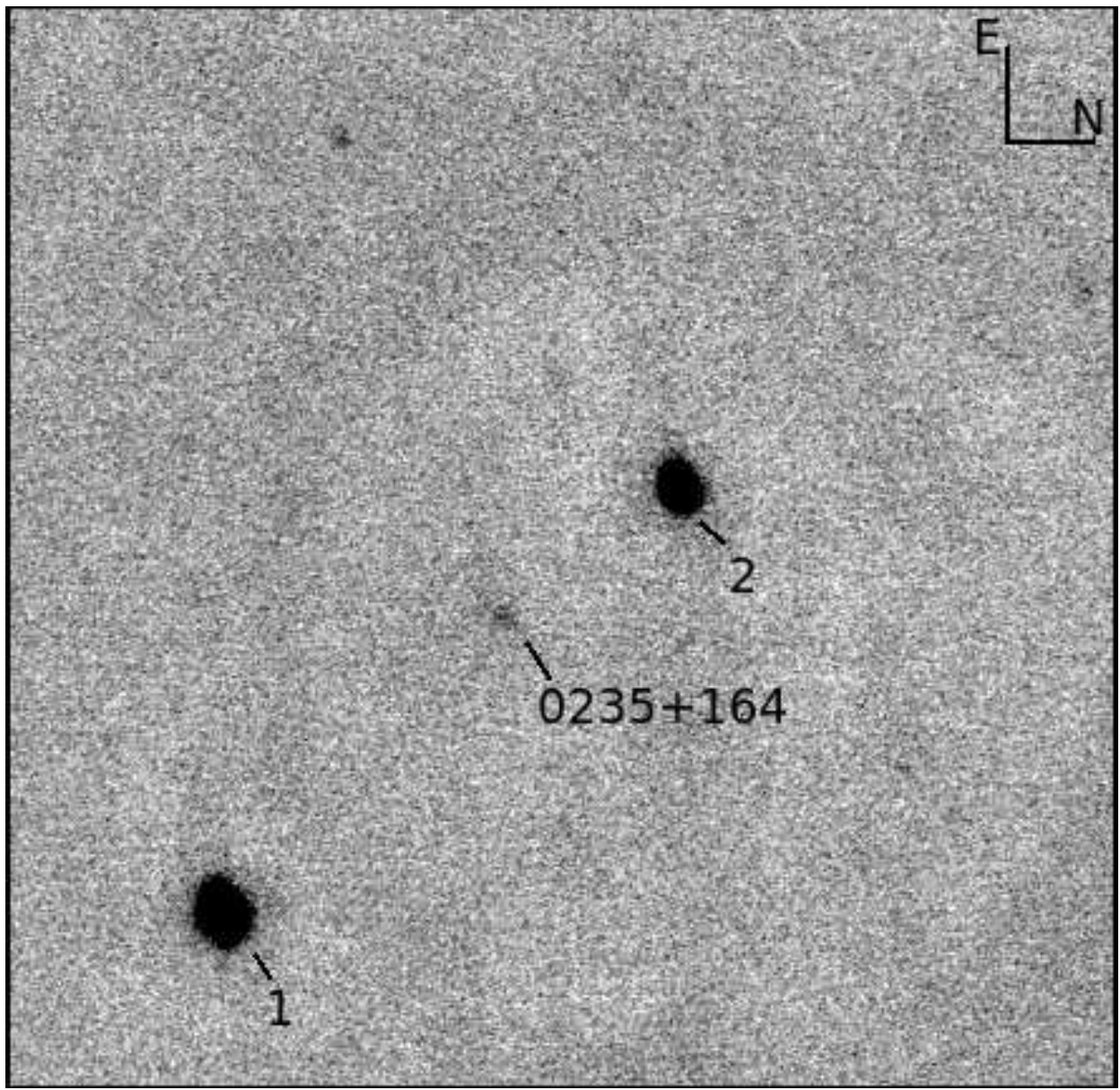} 
\caption*{B.2: AO~0235+164.  }\label{fig:0235irfc}
\end{center}
\end{figure}

\begin{figure}[]
\begin{center}
\plotone{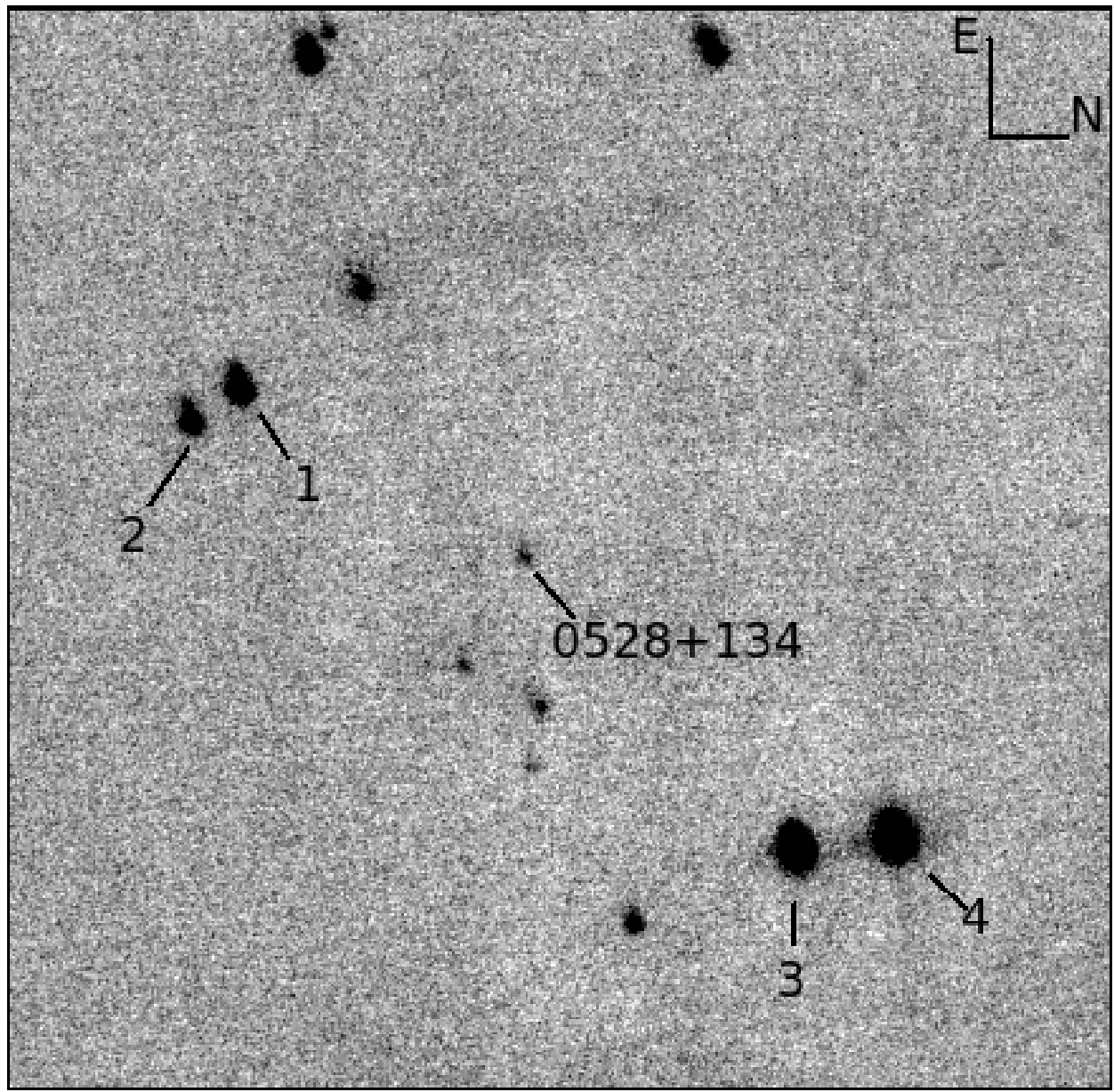} 
\caption*{B.3: PKS~0528+134.  }\label{fig:0528irfc}
\end{center}
\end{figure}

\begin{figure}[]
\begin{center}
\plotone{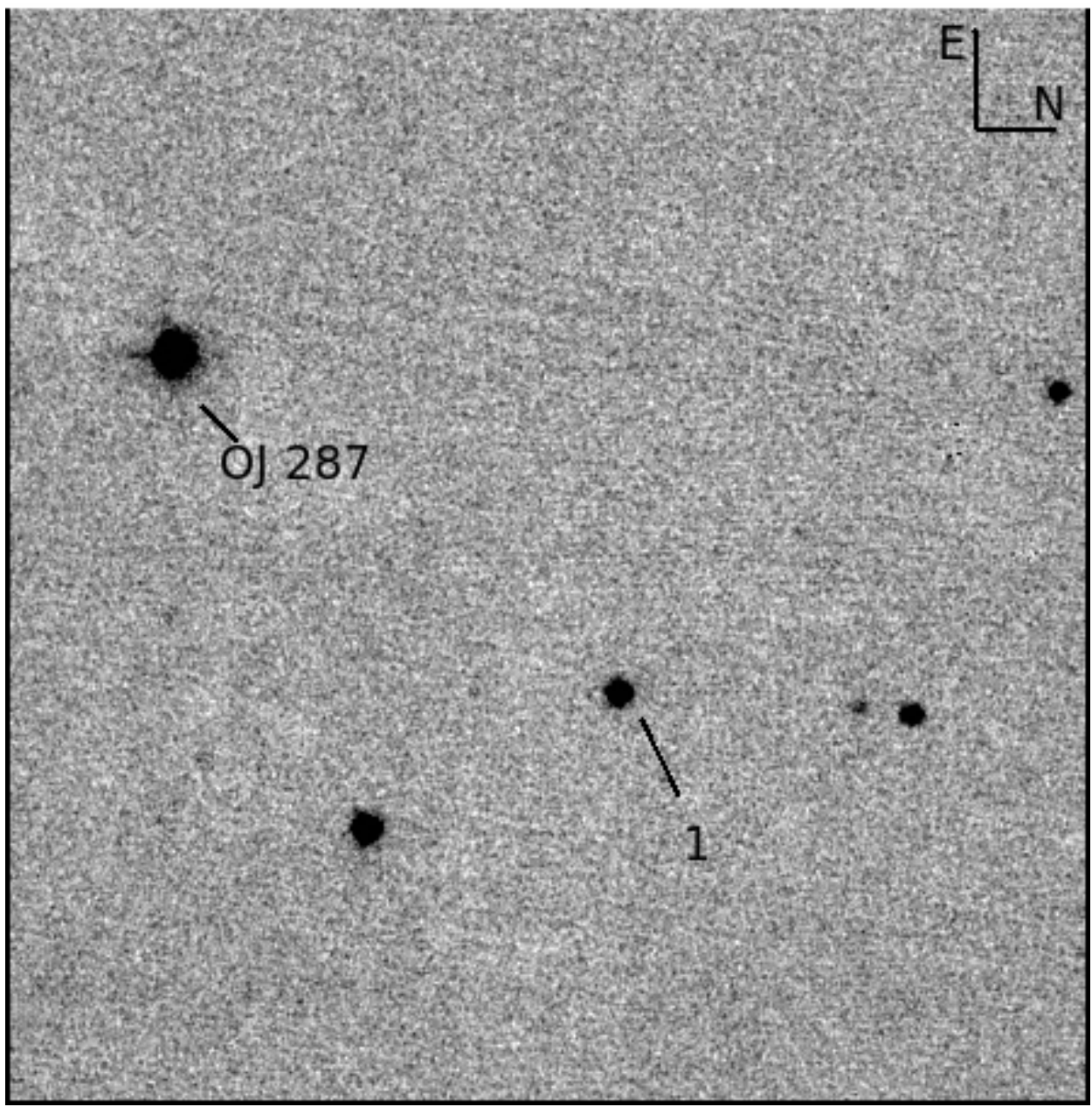} 
\caption*{B.4: OJ~287.  }\label{fig:OJ287irfc}
\end{center}
\end{figure}

\begin{figure}[]
\begin{center}
\plotone{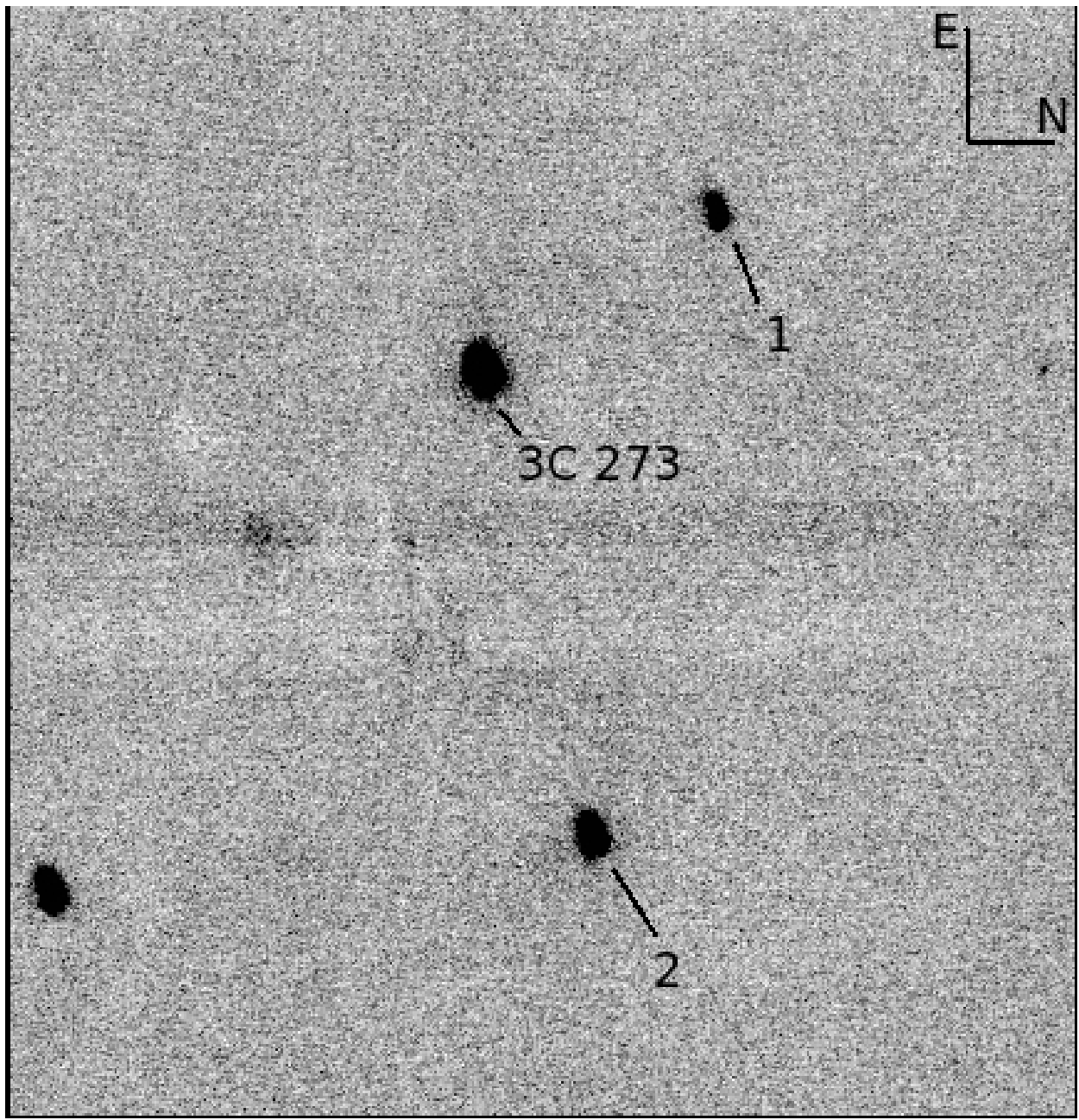} 
\caption*{B.5: 3C~273.  }\label{fig:3C273irfc}
\end{center}
\end{figure}

\begin{figure}[]
\begin{center}
\plotone{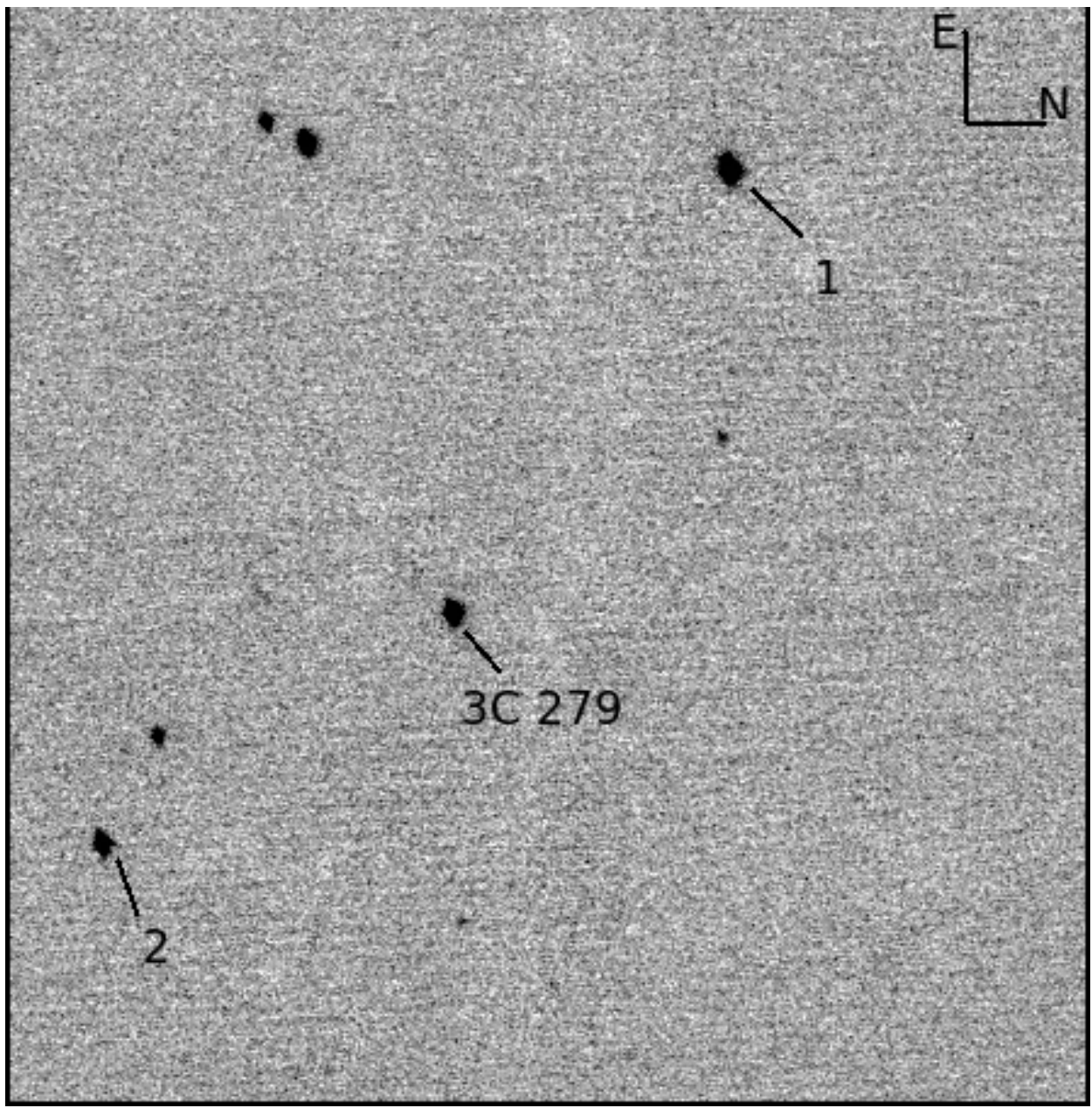} 
\caption*{B.6: 3C~279.  }\label{fig:3C279irfc}
\end{center}
\end{figure}

\begin{figure}[]
\begin{center}
\plotone{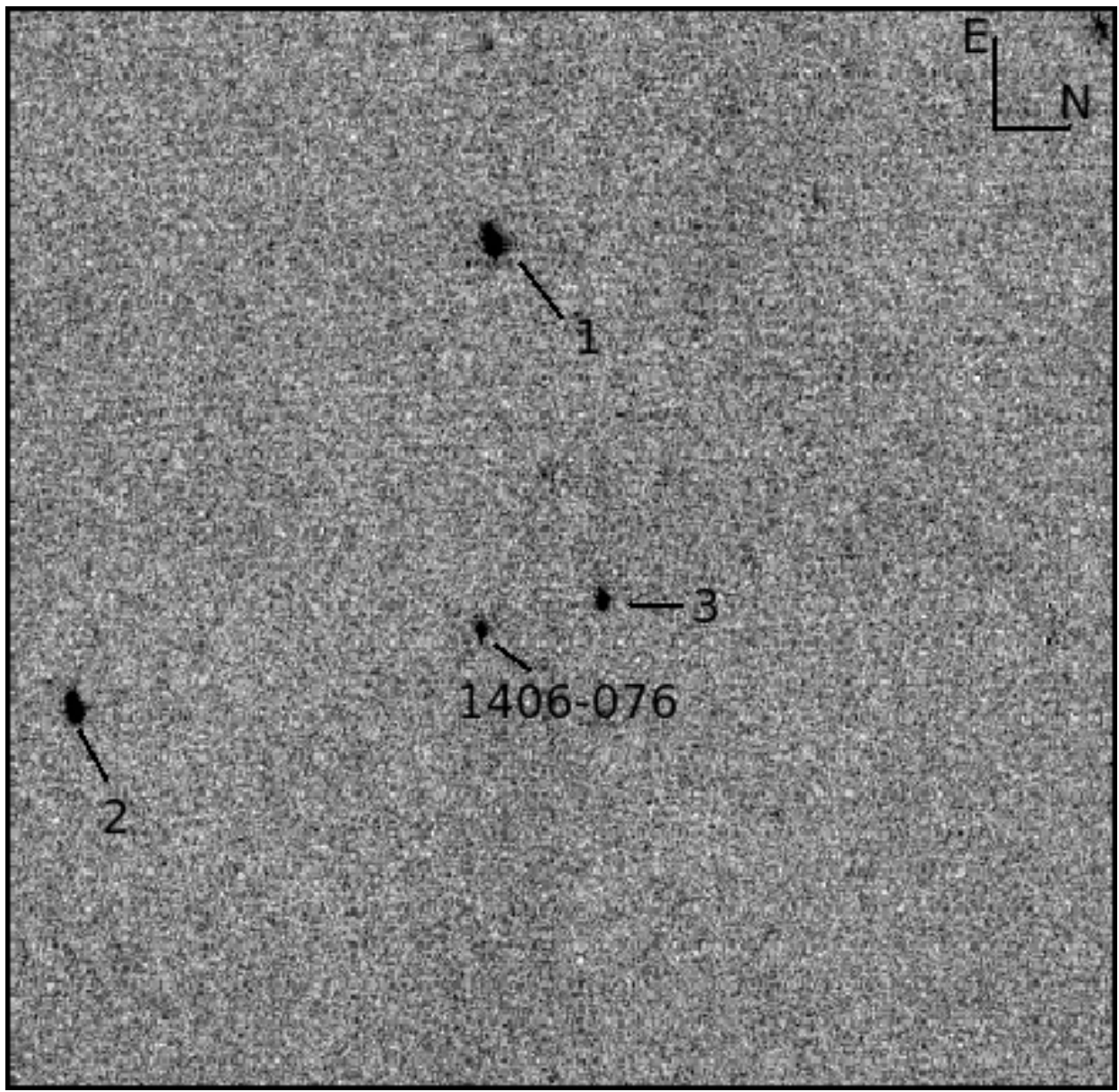} 
\caption*{B.7: PKS~1406$-$076.  }\label{fig:1406irfc}
\end{center}
\end{figure}

\begin{figure}[]
\begin{center}
\plotone{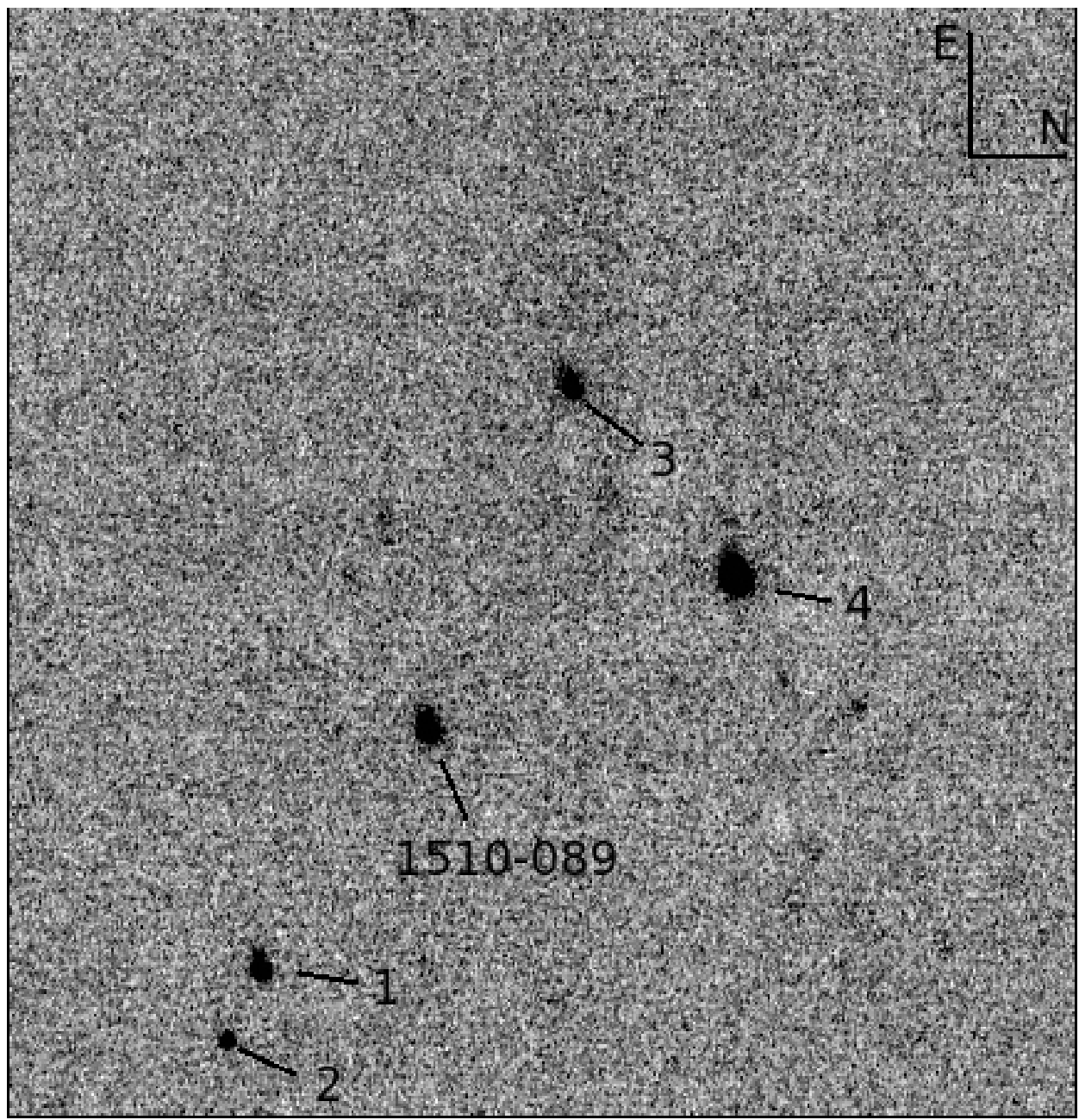} 
\caption*{B.8: PKS~1510$-$089. }\label{fig:1510irfc}
\end{center}
\end{figure}

\begin{figure}[]
\begin{center}
\plotone{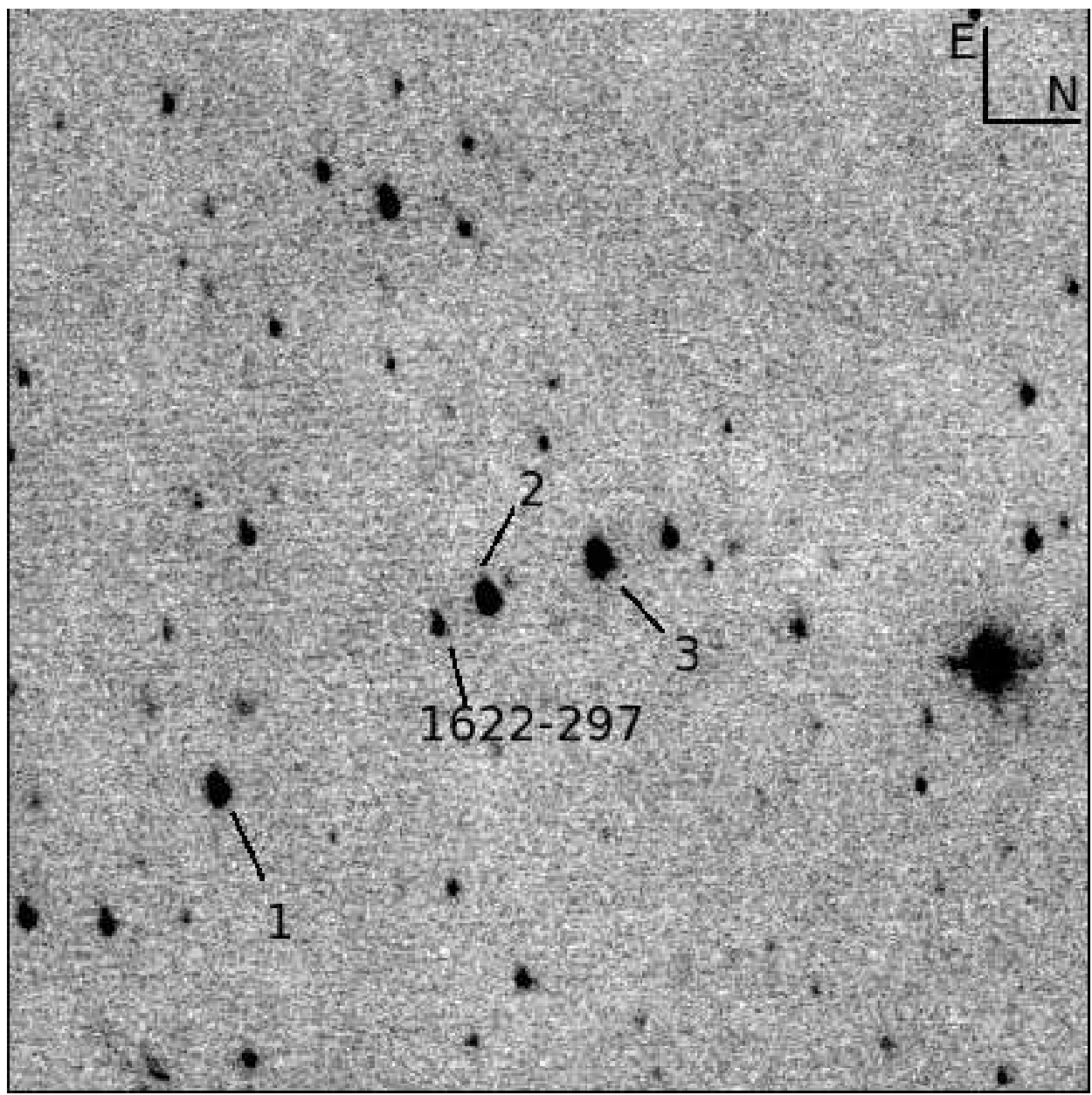} 
\caption*{B.9: PKS~1622$-$297.  }\label{fig:1622irfc}
\end{center}
\end{figure}

\begin{figure}[]
\begin{center}
\plotone{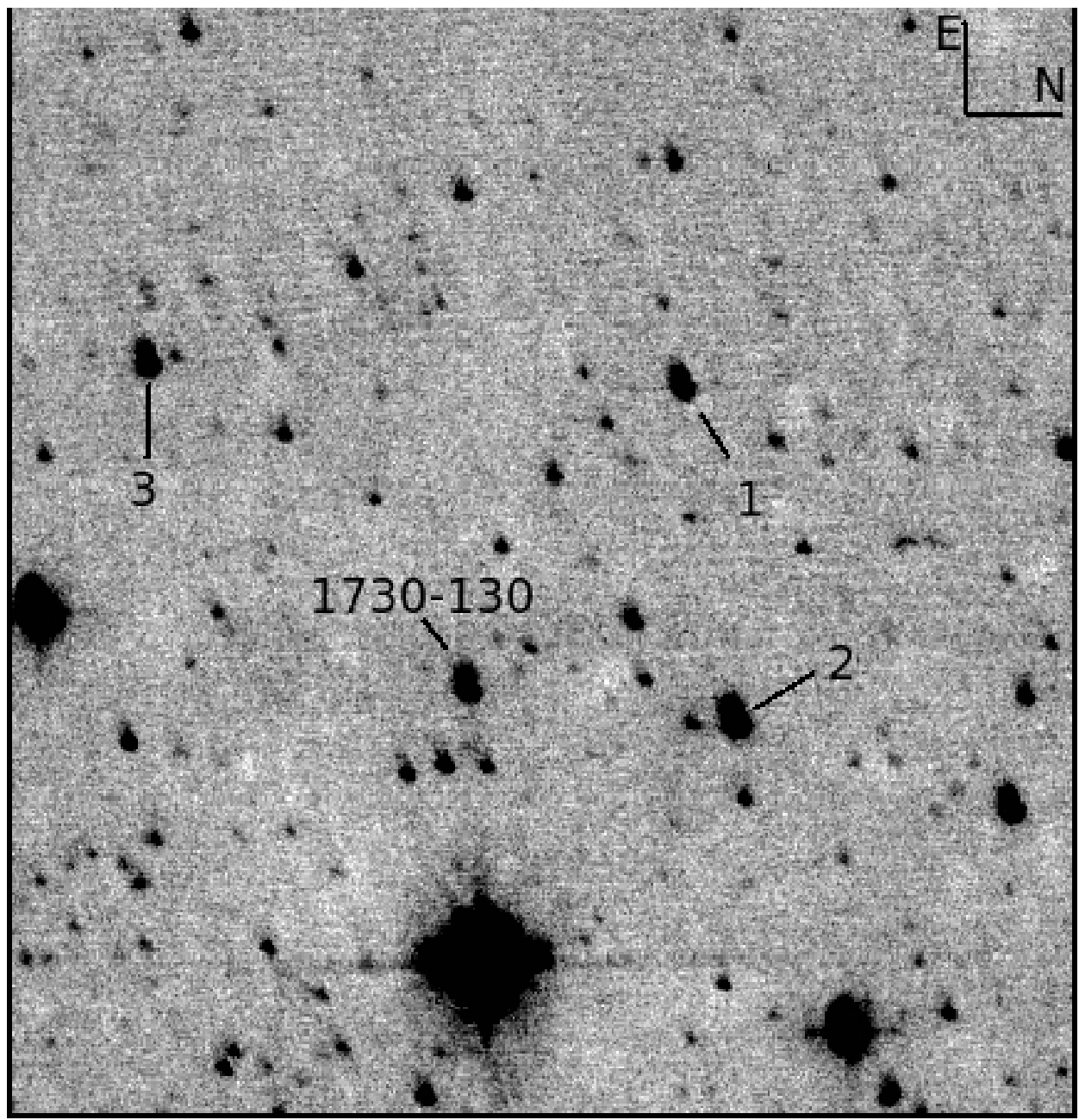} 
\caption*{B.10: PKS~1730$-$130.  }\label{fig:1730irfc}
\end{center}
\end{figure}

\begin{figure}[]
\begin{center}
\plotone{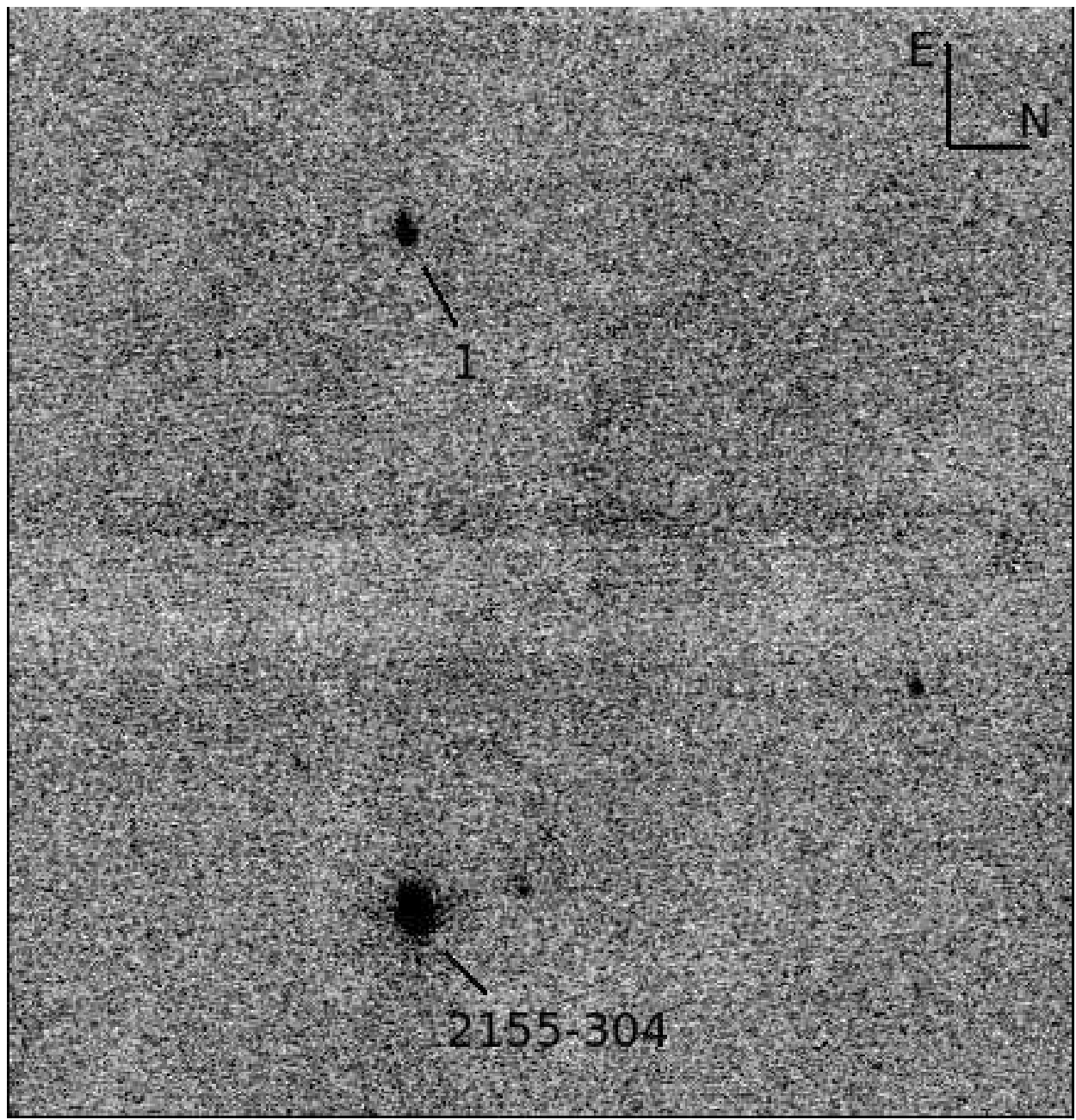} 
\caption*{B.11: PKS~2155$-$304.  }\label{fig:2155irfc}
\end{center}
\end{figure}

\begin{figure}[]
\begin{center}
\plotone{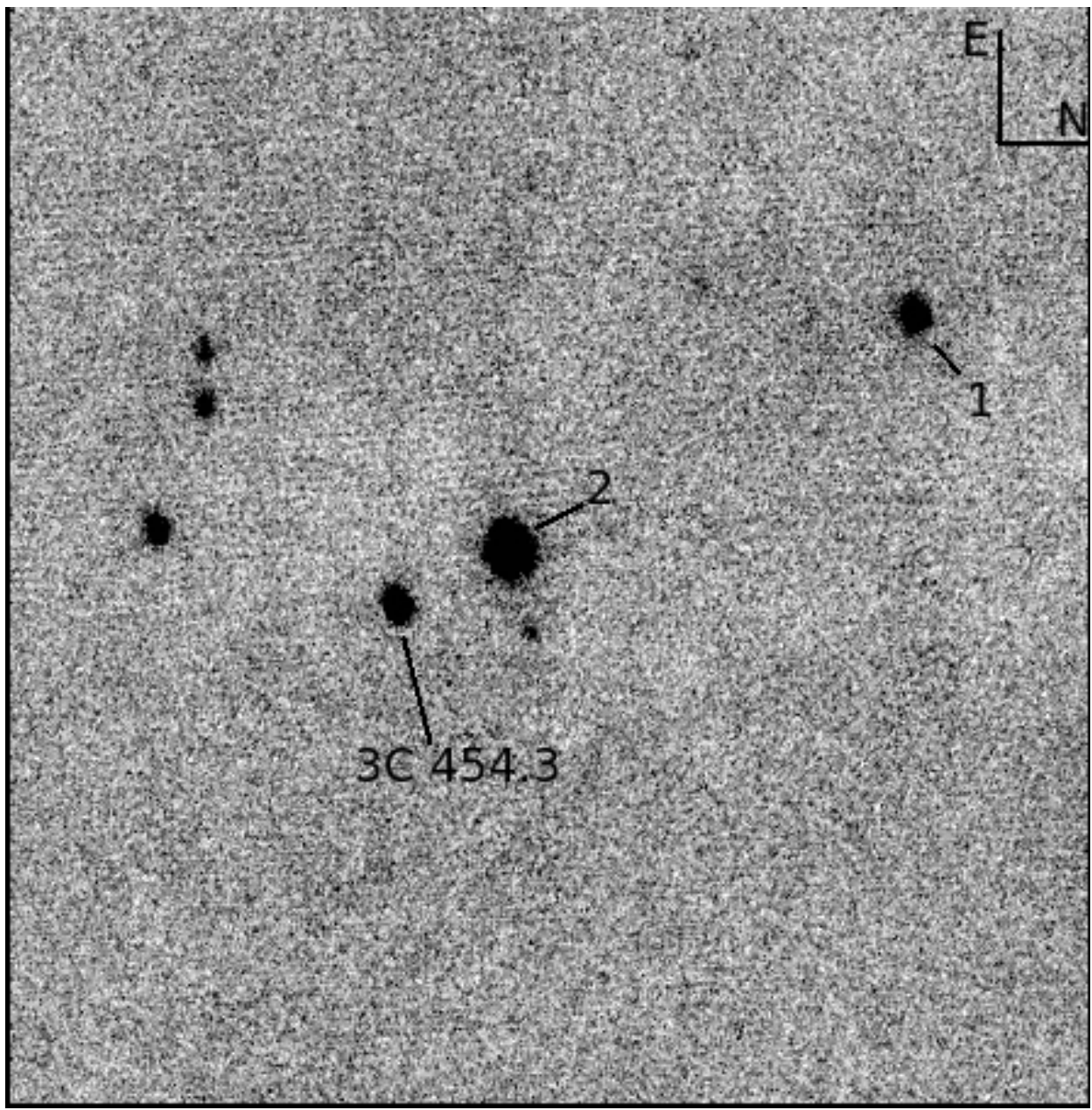} 
\caption*{B.12: 3C~454.3.  }\label{fig:3C454irfc}
\end{center}
\end{figure}

\clearpage

\section{Comparison Star Magnitudes}

Here we present calibrated magnitudes for comparison stars indicated
in the finding charts presented in Appendixes A and B used as
secondary standards for our differential photometry. 

The error in calibrating the secondary star magnitudes was found by
calculating the 1-$\sigma$ standard error of the mean over the number
of photometric nights on which data for the star were taken. Results
that were greater than $\pm$ 3$\sigma$ from the mean were rejected and
the mean and $\sigma$ were recalculated. This procedure was repeated
until no more rejections were made. The resulting 1-$\sigma$ errors
are given in the following tables. Table C.1 gives BVR magnitudes for
comparison stars used in optical photometry. Table C.2 gives JK
magnitudes for comparison stars used in the infrared photometry.

\begin{table}[htbp]
\caption{ Comparison star optical magnitudes.  Errors are 1-$\sigma$.} 
\hspace{-3mm}
  \begin{tabular}{llccc}
\hline
Source &  Star & $B$ & $V$ & $R$ \\
\hline
{\bf PKS~0208-512 } & 1 & 16.43 $\pm$ 0.03 & 15.72 $\pm$ 0.03 & 15.27 $\pm$ 0.04 \\
& 2 & 16.90 $\pm$ 0.04 & 16.24 $\pm$ 0.04 & 15.79 $\pm$ 0.03 \\
& 3 & 15.43 $\pm$ 0.04 & 14.69 $\pm$ 0.03 & 14.29 $\pm$ 0.03 \\
\hline
{\bf AO~0235+164 } & 1 & 13.61 $\pm$ 0.03 & 13.05 $\pm$ 0.03 & 12.60 $\pm$ 0.02 \\
& 2 & 13.55 $\pm$ 0.04 & 12.75 $\pm$ 0.03 & 12.26 $\pm$ 0.03 \\
& 3 & 13.72 $\pm$ 0.04 & 12.98 $\pm$ 0.03 & 12.55 $\pm$ 0.03 \\
& 4 & 14.70 $\pm$ 0.04 & 14.05 $\pm$ 0.03 & 13.64 $\pm$ 0.03 \\
& 5 & 15.74 $\pm$ 0.03 & 14.80 $\pm$ 0.03 & 14.22 $\pm$ 0.03 \\
\hline
{\bf PKS~0528+134 }& 1 & 15.90 $\pm$ 0.05 & 14.79 $\pm$ 0.03 & 13.99 $\pm$ 0.05 \\
& 2 & 16.81 $\pm$ 0.03 & 15.80 $\pm$ 0.03 & 15.09 $\pm$ 0.03 \\
\hline
{\bf OJ~287} & 1 & 15.03 $\pm$ 0.05 & 14.25 $\pm$ 0.04 & 13.69 $\pm$ 0.05 \\
& 2 & 15.18 $\pm$ 0.04 & 14.65 $\pm$ 0.04 & 14.26 $\pm$ 0.04 \\
& 3 & 15.53 $\pm$ 0.05 & 14.97 $\pm$ 0.04 & 14.55 $\pm$ 0.04 \\
& 4 & 16.67 $\pm$ 0.05 & 15.94 $\pm$ 0.04 & 15.43 $\pm$ 0.04 \\
\hline
{\bf 3C~273} & 1 & 14.14 $\pm$ 0.04 & 13.53 $\pm$ 0.04 & 13.14 $\pm$ 0.03 \\
\hline
{\bf 3C~279} & 1 & 16.76 $\pm$ 0.04 & 15.92 $\pm$ 0.03 & 15.35 $\pm$ 0.03 \\
\hline
{\bf PKS~1406-076 } & 1 & 17.67 $\pm$ 0.05 & 16.89 $\pm$ 0.03 & 16.31 $\pm$ 0.03 \\
& 2 & 17.96 $\pm$ 0.05 & 17.23 $\pm$ 0.04 & 16.71 $\pm$ 0.03 \\
& 3 & 16.55 $\pm$ 0.05 & 15.74 $\pm$ 0.03 & 15.22 $\pm$ 0.03 \\
& 4 & 16.04 $\pm$ 0.04 & 15.40 $\pm$ 0.05 & 14.96 $\pm$ 0.03 \\
\hline
{\bf PKS~1510-089 }& 1 & 13.82 $\pm$ 0.04 & 13.29 $\pm$ 0.03 & 12.96 $\pm$ 0.02 \\
& 2 & 15.16 $\pm$ 0.05 & 14.44 $\pm$ 0.03 & 13.98 $\pm$ 0.03 \\
& 3 & 15.32 $\pm$ 0.04 & 14.66 $\pm$ 0.03 & 14.24 $\pm$ 0.03 \\
& 4 & 15.55 $\pm$ 0.04 & 14.82 $\pm$ 0.03 & 14.36 $\pm$ 0.03 \\
& 5 & 16.18 $\pm$ 0.04 & 15.23 $\pm$ 0.03 & 14.59 $\pm$ 0.02 \\
\hline
{\bf PKS~1622-297 }& 1 & 17.79 $\pm$ 0.06 & 16.66 $\pm$ 0.05 & 16.00 $\pm$ 0.03 \\
& 2 & 17.83 $\pm$ 0.06 & 16.68 $\pm$ 0.05 & 15.89 $\pm$ 0.05 \\
& 3 & 18.21 $\pm$ 0.06 & 17.06 $\pm$ 0.05 & 16.24 $\pm$ 0.04 \\
& 4 & 18.16 $\pm$ 0.06 & 17.04 $\pm$ 0.04 & 16.20 $\pm$ 0.03 \\
& 5 & 17.41 $\pm$ 0.05 & 16.59 $\pm$ 0.05 & 16.04 $\pm$ 0.03 \\
\hline
{\bf PKS~1730-130 }& 1 & 16.11 $\pm$ 0.04 & 14.50 $\pm$ 0.03 & 13.46 $\pm$ 0.03 \\
& 2 & 15.71 $\pm$ 0.05 & 14.28 $\pm$ 0.03 & 13.37 $\pm$ 0.03 \\
& 3 & 16.83 $\pm$ 0.05 & 15.36 $\pm$ 0.03 & 14.41 $\pm$ 0.03 \\
& 4 & 15.76 $\pm$ 0.05 & 14.68 $\pm$ 0.03 & 13.98 $\pm$ 0.03 \\
& 5 & 16.60 $\pm$ 0.05 & 15.05 $\pm$ 0.04 & 14.04 $\pm$ 0.02 \\
\hline
{\bf PKS~2155-304 } & 1 & 15.97 $\pm$ 0.03 & 15.33 $\pm$ 0.04 & 14.94 $\pm$ 0.03 \\
\hline
{\bf 3C~454.3} & 1 & 16.76 $\pm$ 0.05 & 15.87 $\pm$ 0.04 & 15.29 $\pm$ 0.04 \\
& 2 & 15.82 $\pm$ 0.05 & 15.17 $\pm$ 0.05 & 14.74 $\pm$ 0.03 \\
& 3 & 16.89 $\pm$ 0.05 & 15.71 $\pm$ 0.05 & 14.92 $\pm$ 0.03 \\
& 4 & 14.57 $\pm$ 0.05 & 13.65 $\pm$ 0.04 & 13.09 $\pm$ 0.03 \\
& 5 & 16.65 $\pm$ 0.05 & 15.77 $\pm$ 0.04 & 15.20 $\pm$ 0.04 \\ 
\hline
  \end{tabular}
 \label{tab:opcompstar}
\end{table}

\clearpage

\begin{table}[htbp]
\caption{Comparison star near-infrared magnitudes.  Errors are 1-$\sigma$.} 
\hspace{-3mm}
  \begin{tabular}{llcc}
\hline
Source & Star & $J$ & $K$ \\
\hline
{\bf PKS~0208-512} & 1 & 12.16 $\pm$ 0.04 & 11.81 $\pm$ 0.07 \\
\hline
{\bf AO~0235+164} & 1 & 11.96 $\pm$ 0.04 & 11.68 $\pm$ 0.06 \\
& 2 & 12.97 $\pm$ 0.04 & 12.14 $\pm$ 0.09 \\
\hline
{\bf PKS~0528+134} & 1 & 14.03 $\pm$ 0.08 &  \\
& 2 & 12.72 $\pm$ 0.06 & \\
& 3 & 12.32 $\pm$ 0.05 & \\
\hline
{\bf OJ~287} & 1 & 14.58 $\pm$ 0.080 & 13.79 $\pm$ 0.10 \\
\hline
{\bf 3C~273} & 1 & 13.64 $\pm$ 0.06  & \\
& 2 & 12.29 $\pm$ 0.05  &\\
\hline
{\bf 3C~279} & 1 & 14.41 $\pm$ 0.07 & 13.86 $\pm$ 0.12 \\
& 2 & 15.47 $\pm$ 0.14 & 13.94 $\pm$ 0.20 \\
\hline
{\bf PKS~1406-076} & 1 & 15.32 $\pm$ 0.07 & 14.84 $\pm$ 0.16 \\
& 2 & 15.77 $\pm$ 0.11 & 15.30 $\pm$ 0.47 \\
\hline
{\bf PKS~1510-089} & 1 & 15.61 $\pm$ 0.13 & 14.71 $\pm$ 0.27 \\
& 2 & 15.57 $\pm$ 0.11 & 14.44 $\pm$ 0.32 \\
& 3 & 13.49 $\pm$ 0.05 & 12.77 $\pm$ 0.12 \\
\hline
{\bf PKS~1622-297} & 1 & 14.43 $\pm$ 0.11 & 13.47 $\pm$ 0.14 \\
& 2 & 13.83 $\pm$ 0.11 & 12.69 $\pm$ 0.12 \\
& 3 & 13.53 $\pm$ 0.08 & 12.67 $\pm$ 0.11 \\
\hline
{\bf PKS~1730-130} & 1 & 14.37 $\pm$ 0.07 & 13.17 $\pm$ 0.10 \\
& 2 & 13.06 $\pm$ 0.07 & 11.96 $\pm$ 0.06 \\
& 3 & 14.60 $\pm$ 0.07 & 13.48 $\pm$ 0.10 \\
\hline
{\bf PKS~2155-304} & 1 & 14.30 $\pm$ 0.06 & 13.72 $\pm$ 0.14 \\
\hline
{\bf 3C~454.3}& 1 & 14.26 $\pm$ 0.10 & 13.49 $\pm$ 0.15 \\
& 2 & 11.88 $\pm$ 0.07 & 11.31 $\pm$ 0.61 \\
\hline
  \end{tabular}
 \label{tab:ircompstar}
\end{table}


\end{document}